\documentclass[11pt]{article} 
\usepackage[utf8]{inputenc}
\usepackage{multirow}
\usepackage{amsfonts}
\usepackage{amsmath}
\usepackage{amssymb}
\usepackage{amsthm}
\usepackage{mathtools}
\usepackage{caption}
\usepackage[dvipsnames]{xcolor}
    \definecolor{darkgreen}{rgb}{0,0.5,0}
    \definecolor{darkblue}{rgb}{0,0,0.6}
    \definecolor{purple}{rgb}{0.4,.2,0.7}
\usepackage[margin = 2.5 cm]{geometry}
\usepackage{graphicx}
\usepackage[hyperfootnotes = false, colorlinks = true, linkcolor = darkblue, citecolor = purple]{hyperref}
\usepackage{subcaption}
\usepackage{ytableau}

\usepackage{cite}

\usepackage{comment}

\usepackage{changepage}

\newenvironment{widerequation}{%
    \begin{adjustwidth}{-1.5cm}{-2cm}\begin{eqnarray*}}
{\end{eqnarray*}\end{adjustwidth}}

\newcommand{\be}{\begin{equation}}
\newcommand{\ee}{\end{equation}}
\newcommand{\bea}{\begin{eqnarray}}
\newcommand{\eea}{\end{eqnarray}}
\def\la{\label}
\def\nref#1{(\ref{#1})}
\def\half{{1 \over 2 }}

\def\tr{\mathrm{tr}}

	\newcommand{\bes}{\begin{equation} \begin{split} }	
	\newcommand{\ees}{\end{split} \end{equation} }

	\newcommand{\lp}{\left (}
	\newcommand{\rp}{\right )}
	\newcommand{\lb}{\left [}
	\newcommand{\rb}{\right ]}
	\newcommand{\RA}{\Rightarrow}

	\newcommand{\Z}{\mathbb{Z}}

	\usepackage{physics}
	
\usepackage{caption}

\usepackage{tikz-feynman}
\usetikzlibrary{patterns,calc}
\usetikzlibrary{decorations.markings,decorations.pathmorphing}
\usepackage{pgfplots}
\usepgfplotslibrary{fillbetween}
\usetikzlibrary{intersections}

\tikzset{snake it/.style={decorate, decoration=snake}}
\usepackage[dvipsnames]{xcolor}

\tikzfeynmanset{ with arrow/.style = {
   decoration={
     markings,
     mark=at position 0.5
          with {\arrow[xshift=2mm]{Stealth[width=2mm,length=3mm]}}
     },
   postaction=decorate},
   with reversed arrow/.style = {
   decoration={
     markings,
     mark=at position 0.5
          with {\arrowreversed[xshift=-2mm]{Stealth[width=2mm,length=3mm]}}
     },
   postaction=decorate}
}

    \def\la{\label}
\def\nref#1{(\ref{#1})}
\def\half{{1 \over 2 }}

\begin{document}

	\thispagestyle{empty}
    \begin{flushright}
        
    LITP-26-01
    \end{flushright}
\begin{center}
    ~\vspace{5mm}

  {\LARGE \bf {A melonic quantum mechanical model without disorder
  \\}}

   \vspace{0.5in}
     
   {\bf    Anna Biggs$^1$, Loki L.\ Lin$^2$, and Juan Maldacena$^3$
   }

    \vspace{0.5in}

  $^1$
  Jadwin Hall, Princeton University,  Princeton, NJ 08540, USA 
   \\
  $^2$
  Leinweber Institute for Theoretical Physics, University of Michigan, Ann Arbor, MI 48109, USA 
   \\ $^3$
  Institute for Advanced Study,  Princeton, NJ 08540, USA 
   \\               
    \vspace{0.5in}

    \vspace{0.5in}

\end{center}

\vspace{0.5in}

\begin{abstract}

 We consider a quantum mechanical model involving interacting fermions without disorder that has the same low energy physics as the supersymmetric SYK model. The model is $SU(2)$ invariant, and the supercharge involves the $ SU(2) $ 3j symbol. We analyze various solvable corners,    conceptually explain why it has a melonic expansion, and perform an exact diagonalization for small values of $N$.  Expanded around the states with maximal angular momentum, the model is approximated by a two dimensional CFT. The BPS states have a simple description in that regime. 

 \end{abstract}
 
\vspace{1in}

\pagebreak

\setcounter{tocdepth}{3}
{\hypersetup{linkcolor=black}\tableofcontents}

\section{Introduction and Motivation}

We consider a quantum mechanical model involving interacting fermions {\it without disorder}  which has a melonic approximation when the number of fermions is large. It also has a nearly conformal region in the infrared. 

The model involves $N = 2j+1$ complex fermions that are in a single spin $j$ representation of $SU(2)$. It has an SU(2) invariant interaction constructed from the Wigner 3j symbol, or Clebsch–Gordan coefficient, 
 of SU(2). The simplest version of the model has  ${\cal N}=2$ supersymmetry, with   supercharges of the schematic form 
\be \la{HamSche}
 Q \sim (3j~{\rm symbol}) \psi^3 , ~~~~~~Q^\dagger \sim (3j~{\rm symbol})  (\psi^\dagger)^3 ~,~~~~~~~H \equiv \{ Q , Q^\dagger \} 
 \ee 
Below, we write these quantities more explicitly. In the melonic approximation, and for $SU(2)$ singlet observables,  the theory becomes the same as the ${\cal N}=2$ supersymmetric SYK model studied in \cite{Fu:2016vas}. The model also features interesting composite operators with $SU(2)$ angular momentum, which can also be studied using melonic diagrams, as discussed below.

One motivation to study this model is that it has a melonic expansion without disorder. In that sense, it is in the same spirit as the tensor models discussed in \cite{Witten:2016iux,Klebanov:2016xxf}. We will find that this model is more amenable to exact numerical diagonalization than the tensor models. Models based on $SU(2)$ 3j symbols were previously discussed by Amit and Roginsky \cite{Amit:1979ev}, and more recently in \cite{Benedetti:2020iku}, which inspired this work. As discussed in \cite{Amit:1979ev, Benedetti:2020iku}, this appears to be a new mechanism for obtaining melonic dominance, and we wanted to explore it further. 

Recently, there have been studies on BPS chaos \cite{Chang:2024zqi,Choi:2023znd,Choi:2023vdm,Chang:2024lxt,Chen:2024oqv,Chen:2025sum}, and this model appears to be an interesting example for studying such questions. 
We have not been able to solve the BPS conditions in general, but there is a particular simple limit of large $SU(2) $ charges which becomes particularly tractable. 
Namely, the BPS states come with various values of the $SU(2) $ charges. When the angular momentum is near the maximal value, of order $J_3 \sim j^2/2$, then the whole theory simplifies and it becomes equivalent to a 1+1 dimensional CFT with two decoupled sectors, one sector contributing to the energy and the other generating the BPS states. 

The remarkable simplification of the model in this sector can be understood by viewing the model as follows. The fermion in the spin $j$ representation can result from a fermion on the sphere in the lowest Landau level. This is also sometimes called a ``fuzzy sphere''. The Hamiltonian in \nref{HamSche} is an interaction among these fermions, which is non-local on the sphere.  
The states with the largest angular momentum correspond to filling just the northern hemisphere of the sphere. There are then edge states along the equator that can be bosonized. The interactions from \nref{HamSche} raise the energies  of some modes of this boson, leaving the rest with zero energy. Though we have an explicit description for these large SU(2) angular momentum BPS states, most of the BPS states occur with relatively small values of the angular momentum, for which we do not have an explicit description, beyond a simple count.

One reason to study models with an $SU(2)$ symmetry is to explore a path towards describing the sphere directions for $AdS_2 \times S^2$. Ultimately, we will find that this model does {\it not} contain operators that can be viewed as Kaluza Klein modes. 

There is a vague similarity between the low energy properties of this model and some black hole systems in string theory, where some BPS states have a simple description as classical solutions or fluctuations of classical solutions, while  most of the zero energy states can be understood in terms of BPS black holes. 

We will comment in the conclusions about natural generalizations to models without supersymmetry or to models based on other groups.

The rest of the paper is structured as follows. In section \ref{sec:defs} we define the model, discuss its symmetries, and count states by their angular momentum and R charge. We do this both on the entire Hilbert space and separately, by computing Witten indices, on the BPS subspace.  The remaining sections are roughly organized by the different soluble regimes we consider. In section \ref{Melonic} we discuss the regime of small angular momentum and R charge, where the model has a large $N$ melonic expansion. We give an intuitive explanation for the mechanism behind melonic dominance, which involves thinking about the degeneracy of angular momentum configurations on the sphere allowed by the interaction \nref{HamSche}. In section \ref{NearConf} we focus more specifically on the regime of low energy and small charges, where the melonic expansion leads to an approximate conformal symmetry in the IR, as in $\mathcal{N} = 2$ SYK. We also discuss the dimensions of operators in nontrivial SU(2) representations which were not present in SYK. In section \ref{sec:largeU1} we discuss states in the large R charge, high energy regime. This includes, for example, the Fock vacuum and simple excitations around the Fock vacuum. In section \ref{lgspinperR} we describe a specific set of states which have the largest SU(2) spin in each fixed R charge sector. In section \ref{SU2Sec} we discuss the limit of large SU(2) charge, highlighted above, where the model becomes an effective CFT$_2$. We give an explicit description of all states, including the BPS states, in this limit. In section \ref{sec:exact}, we present various results on exact numerical diagonalization of the Hamiltonian, which will be referenced throughout the preceding sections.

\section{Definitions and Basic Properties}\label{sec:defs}

We consider a complex fermion in the spin $j$ representation of $SU(2)$ which we denote as $\psi_m$ with $\{ \psi_m , (\psi_{m'})^\dagger \} = \delta_{m,m'}$\footnote{We could also define $ \bar \psi_m = (-1)^{ j -m} \psi^{\dagger}_{-m} $, which  is an object that transforms as $\psi_m$ under $SU(2) $. Note also that the $SU(2)$ invariant metric is $\{  \bar \psi_m , \psi_{m'} \} = (-1)^{j-m} \delta_{m+m'}$. This metric and similar phase factors could be introduced in some of the formulas below in order to make them manifestly $SU(2) $ invariant. We also define $\psi^\dagger_m = (\psi_m)^\dagger$.}.   
We denote the $SU(2) $ 3j symbol as 
\begin{align} \la{3jDefi}
    C^{j}_{m_1 m_2 m_3} \equiv \begin{pmatrix}
        j  & j & j \\ m_1 & m_2 & m_3
    \end{pmatrix}\equiv \frac{(-1)^{j_1-j_2-m_3}}{\sqrt{2j_3+1}}\langle j m_1 ~j m_2 | j ~(-m_3)\rangle 
\end{align}
where $\langle j m_1 ~j m_2 | j ~(-m_3)\rangle$ is the Clebsh-Gordan coefficient.  
We then define the supercharge and the Hamiltonian 
\begin{align}
   \la{SuperCharge} Q &=\frac{1}{3!} \sqrt{2 \mathsf{J}N } \sum_{-j \leq m_1, m_2, m_3 \leq j}C^{j}_{m_1 m_2 m_3}\psi_{m_1}\psi_{m_2}\psi_{m_3}\\
    Q^{\dagger} &=-\frac{1}{3!} \sqrt{2 \mathsf{J} N}\sum_{-j \leq m_1, m_2, m_3 \leq j}C^{j}_{m_1 m_2 m_3}\psi^{\dagger}_{m_1}\psi^{\dagger}_{m_2}\psi^{\dagger}_{m_3}%
    \\
    H& = \{Q, Q^{\dagger}\}  \la{HamFi}
\end{align}
Note that $C^{j}_{m_1 m_2 m_3}$ is symmetric in the indices for even $j$ and antisymmetric for odd $j$, and it vanishes for half-integer $j$. Since the fermions in \nref{SuperCharge} are antisymmetric, we restrict $j$ to be an odd integer.  So the symmetry of our problem is really $SO(3)$ rather than $SU(2)$. 
We have also used that the $C$s are real and that $C^{j}_{-m_1, -m_2, -m_3} = - C^{j}_{m_1 m_2 m_3}$. Note also that $\mathsf{J}$ is a constant with dimensions of energy.

It can be convenient to view the fermions $\psi_m$ as the modes of a 2+1 dimensional fermion moving on $S^2\times$(time) with a magnetic field on $S^2$, with $N=2j+1$ units of magnetic flux. From that point of view the the interaction in 
\nref{HamFi} is not local on the sphere. However, it will still be useful to consider this perspective.

The supercharge acts on the fermion as
\begin{align}
    \{ Q, \psi_{n}^{\dagger} \} =  \sqrt{\frac{ \mathsf{J}N}{2}}\sum_{-j \leq m_2, m_3 \leq j} C_{n m_2 m_3}\psi_{m_2}\psi_{m_3}~~~~~~~~\{Q, \psi_{n}\} = 0
\end{align}

We introduce complex (composite) bosonic fields, $b_{n} \equiv \{Q^{\dagger}, \psi_{n}\}$ and $b^{\dagger} \equiv \{Q, \psi_{n}^{\dagger}\}$ to linearize the action of the supersymmetry transformations.

We can also write a Lagrangian of the form 
\begin{align}
    L =\sum_{-j \leq m_1\leq j} \lb  i \psi_{m_1}^{\dagger}\partial_t\psi_{m_1} - b_{m_1}^{\dagger}b_{m_1} +\sqrt{\frac{ \mathsf{J} N}{2}} \sum_{ -j \leq m_2, m_3\leq j} \lp C_{m_1 m_2 m_3}b_{m_1}\psi_{m_2}\psi_{m_3} -C_{m_1 m_2 m_3}b^{\dagger}_{m_1}\psi^{\dagger}_{m_2}\psi^{\dagger}_{m_3}\rp \rb 
\end{align}

We define the fermion number operator $N_\psi$ as
\begin{align}\la{Npsi}
    N_\psi=\sum_{m=-j}^j\psi^\dagger_m\psi_m-\frac{2j+1}{2}
\end{align}
The constant shift makes $N_{\psi}$ charge conjugation symmetric. More precisely, the model has charge conjugation symmetry, under which $\psi_{m} \to (-1)^{m}\psi_{-m}^{\dagger}$ and $N_{\psi}$ transforms as $N_{\psi} \to - N_{\psi}$.  The definition \nref{Npsi} means that the Fock vacuum, defined by $\psi_m |0 \rangle =0$ for all $m$,  has fermion number $-(j+\frac{1}{2})$.
The theory has a  U(1)$_R$ charge which is simply related to fermion number by
\begin{align}
    R=\frac{N_\psi}{3}\;.
\end{align}
with the proportionality constant chosen so that  the supercharges $Q$ and $Q^\dagger$ have $R$ charges $\pm 1$. 

The SU(2) currents are
\begin{align}\la{SUTwoCharges}
    &J_3=\sum_{m=-j}^jm\,\psi^\dagger_m\psi_m\;,\\
    &J_+=\sum_{m=-j}^{j-1}\sqrt{(j-m)(j+m+1)}\psi_{m+1}^\dagger\psi_m\;, \la{Jp}\\
    &J_-=\sum_{m=-j+1}^{j}\sqrt{(j+m)(j-m+1)}\psi_{m-1}^\dagger\psi_m\;.\la{Jm}
\end{align}
These are normalized such that 
\begin{align} \la{J3ang}
    [J_3,\psi_m]=-m\,\psi_m\;,\qquad [J_3,\psi^\dagger_m]=m\,\psi^\dagger_m\;.
\end{align}

The Hamiltonian can be normal-ordered into the following form:
\begin{align} \la{HamNor}
    H=\frac{\mathsf{J}}{3}\Big[(2j+1)-3(N_\psi+j+\tfrac{1}{2})+3\sum_m O^\dagger_{j,m}O_{j,m}\Big]
\end{align}
where 
\begin{align}
    O_{\ell,m}=\sqrt{2(2\ell+1)}\sum_{-j\leq m_1<m_2\leq j}\begin{pmatrix}
        j & j & \ell\\
        m_1 & m_2 & -m
    \end{pmatrix}\psi_{m_1}\psi_{m_2}\;.
\end{align}
The simplicity of \nref{HamNor} is due to the presence of a single term (the one with spin $j$) in the Haldane pseudo-potential \cite{Haldane:1983xm}.

\subsection{Hilbert space decomposition by spin and fermion number}\label{sec:hilbertspace}

The $SU(2)$ decomposition of the Hilbert space is given by the partition function
\begin{align}\la{J3count}
    \tr \lb e^{i \theta J_{3}} \rb =  \prod_{m = -j}^{j}(1 + e^{i \theta m}) = \sum_{n} d_{n}e^{i\theta n} 
\end{align}
where $d_{n}$ is the number of states with $J_3$ eigenvalue $n$. $d_n$ is computed by the fourier transform
\begin{align}\la{dnexact}
    d_{n} = \frac{1}{2\pi}\int_{-\pi}^{\pi}d\theta ~ e^{-i n \theta}\prod_{m=-j}^{j}(1+e^{i \theta m})
\end{align}
To find the number of spin $\ell$ representations $D_{\ell}$,  recall that states in a spin $\ell$ multiplet have $J_3$ eigenvalues that run between $-\ell $ and $\ell$. So the number of spin $\ell$ representations in the Hilbert space is given by the difference in the number of states with $J_3$ eigenvalue $\ell$ and $\ell+1$. 
\begin{align}\la{Drel}
    D_{\ell} = d_{\ell}-d_{\ell+1} 
\end{align}
The largest spin representation in the Hilbert space is
\begin{align}
    \ell_{max} = \sum_{m=1}^{j} m = \frac{j(j+1)}{2}
\end{align}
We can see from \nref{J3count} that there will always be 2 such irreps. These two states are in fact always BPS, as we will discuss further in Section \ref{SU2Sec}. 

The integral formula \nref{J3count} is exact in $j$. It can be approximated at large $j$ by a saddle point approximation,  
see Appendices \ref{SPapp} and \ref{SU2plts}.  

  For $n$ small compared to $\ell_{max}$, $d_{n}$ takes the form of a gaussian with standard deviation $\sigma = \sqrt{j^{3}/6}$. 
\begin{align}\la{gauss}
    d_{n} &\approx 2^{2j+1}\frac{1}{ \sigma\sqrt{2 \pi}}\exp(-\frac{1}{2}\frac{n^{2}}{\sigma^{2}})~~~~~~~~~ \sigma = \sqrt{\frac{j^{3}}{6}}~~~~~~ |n| \ll \ell_{max}
\end{align}
For larger values of $n$, $d_{n}$ has a Cardy-like expression.
\begin{align}
    d_{n} &\approx \frac{1}{2}\frac{1}{6^{1/4}}\hat N^{-3/4}\exp(2 \pi \sqrt{\frac{\hat N}{6}})~~, ~~~~~~\hat N = \ell_{max} - |n|~ , ~~~~~ 1 \ll \hat N \ll \ell_{max}\la{cardyd}
\end{align}
It is noteworthy that in the regime where $\ell_{max}-|n| \ll \ell_{max}$, $d_{n}$ resembles the density of states of a 2d CFT with central charge $c = 1$. This will be explained further in section \ref{SU2Sec}.

Mostly, we will be interested in counting states by their angular momentum quantum numbers. Of course, we could also consider the generating function for angular momentum and fermion number together, which is
\begin{align}
    \tr \lb e^{i \theta J_3} y^{N_{\psi}}\rb =\prod_{m=-j}^{j}\lp \frac{1}{\sqrt{y} } + \sqrt{y} e^{i\theta m}\rp  
\end{align}

 \subsection{Witten Index}\label{sec:witten}
 
As in $\mathcal{N} = 2$ SYK, there is a $\Z_{3}$ subgroup of the $U(1)_{R}$ symmetry that commutes with the supercharges. This allows us to compute a nonzero index graded by $\Z_{3}$ charge, which is the same as the one computed in \cite{Fu:2016vas}.
\begin{align}
    W_{r} = \tr \lb (-1)^{F}e^{2 \pi i r R}\rb =\omega^{-{( 2j +1 )\over 2 }} \lp 1-\omega^{r}\rp^{2j+1} ~~~~~~~~~\omega = e^{2 \pi i/3}~~~~~r = 1,2\la{WI1}
    \end{align}
By virtue of the model's $SU(2)$ symmetry, we can also insert a chemical potential for $J_{3}$.
\begin{align}\la{WI2}
     W_{r}(\theta) = \tr \lb (-1)^{F}e^{2 \pi i r R} e^{i \theta J_{3}}\rb 
\end{align}
Exact diagonalization for small values of $j$ suggests that the BPS states are concentrated at small values of the R charge. In particular, we find numerically\footnote{At $j = 1,3,5$ all BPS states have  $R = \pm 1/6$. At $j = 7$ we find that all BPS states have $R = \pm 1/6$ except for two $\ell = 7$ multiplets with $R = 1/2$ and $R = -1/2$, respectively. At $j=9$ the exceptions are four $\ell=9$ multiplets with $R=\pm1/2$ and four $\ell=0$ states with $R=\pm5/6$. 
At $j=11$ all BPS states have $R=\pm1/6$. 
} that the number of BPS states with $|R| = 1/6$ is exponentially large in $j$, while the number of states with larger values of $|R|$ is $O(1)$. 
If we {\it assume} that all the BPS states have $R = \pm 1/6$, then we can solve for their degeneracy $D^{BPS}(j, R)$ using \nref{WI1}, and we find
\begin{align}\la{DBPSR}
    D^{BPS}\lp j, \frac{1}{6}\rp = D^{BPS}\lp j,  -\frac{1}{6}\rp  = 3^{j}
\end{align}

A similar approach using \nref{WI2} allows us to compute the degeneracy $d_{n}^{BPS}$ of BPS states with $J_{3}$ eigenvalue $n$ (and by extension the number $D_{\ell}^{BPS}$ in a given spin $\ell$ representation.)
\begin{align}\la{DBPSL}
    d_{n}^{BPS} &= \frac{1}{2 \pi}\int_{-\pi}^{\pi}d\theta e^{-i n \theta}Z_{BPS}(\theta)~~~~~~~~~~~Z_{BPS}(\theta) = 2e^{-i \theta \frac{j(j+1)}{2}}\prod_{m=1}^{j}(1+ e^{i \theta m}+e^{2 i \theta m} )
\end{align}
As in the previous section, these integral formulas can be approximated by their saddle point in the large $j$ limit, see appendix \ref{SPapp}. 
\begin{align}\la{BPSspa}
    d_{n}^{BPS}\approx  \begin{cases} \frac{2 \times 3^{j}}{\sigma \sqrt{2 \pi}}\exp \lp -\frac{1}{2}\frac{n^{2}}{\sigma^{2}} \rp ~,~~~~ \sigma = \sqrt{\frac{2j^{3}}{9}}~~~~~~~~~~~~~~~~~~~~~~~~~~~~~~~|n| \ll \ell_{max}\\
\frac{1}{3 \sqrt{3}  \hat N ^{3/4}} \exp \lp 2 \pi \sqrt{\frac{\hat N }{9}}\rp ~, ~~~{\hat N } =\ell_{max} - |n| ~,   ~~~~~~~~~ 1 \ll \hat  N \ll \ell_{max} 
    \end{cases}
\end{align}
Note that \nref{DBPSR} and \nref{DBPSL} are the degeneracies we get when we ignore the $O(1)$ number of BPS states with R charge larger than $|R| = 1/6$, which are of course negligible in the large $j$ limit. In the future, it would be nice to compute a refined index sensitive to the BPS states with exceptional R charge.

\section{Melonic approximation -- small charges $(R\ll 1, J \ll 1)$}
\la{Melonic}

In this section, we explain why the theory has a melonic approximation. Let us consider the case that we have no chemical potential for the $SU(2)$ symmetry. Similarly, let us also assume that we have zero chemical potential for the $U(1)_R$ symmetry. Then we expect that only $SU(2)$   invariant correlators  are non-vanishing. We can write these correlators as  
\be 
G(t,t') = { 1 \over N } \sum_{m =-j}^j\langle \psi^\dagger_m(t) \psi_m(t') \rangle ~,~~~~~ N= 2 j+1 ~,~~~~~~~~~\langle \psi^\dagger_m(t) \psi_{m'}(t') \rangle =  G(t,t') \delta_{m,m'}
\ee 
and similarly for $\langle b^\dagger_m  b_{m' }\rangle $.

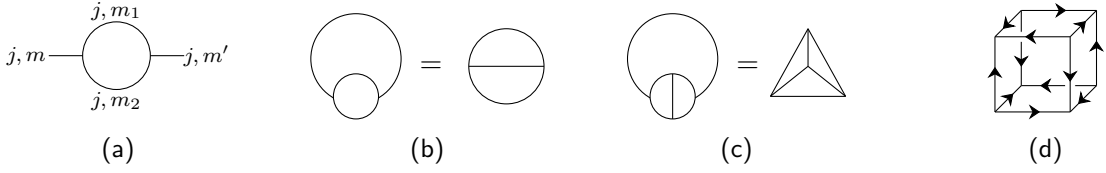
\begin{figure}
    \centering
    \begin{subfigure}[b]{0.24\textwidth}
        \centering
        \begin{tikzpicture}
    \draw(0,0) circle (.45);
    \draw (-.9,0) -- (-.45,0);
    \node at (-1.2, 0) {\scriptsize $j,m$};
    \draw (.9,0) -- (.45,0);
    \node at (1.2, 0) {\scriptsize $j,m'$};
    \node at (0,.6) {\scriptsize $j,m_1$};
    \node at (0,-.6) {\scriptsize $j,m_2$};
\end{tikzpicture}
    \caption{}
    \end{subfigure}
    \begin{subfigure}[b]{0.24\textwidth}
        \centering
        \begin{tikzpicture}
            \draw (0,.5) circle (.6);
            \draw[fill=white] (0,0) circle (0.3);
            \node at (1,.4) {$=$};
            \draw (2,0.4) circle (.5);
            \draw (1.5,.4) -- (2.5, .4);
        \end{tikzpicture}
        \caption{}
    \end{subfigure}
    \begin{subfigure}[b]{0.24\textwidth}
        \centering
        \begin{tikzpicture}
            \draw (0,.5) circle (.6);
            \draw[fill=white] (0,0) circle (0.3);
            \draw (0,-.3) -- (0,.3);
            \node at (1,.4) {$=$};
            \draw (1.3,0) -- (2.3,0);
            \draw (1.3,0) -- (1.8, .9);
            \draw (2.3,0) -- (1.8, .9);
            \draw (1.3,0) -- (1.8, .4);
            \draw (2.3,0) -- (1.8, .4);
            \draw (1.8,.4) -- (1.8, .9);
        \end{tikzpicture}
        \caption{}
    \end{subfigure}
    \begin{subfigure}[b]{0.24\textwidth}
        \centering
        \begin{tikzpicture}
\coordinate (A) at (0,0);
\coordinate (B) at (1,0);
\coordinate (C) at (1,1);
\coordinate (D) at (0,1);

\coordinate (E) at (0.35,0.35);
\coordinate (F) at (1.35,0.35);
\coordinate (G) at (1.35,1.35);
\coordinate (H) at (0.35,1.35);

\draw[decoration={markings, mark=at position 0.75 with {\arrow{Stealth[width=2mm,length=1.5mm]}}},
        postaction={decorate}] (F) -- (E);
\draw[decoration={markings, mark=at position 0.6 with {\arrow{Stealth[width=2mm,length=1.5mm]}}},
        postaction={decorate}] (F) -- (G);
\draw[decoration={markings, mark=at position 0.6 with {\arrow{Stealth[width=2mm,length=1.5mm]}}},
        postaction={decorate}] (H) -- (G);
\draw[decoration={markings, mark=at position 0.75 with {\arrow{Stealth[width=2mm,length=1.5mm]}}},
        postaction={decorate}] (H) -- (E);

\draw[white,fill=white] (0.35,1) circle (0.06);
\draw[white,fill=white] (1,.35) circle (0.06);

\draw[decoration={markings, mark=at position 0.6 with {\arrow{Stealth[width=2mm,length=1.5mm]}}},
        postaction={decorate}] (A) -- (B);
\draw[decoration={markings, mark=at position 0.6 with {\arrow{Stealth[width=2mm,length=1.5mm]}}},
        postaction={decorate}] (C) -- (B);
\draw[decoration={markings, mark=at position 0.6 with {\arrow{Stealth[width=2mm,length=1.5mm]}}},
        postaction={decorate}] (C) -- (D);
\draw[decoration={markings, mark=at position 0.6 with {\arrow{Stealth[width=2mm,length=1.5mm]}}},
        postaction={decorate}] (A) -- (D);

\draw[decoration={markings, mark=at position 0.7 with {\arrow{Stealth[width=2mm,length=1.5mm]}}},
        postaction={decorate}] (A) -- (E);
\draw[decoration={markings, mark=at position 0.7 with {\arrow{Stealth[width=2mm,length=1.5mm]}}},
        postaction={decorate}] (F) -- (B);
\draw[decoration={markings, mark=at position 0.6 with {\arrow{Stealth[width=2mm,length=1.5mm]}}},
        postaction={decorate}] (C) -- (G);
\draw[decoration={markings, mark=at position 0.7 with {\arrow{Stealth[width=2mm,length=1.5mm]}}},
        postaction={decorate}] (H) -- (D);
\end{tikzpicture}
        \caption{}
        \label{fig:6j}
    \end{subfigure}
    \caption{(a) Simplest melon diagram. (b) Melon vacuum diagram.  (c) Simplest non-melonic vacuum diagram. (d) Simplest non-melonic diagram that appears in our theory, which involves oriented lines (we have not distinguished the bosonic vs fermionic lines). It gives a correction of order $(\log{j})/j$ and is proportional to the 12j symbol of the second kind. 
   }
    \label{Melons}
\end{figure}
Let us first look at the simplest melonic diagram which is a correction to the propagator, see figure \ref{Melons}(a). Let us focus on the part of the diagram that involves the $SU(2)$ algebra. This involves a particular case ($\ell = \ell' = j$) of the identity\footnote{An equivalent, but manifestly $SU(2) $ invariant way to write this equation, for $\ell = \ell' =j$, is 
$$ \sum_{m_1,m_2 } (-1)^{ 2 j -m_1 - m_2} C^j_{m,m_1,m_2 } C^j_{m',-m_1,-m_2 } =  (-1)^{j-m} { 1 \over N } \delta_{m+m' } $$.
}  
\be \la{BubId}
\sum_{m_1,m_2 } \begin{pmatrix}
        \ell & j & j\\
        m & m_1 & m_2  
    \end{pmatrix} \begin{pmatrix}
        \ell' & j & j\\
         m' & m_1 & m_2  
    \end{pmatrix} = { 1 \over 2 \ell +1  } \delta_{\ell, \ell'} \delta_{m,m' } 
\ee  
 which follows from the orthogonality relations for the 3j symbols \nref{3jDefi}. 
Remembering that each three point vertex has a factor of $\sqrt{N } \sqrt{\mathsf{J}}$, see \nref{SuperCharge}, we find that the bubble simply gives a factor of ${\mathsf{J}}$,  independent of $N$. Therefore, the single melon diagram of figure \ref{Melons}(a) gives us the same result as we would have obtained in SYK after averaging over couplings.  Note that this implies that the same is true for all other melon diagrams, since melon diagrams are obtained by substituting the simplest melon in all lines of a melon diagram. 

In conclusion, if we restrict to melon diagrams we get the same expressions as for the ${\cal N}=2$ supersymmetric SYK model in \cite{Fu:2016vas}\footnote{We have chosen conventions so that the final Dyson-Schwinger equations are the same as in \cite{Fu:2016vas}, where $\mathsf{J}_{\rm here} = J_{\rm there}$.}. In other words, the melonic  Dyson-Schwinger equations are exactly the same as the ones for the ${\cal N}=2$ SYK model in  \cite{Fu:2016vas}.

\begin{figure}[h]
    \begin{center}
    \includegraphics[scale=.5]{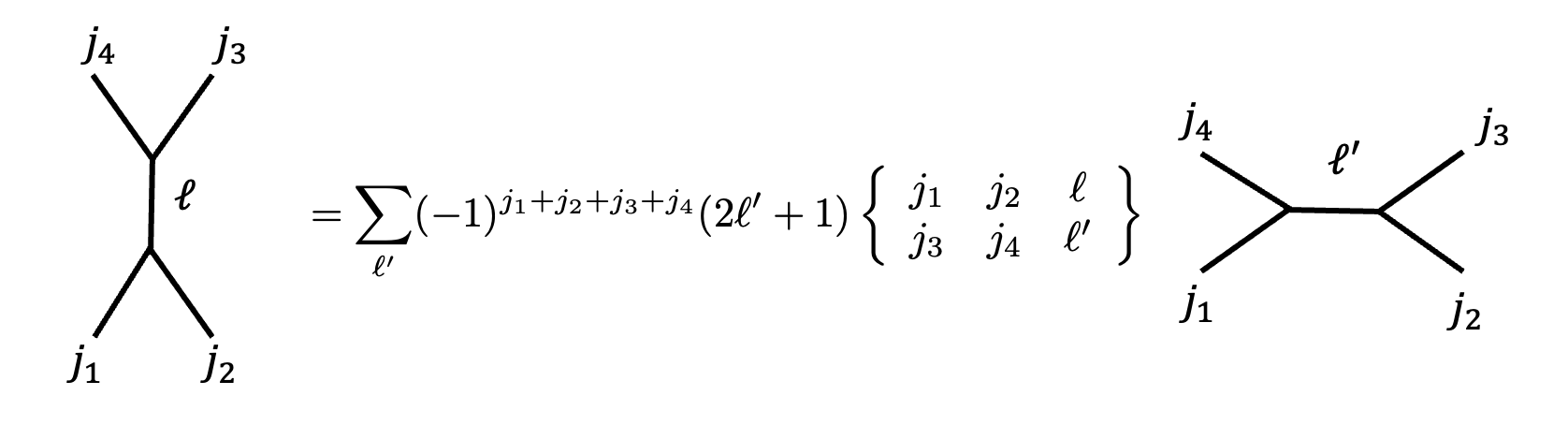}
    \end{center}
    \vspace{-7mm}
    \caption{The 6j symbol appears when we use a ``crossing relation'' to rewrite the contractions of 3j symbols that appear in these diagrams. In other words, this diagram should be read as a relation between two sums that are quadratic in the 3j symbols.  
    }
    \label{Crossing}
\end{figure}

\begin{figure}[h]
\begin{center}
    \includegraphics[width=.6\linewidth]{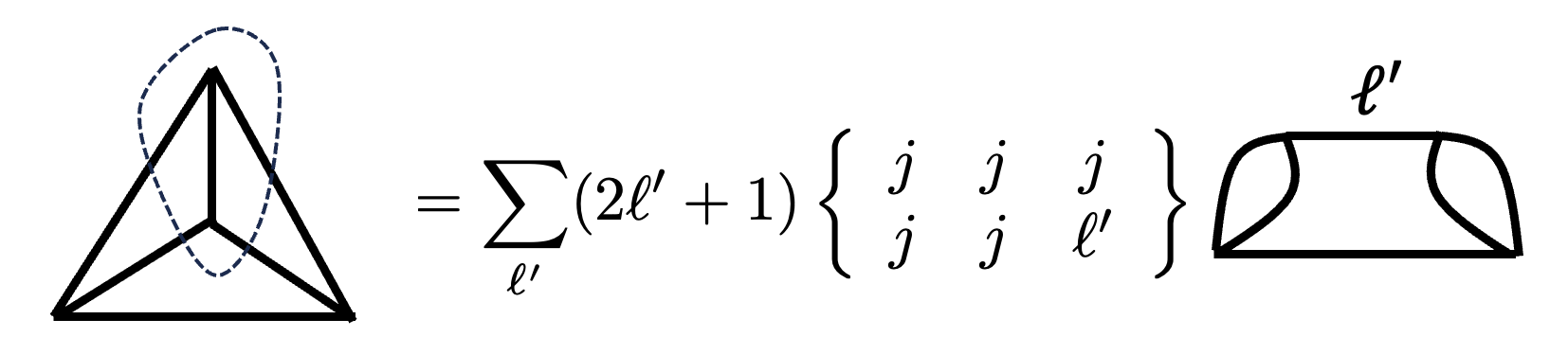}
\end{center}  
\vspace{-5mm}
\caption{We use the crossing equation of figure \ref{Crossing} to simplify the tetrahedron diagram. We apply crossing to the subdiagram inside the dotted-lined circle. Then the bubble identity \nref{BubId} implies that $\ell' =j$ and leads to a final expression involving a 6j symbol with all entries equal to $j$.  All lines without a label have angular momentum $j$.  }
    \label{Tetrahedron}
\end{figure}
However, we should still worry about the non-melonic diagrams and show that they are subleading to the melons. 
A simple vacuum non-melonic diagram, a tetrahedron, is displayed in figure \ref{Melons}(c). This diagram does {\it not} actually appear in our theory because the our lines are oriented and each vertex has either all incoming or all outgoing lines. However, it is a good diagram to practice the SU(2) algebra. Therefore,  let us evaluate the SU(2) part of the algebra for such a diagram. To achieve this, we can use the properties of the 6j symbol. Recall that a 6j symbol is defined in terms of a change of basis for the expansion of wavefunctions in the product of three angular momentum representations. We can think of them as a crossing equation of the form displayed in figure \ref{Crossing}\footnote{To be more precise, the identity in figure \nref{Crossing} is \begin{align*}\sum_{\mu}(-1)^{\ell-\mu}&\begin{pmatrix}
        j_1 & j_2 & \ell \\ m_1 & m_2 & -\mu
    \end{pmatrix}\begin{pmatrix}
         \ell & j_3 & j_4 \\ \mu & m_3 & m_4
    \end{pmatrix}\\
  &= \sum_{\ell'}\sum_{\mu'}(-1)^{j_1 + j_2 + j_3 + j_4}(2 \ell'+1)\begin{Bmatrix}
        j_1 & j_2 & \ell \\ j_3 & j_4 & \ell'
    \end{Bmatrix}(-1)^{\ell'-\mu'}\begin{pmatrix}
        j_2 & j_3 & \ell'\\ m_2 & m_3 & -\mu'
    \end{pmatrix}\begin{pmatrix}
        j_1 & \ell' & j_4 \\ m_1 & \mu' & m_4
    \end{pmatrix}\end{align*}}. 
We can apply this equation to evaluate the tetrahedron diagram as shown in figure \ref{Tetrahedron} diagram. After using the bubble identity \nref{BubId}, we find that $\ell'=j$ in figure 
\nref{Tetrahedron}, so that we get a single 6j symbol. 
 We now use the large $j$ asymptotic expansion\footnote{The asymptotic expression for the 6j symbol has a well-known geometric interpretation. The phase of \nref{6jasym} is the Regge action of 3d euclidean gravity evaluated on a single tetrahedron with sides of length $j+1/2$. The Regge action is a certain discretization of the Einstein Hilbert action, and the cosine  comes from the sum of two saddles with opposite orientation \cite{PonzanoRegge , Roberts:1998zka , ReggeCalc}. For a bibliography on Regge calculus, see  \cite{Williams:1991cd}.} \cite{PonzanoRegge, Roberts:1998zka}
\begin{align}\la{6jasym}
  N^2  \left\{ \begin{array}{ccc} j & j & j\cr  j & j & j \end{array} \right\}  \sim \frac{N^2}{2^{1/4}\sqrt{\pi}j^{3/2}}\cos\lp 6\lp j+\frac{1}{2}\rp \arccos(\frac{1}{3})+3\pi/4\rp \propto \sqrt{j} \propto j/\sqrt{j}
\end{align}

In conclusion, the tetrahedron diagram is suppressed by a factor of $1/\sqrt{j}$ relative to the melonic vacuum diagram in figure \ref{Melons}(b). It turns out that other non melonic diagrams are similarly suppressed, though there is no proof \cite{Amit:1979ev, Benedetti:2020iku}.

The first non-melonic diagram that can actually appear in our model is the 6-vertex diagram 
\begin{align}\la{9jdiag}
    \vcenter{\hbox{\begin{tikzpicture}
        \draw[white,decoration={markings, mark=at position 1 with {\arrow{Stealth[width=2mm,length=1.5mm,black]}}},
        postaction={decorate}] (-0.1,0.9) -- (.1,0.9);
        \draw (0,.3) circle (.6);
        \draw[white,fill=white] (0,0) circle (0.5);
        \draw[decoration={markings, mark=at position 0.8 with {\arrow{Stealth[width=2mm,length=1.5mm]}}},
        postaction={decorate}] (.354,0.354) to (-0.354,-0.354);
        \draw[white,fill=white] (0,0) circle (0.05);
        \draw[decoration={markings, mark=at position 0.8 with {\arrow{Stealth[width=2mm,length=1.5mm]}}},
        postaction={decorate}] (.354,-0.354) to (-0.354,0.354);
        \draw[white,decoration={markings, mark=at position 1 with {\arrow{Stealth[width=2mm,length=1.5mm,black]}}},
        postaction={decorate}] (0.1,0.5) -- (-.1,0.5);
        \draw[white,decoration={markings, mark=at position 1 with {\arrow{Stealth[width=2mm,length=1.5mm,black]}}},
        postaction={decorate}] (0.485,0.15) -- (0.49,0.14);
        \draw[white,decoration={markings, mark=at position 1 with {\arrow{Stealth[width=2mm,length=1.5mm,black]}}},
        postaction={decorate}] (0.485,-0.15) -- (0.49,-0.14);
        \draw[white,decoration={markings, mark=at position 1 with {\arrow{Stealth[width=2mm,length=1.5mm,black]}}},
        postaction={decorate}] (-0.42,0.27) -- (-0.415,0.28);
        \draw[white,decoration={markings, mark=at position 1 with {\arrow{Stealth[width=2mm,length=1.5mm,black]}}},
        postaction={decorate}] (-0.42,-0.27) -- (-0.415,-0.28);
        \draw[white,decoration={markings, mark=at position 1 with {\arrow{Stealth[width=2mm,length=1.5mm,black]}}},
        postaction={decorate}] (0.1,-0.5) -- (-.1,-0.5);
        \draw (0,0) circle (0.5);
    \end{tikzpicture}}}\;= N^3 \left\{ \begin{array}{ccc} j & j & j\cr  j & j & j \cr j & j & j\end{array} \right\}\;.
\end{align}
 It is proportional to the 9j symbol with all arguments equal to $j$. However, the 9j symbol satisfies the (anti)symmetry property under odd permutations of its columns \cite{PhysRev.93.318},
\begin{align}
    \left\{ \begin{array}{ccc} j_1 & j_2 & j_3\cr  j_4 & j_5 & j_6 \cr j_7 & j_8 & j_9\end{array} \right\}=(-1)^{\sum_{i=1}^9j_i}\times\left\{ \begin{array}{ccc} j_2 & j_1 & j_3\cr  j_5 & j_4 & j_6 \cr j_8 & j_7 & j_9\end{array} \right\}
\end{align}
so that when all $j_i=j$ odd, the symbol vanishes identically. We explain this more explicitly in appendix \ref{9jZero}.

The next non-melonic diagram that we can consider is the 8-vertex (or cube) diagram of Figure \ref{Melons}(d), corresponding to the 12j symbol of the second kind.\footnote{The diagram corresponding to the 12j symbol of the first kind, like the tetrahedral diagram, is not compatible with our oriented lines.} 
The asymptotics of this symbol for all equal $j$ were computed in \cite{Garoufalidis:2009vi}\footnote{ The normalization of the 12j symbol used in \cite{Garoufalidis:2009vi} differs from the one here; see appendix \ref{12jnorm}.}
\begin{equation}
\begin{split}\la{12jasmtot}
N^4 \left\{ \begin{array}{cccc} j & j & j & j\cr  j & j & j & j\cr j & j & j & j\end{array} \right\}_{(II)} & \sim N^4\frac{\log(2^{7/4}3j) + \gamma}{3\pi^{2}j^{4}}+ \frac{N^4}{2^{\frac{13}{4}}\pi^{\frac{3}{2}}j^{\frac{9}{2}}}\cos(24\lp j+\frac{1}{2}\rp \arccos(\frac{1}{\sqrt{3}})+\frac{\pi}{4})+\mathcal{O}(j^{-5})\\
&\propto j \times \lp \frac{\log{j}}{j} + \frac{\#}{j} + \frac{\#'}{j^{3/2}}\rp 
\end{split}
  \end{equation}

The conclusion is that the first non-melonic correction is suppressed by an extra factor of $(\log j)/j$ relative to the melonic diagrams. We suspect that higher order correction will lead to an expansion in further powers of $1/\sqrt{j}$ (with perhaps some logarithms). 

\subsection{A simple explanation for melonic dominance  }

\la{ArrowsSec}

We would now like to give another, more intuitive, argument for the suppression of non-melon diagrams. 
This involves a couple of observations. The first is that a fermion in the spin $j$ representation of $SU(2) $ is a familiar object when we analyze fermions moving on a sphere in the presence of a magnetic field, and we restrict to the lowest Landau level. In other words, a lowest Landau level fermion on a sphere, with $N=2j+1$ flux quanta for the magnetic field, transforms in the spin $j$ representation of $SU(2)$\footnote{This is true for a fermion with spin. For a spinless fermion  we have $N=2j$. We ignore this small difference since we focus on $j\gg 1$.}.  For large $j$, the sphere is very big,  and we can consider wavefunctions that are localized within a flux quantum on the sphere. One such wavefunction sitting at the north pole would have angular momentum $+j$ in the $J_3$ direction, or $m=+j$. One at the south pole would have angular momentum $J_3 = - j$. And one in some direction $\vec n$ would have momentum $\vec J = j \vec n$. So for large $j$ we can think of the fermions as carying a vector $\vec n$ which indicates the direction of the angular momentum. Of course, this is not a basis of the Hilbert space, but we could say that it is close to a basis if we pick a set of $N$ vectors $\vec n$ that are roughly uniformly distributed on the $S^2$. 
Of course, we need not appeal to the quantum Hall picture, since all we are saying is that a representation with high angular momentum is similar to a classical spin that points in a definite direction. However, the quantum Hall picture will also be useful for us later. 

\begin{figure}[h]
   \begin{center}
    \begin{subfigure}[b]{0.24\textwidth}
        \centering
\begin{tikzpicture}
    \draw (0.05,0.3) to[bend left=60] (0.26,-.1);
    \node at (0.42, 0.17) {\tiny $\frac{2\pi}{3}$};
    \draw (-0.05,0.25) to[bend right=60] (-.24,-.07);
    \node at (-0.38, 0.17) {\tiny $\frac{2\pi}{3}$};
    \draw (.155,-.145) to[bend left=50] (-.155,-.145);
    \node at (0, -0.4) {\tiny $\frac{2\pi}{3}$};
    \draw[-to,line width=1pt] (0,0) -- (0,1);
    \node at (0,1.2) {\small $\vec n_1$};
    \draw[-to,line width=1pt] (0,0) -- (-.866,-.5);
    \node at (-1.1,-.6) {\small $\vec n_2$};
    \draw[-to,line width=1pt] (0,0) -- (.866,-.5);
    \node at (1.15,-.6) {\small $\vec n_3$};
\end{tikzpicture}
    \caption{}
    \end{subfigure}
 \begin{subfigure}[b]{0.24\textwidth}
        \centering
        \begin{tikzpicture}
    \draw[gray,dashed] (-1,0) .. controls +(0.1,.3) and +(-.1,.3) .. (1,0);
    \draw[decoration={markings, mark=at position .8 with {\arrow{to[width=2mm,length=1.5mm]}}},
        postaction={decorate},line width=.8pt,black] (0,.02) -- (.23,-.56);
    \draw[blue] (.5,-.55) to[in=0,out=150] (.26,-.5);
    \draw[white,decoration={markings, mark=at position 1 with {\arrow{Stealth[blue,width=.7mm,length=1mm]}}},
        postaction={decorate}] (.26,-.5) -- (.24,-.5);
    \draw[red,dashed,line width=.8pt] (-.7,-.7) .. controls +(0.1,.2) and +(-.1,.2) .. (.7,-.7);
    \draw[decoration={markings, mark=at position 0.6 with {\arrow{to[width=2mm,length=1.5mm]}}},
        postaction={decorate},line width=1pt] (0,0) -- (0,1);
    \draw[decoration={markings, mark=at position 0.6 with {\arrow{to[width=2mm,length=1.5mm]}}},
        postaction={decorate},line width=1pt] (0,0) -- (-.71,-.71);
    \draw[decoration={markings, mark=at position 0.6 with {\arrow{to[width=2mm,length=1.5mm]}}},
        postaction={decorate},line width=1pt] (0,0) -- (.71,-.71);
    \draw[decoration={markings, mark=at position 0.6 with {\arrow{to[width=2mm,length=1.5mm]}}},
        postaction={decorate},line width=1pt,black] (0,0) -- (-.3,-.85);
    \draw[white, line width=1.5pt] (-1,0) .. controls +(0.1,-.3) and +(-.1,-.3) .. (1,0);
    \draw[gray] (-1,0) .. controls +(0.1,-.3) and +(-.1,-.3) .. (1,0);
    \draw (0,0) circle (1);
        postaction={decorate}] (-.18,-.75) to[bend right=15] (.18,-.75);
        postaction={decorate}] (.18,-.75) to[bend left=15] (-.18,-.75);
    \draw[blue] (-.58,-.68) .. controls +(-0.11,-.01) and +(-.1,-.01) .. (-.36,-.76);
    \draw[white,decoration={markings, mark=at position 1 with {\arrow{Stealth[blue,width=.7mm,length=1mm]}}},
        postaction={decorate}] (-.36,-.76) -- (-.3,-.76);
    \draw[red,line width=.8pt] (-.7,-.7) .. controls +(0.1,-.2) and +(-.1,-.2) .. (.7,-.7);
    \node at (0.95, 0.95) {\small $S^2$};
    \node[white] at (-0.95, 0.95) {\small $S^2$};
    \node at (0.2, 0.5) {\tiny 1};
    \node at (0.55, -0.35) {\tiny 2};
    
    \node at (-0.55, -0.35) {\tiny 3};
\end{tikzpicture}
    \caption{}
    \end{subfigure}
    \begin{subfigure}[b]{0.24\textwidth}
        \centering
        \begin{tikzpicture}
    \draw(0,0) circle (.45);
    \draw (-.9,0) -- (-.45,0);
    \node at (-1.1, 0) {\scriptsize $\vec n_1$};
    \draw (.9,0) -- (.45,0);
    \node at (1.1, 0) {\scriptsize $\vec n_1$};
    \node at (0,.6) {\scriptsize $\vec n_2$};
    \node at (0,-.6) {\scriptsize $\vec n_3$};
\end{tikzpicture}
    \caption{}
    \end{subfigure}
\begin{subfigure}[b]{0.24\textwidth}
        \centering
        \begin{tikzpicture}
    \draw (-1,0) -- (-.45,0);
    \node at (-1.2, 0) {\scriptsize $\vec n_1$};
    \draw (1,0) -- (.45,0);
    \node at (1.2, 0) {\scriptsize $\vec n_1$};
    \node at (0.55,.5) {\scriptsize $\vec n_5$};
    \node at (-0.6,.5) {\scriptsize $\vec n_2$};
    \node at (0.55,-.5) {\scriptsize $\vec n_6$};
    \node at (-0.5,-.5) {\scriptsize $\vec n_3$};
    \draw[fill=white](0,0) circle (.55);
    \draw (0,.55) -- (0,-.55);
    \node at (.2,0) {\scriptsize $\vec n_4$};
\end{tikzpicture}
    \caption{}
    \end{subfigure}
    \end{center}
    \caption{(a) Three unit vectors summing to zero determine an equilateral triangle. (b,c) In a bubble diagram, taking $\vec n_1$ fixed, the angular-momentum-conserving configurations for $\vec n_2$ and $\vec n_3$ sweep out a circle (highlighted \textcolor{red}{red}) on the surface of $S^2$, leading to an order $\sqrt j$ enhancement for such a bubble appearing in any diagram. (d) The non-melonic diagrams do not have as much freedom for the angular momenta of the internal lines and thus do not receive an enhancement compared to melonic diagrams at the same order in the coupling $\mathsf{J}$.}
    \label{Arrows}
\end{figure}
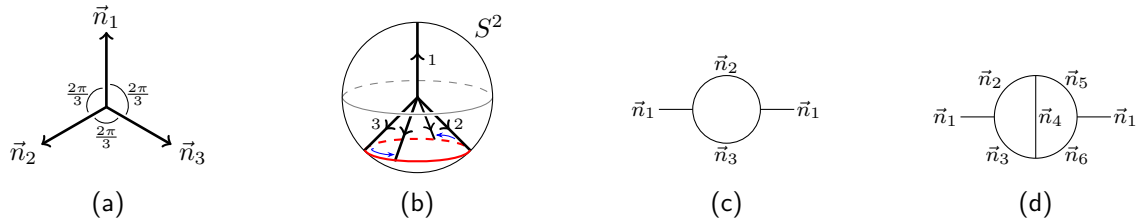

The expression for the supercharge involves an $SU(2)$ invariant interaction between three fields. This implies that the angular momenta of the fermions, or the vectors $\vec n_i$, should sum up to zero 
\be \la{MomCon}
 \vec n_1 + \vec n_2 + \vec n_3 \sim 0
 \ee 
 This equation should hold at any vertex. This resembles momentum conservation equations in Feynman diagrams. However, each vector is on the unit sphere,   $\vec n_1^2 =1$, etc. 
 Since they are points on $S^2$ these three vectors should have a relative angle of $2\pi/3$, see figure \ref{Arrows}(a). But, of course, they can be rotated inside the sphere. 

 From the quantum Hall point of view, the interaction is very strange and non-local, and not particularly natural. Any two fermion interaction can be decomposed in angular momentum channels called Haldane pseudopotentials \cite{Haldane:1983xm}. For our particular interaction, we have only the spin $j$ Haldane pseudopotential, and it is associated to an oscillatory potential between the fermions.

 Something special happens for the simplest bubble diagram, see figure \ref{Arrows}(b,c). In this case, we see that the orientation of the incoming and outgoing arrows should be trivially the same. However the arrows for the particles in the bubble could be rotated around the axis defined by the external arrows. This means that there is an enhancement for this diagram which is proportional to the total number of states that we sweep out by this rotation. This number scales linearly with the size of the sphere, so we expect that the enhancement factor scales like $\sqrt{j}$.  This is the enhancement factor relative to a diagram where the arrows cannot be rotated.

An example of a diagram where the arrows cannot be rotated as much as in a melonic diagram is the one in figure \ref{Arrows}(d). (This is not a diagram that actually appears in our theory, but it is useful to consider it to understand the $SU(2)$ algebra and the extra suppression due to non-melonic diagrams.)  In this case, we can check that the equations for angular momentum conservation have only a discrete set of solutions.  
To see this, note that we can first use our freedom to rotate all vectors uniformly inside the sphere to fix $\vec{n}_1, \vec{n}_2,$ and $\vec{n}_3$. 
We now use the following fact. Consider three unit vectors $\hat{a}, \hat{b}, \hat{c}$ where $\hat{a}$ and $\hat{b}$ are fixed and non-collinear. Fixing the angle subtended by $\hat{a}$ and $\hat{c}$ and the angle subtended by $\hat{b}$ and $ \hat{c}$, then there are at most two possible discrete solutions for $\hat{c}$. 
In our diagram, $\vec{n}_4$  shares a vertex with two of the fixed vectors, namely  $ \vec{n}_2$ and  $ \vec{n}_3$, which means that the angle between $\vec n_4$ and these two vectors is fixed at $2\pi/3$. So there is only a discrete set of solutions for $\vec {n}_4$. Similarly, $\vec{n}_5,$ and $\vec{n}_6$ also share two vertices with the fixed vectors. 
Therefore, the vectors $\vec n_4, ~ \vec n_5$ and $\vec n_6$ are all fixed, and the only freedom we have is an overall rotation around $\vec n_1$ for all the vectors. This means that we have introduced an extra loop, relative to the diagram in figure \ref{Arrows}(c), but not an extra freedom to rotate some of the arrows. Therefore, the diagram is suppressed relative to melon diagrams by a factor of $1/\sqrt{j}$. In fact, we can analyze this diagram using the crossing relation, as we did for the tetrahedron diagram in figure \ref{Tetrahedron} and reach the same conclusion.

Let us now consider general vacuum diagrams. These diagrams have a number of lines $L$ and a number of vertices $V$ related by 
$ 2 L = 3 V $. The number of free parameters associated to the orientations of the arrows is just $ 2L$ (a point on the sphere for each line), and the number of constraints is naively $ 3V$ since there is conservation condition as in \nref{MomCon} at each vertex. However the real number is really $3 (V-1)$ since the conservation condition will be automatically obeyed at the last vertex if it is obeyed in all previous ones. So we find that the number of free parameters is 
\be \la{FreePar}
N_{\rm free} = 2L - 3( V-1) = 3 
\ee 
which is accounted for by the possibility of performing an overall rotation of all the arrows. 
Therefore, up to this trivial overall rotation we expect that for a generic diagram we have no free parameters. This is what we have checked explicitly in the case of the simplest non melonic diagram in figure \ref{Arrows}(d) (if we close the $\vec n_1$ line to make it into a vacuum diagram). 
In contrast,  for melon diagrams, we have seen that we get a free parameter per pair of vertices.  
We have not really proven that melons are the {\it only} kind of diagrams that lead to this large number of free parameters, but this seems likely.

This argument suggests that melons dominate because they are diagrams for which there is an enhancement for some purely kinematic reasons. 
 This argument is reminiscent of the usual BCS theory of superconductivity, where the interaction is enhanced for opposite momenta.

This argument also enables us to understand the powers of $j$ given in \nref{12jasmtot} for the diagram of figure \ref{fig:6j}.
If we were to use the above counting argument \nref{FreePar}, we would naively think that, after fixing the three arrows at one vertex, there would be no freedom left. This would then give a diagram scaling like $1/\sqrt{j}$. However, the diagram is somewhat special and does have one degree of freedom to move the arrows, see appendix \ref{CubeDiag}. This leads us to expect a contribution of order $j^0$. This is essentially what we had from a more explicit computation in \nref{12jasmtot}, except that in \nref{12jasmtot} we had an extra $\log j$ factor. We have not understood the origin of this logarithm from the present point of view. Note, however,  that it is a relatively smaller factor compared to powers of $j$. In conclusion, it is possible to understand the powers of $j$ for a given diagram by counting the number of free parameters in a configuration of arrows (one for each line) which obeys the angular momentum conservation conditions \nref{MomCon} at all the vertices.   

These considerations suggest that further diagrams might lead to subleading powers of the form $1/\sqrt{j}$. In other words, it seems to be an accident that the first $1/\sqrt{j}$ correction was zero and that the first correction is down by $(\log j)/j$ relative to the melons. Furthermore, we also explicitly see that the diagram in \nref{12jasmtot} contains a further correction with a suppression of the form  $1/j^{3/2} $ relative to melons. 

\subsection{Turning on a chemical potential for the $R$ charge}\label{sec:chempot}

If we turn on a chemical potential $\mu_R$ for the $R $ charge, keeping the $SU(2)$ chemical potential zero, we also find that melon diagrams dominate and they reduce to the same equations as for the case of the ${\cal N}=2$ SYK model with chemical potential analyzed in \cite{Heydeman:2022lse}. In fact, \cite{Heydeman:2022lse} showed that there is a phase transition at a particular value of the chemical potential or the charge.   The value of the R charge is
\be \la{RcMu}
R_c = { ( \sqrt{2} -1 ) \over 3} N , ~~~~~~N = 2 j +1 ~,~~~~~~~~{\rm or }
~~~~~{ R_c \over R_{\rm max} } \sim 0.828\ee 
in our normalizations. Here $R_{\rm max}= (2 j+1)/6$ is the maximal $R$ charge.  At this transition, we lose the approximately conformal regime at low energies \cite{Heydeman:2022lse}.

 We expect that it should be possible to extend the melonic (or dynamical mean field theory) treatment to the regime with a relatively small chemical potential for the $SU(2)$ symmetry, but we will not attempt to do this here.

\section{The nearly conformal regime: low energies and small  charges
}
\la{NearConf}

Since the melon diagrams are the same as in the supersymmetric SYK model discussed in \cite{Fu:2016vas}, we can see that the model has a low energy limit where it develops an approximate conformal symmetry. The dimensions of the fermions and the bosons are \cite{Fu:2016vas}
\be 
\Delta_\psi = { 1 \over 6 } ~,~~~~~~~~\Delta_{b} = { 2 \over 3 } = \Delta_\psi + \half 
\ee 
We can understand these dimensions also from the point of view of an enhanced superconformal symmetry in the IR, once we say that the $R$ charge of the fermion is $1/3$. In other words, the operator pair $\psi$ and $b$ are in a short multiplet of the superconformal symmetry of the IR theory. Note that the full model does not have a superconformal symmetry, only the IR theory has such a symmetry, in an approximate manner.

\begin{figure}[h]
   \begin{center}
    \includegraphics[scale=.4]{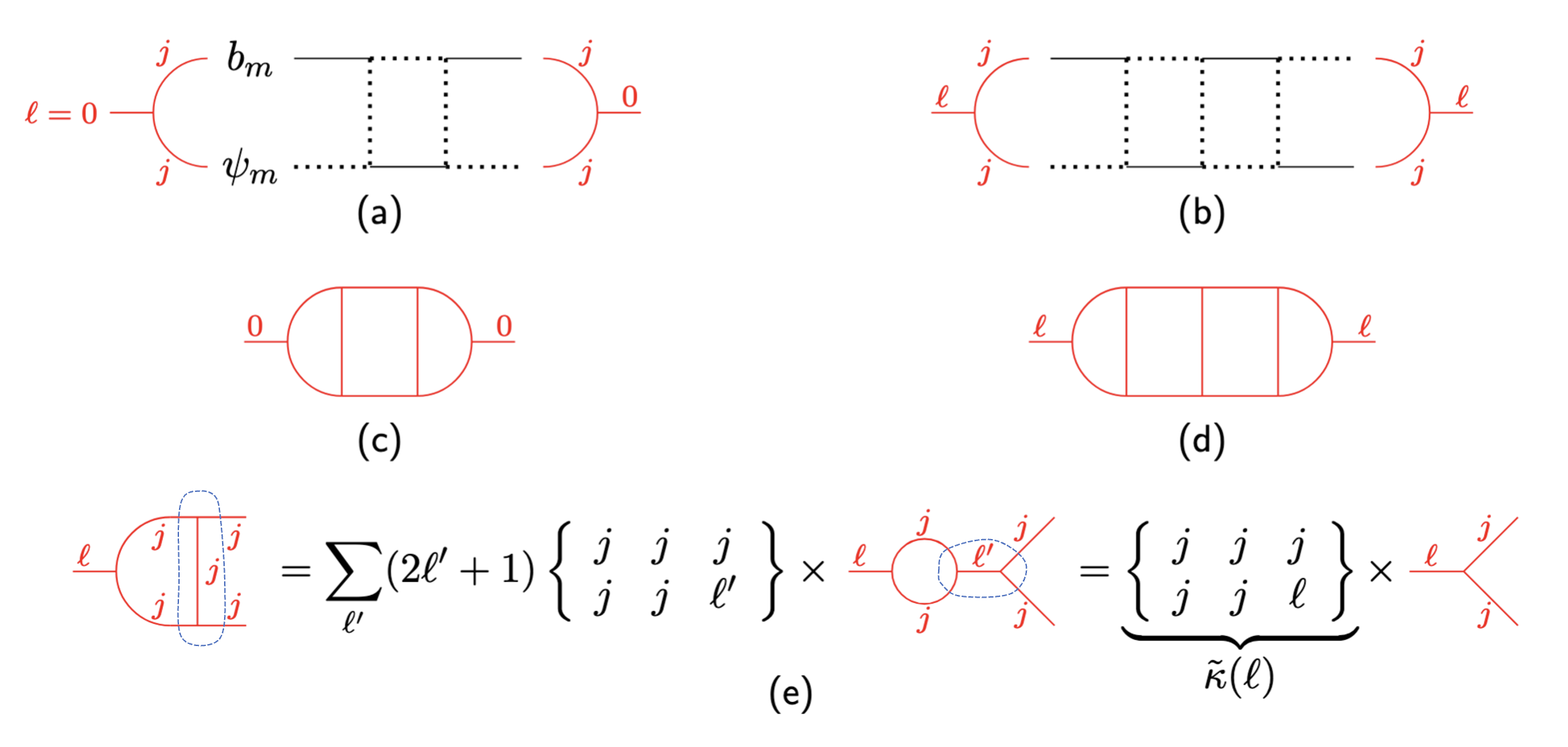}
    \end{center}
    \vspace{-5mm}
    \caption{  (a) In black, a diagram that contributes to the four point function. The red lines indicate how the $SU(2)$ indices are contracted. In this case, they are contracted to a singlet with spin zero, $\ell=0$. The dotted lines represent fermions, and the solid lines represent bosons.   
In (b), we see another diagram. Here, the $SU(2) $ indices of the external lines are contracted into a spin $\ell$ representation of $SU(2)$. In (c) we  see the pattern of contractions of $SU(2) $ indices for the diagram in (a). All the lines whose spins are not indicated carry spin $j$.  In (d), the SU(2) pattern of contractions that arises from (b). In (e), we see the basic identity that we use to reduce the diagrams. After we add a rung, we can use the crossing relation in figure \ref{Crossing} to express the circled part of the diagram in the other channel, which produces the 6j symbol. Then we use the identity \nref{BubId} which gives the result for the bubble. In the end, the addition of the rung just results in a multiplication by an $\ell$-dependent numerical factor \nref{TilK}.      }
    \label{4ptFunction}
\end{figure}

\begin{figure}[h]
   \begin{center}
   \includegraphics[scale=.6]{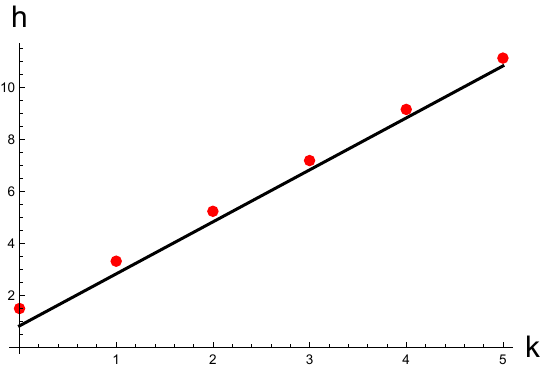} ~~~~~~~~~\includegraphics[scale=.6]{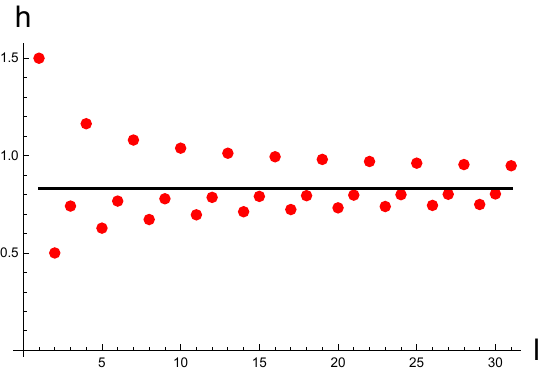}
    \\
    ~~~~~~~~~(a)~~~~~~~~~~~~~~~~~~~~~~~~~~~~~~~~~~~~~~~~~~~~~~~~~~~~~~~~(b)
    \end{center}
    \caption{ (a) In red, values of $h$ for the solutions of the first equation in \nref{Kernel2} for the $\ell=0$ case. The black line is the asymptotic limit \nref{AsyDim} with $\epsilon$ set to zero. The first red point is at $h=3/2$. (b) The dimension of the first operator as a function of $\ell$. Note that for $\ell=0$ we get 3/2 and for $\ell =1$ we get $h=1/2$. For large $\ell$ we asymptote to the value in \nref{AsyDim} with $k=0,~\epsilon=0$.      }
    \label{Dimensions}
\end{figure}

By studying the four point function we can determine the dimensions of other operators. These are operators that are made by pairs of fermions or bosons. 
 The simplest four point function involves operators that are in the $SU(2)$ singlet channel, see figure \ref{4ptFunction}(a). 
  This gives rise to fermionic operators  of the schematic form $\psi_m^\dagger \partial^k b_m $ or bosonic operators that are linear combinations of 
$ b^\dagger_m \partial^k b_m $ and $ \psi^\dagger_m \partial^k \psi_m $. For these cases, the diagrams are exactly the same as in the supersymmetric SYK case \cite{Fu:2016vas,Peng:2017spg,Yoon:2017gut,Bulycheva:2018qcp}. 
The fermionic operators have dimensions given by the solutions of 
\bea \la{Kernel}
&~& 1 =  \kappa^s(h) ~,~~~~~~~~ \kappa^s(h) \equiv  - { \Gamma({5\over 3 } ) \Gamma( { 5\over 12 } - { h \over 2 } ) \Gamma({5 \over 12 } + { h \over 2 } ) \over 2^{1/3} \Gamma( { 4 \over 3 } ) \Gamma({13\over 12 } - { h \over 2 } ) \Gamma( { 1 \over 12 } + { h \over 2 } ) } 
\cr 
&~& 1 = \kappa^a(h) ~,~~~~~~~~ \kappa^a(h) \equiv - { \Gamma({5\over 3 } ) \Gamma( { 11 \over 12 } - { h \over 2 } ) \Gamma( -{ 1 \over 12 } + { h \over 2 } ) \over 2^{1/3} \Gamma( { 4 \over 3 } ) \Gamma({7\over 12 } - { h \over 2 } ) \Gamma( { 7 \over 12 } + { h \over 2 } ) }  
\eea
For each value, there are two operators that are complex conjugates of each other. In addition, there are two other bosonic operators that have dimensions $h+\half$ and $h-\half$. Note that there are four operators in each multiplet, so that these are long multiplets under the superconformal algebra\footnote{  Note that the relation $\kappa^a(h)= \kappa^s( 1-h)$   implies a relation between the roots. However, we will be interested in roots of both equations that are positive, for which there is no obvious relation between the two sequences of roots.}.  See figure \ref{Dimensions}(a) for some numerical values of the solutions. 

In particular, we can check that $h={3\over 2} $ is a solution of both of the equations in \nref{Kernel}. One of these,  together with their bosonic partners, leads to 
 an ${\cal N}=2$ super Schwarzian that produces an action for super-reparametrization modes. This action gives the leading quantum corrections away from the conformal limit. This action is of the form 
 \be \la{SchAct}
 S = - { N \over { \mathsf{J} } } \alpha_S \int dt \{ f(t), t \} + {\text{susy-partners}} ~,~~~~~~{\rm with } ~~~\alpha_S \approx 0.00842
\ee 
where we have quoted the value of $\alpha_S$  which was computed  in \cite{Heydeman:2022lse}  by numerically solving the large $N$ equations.

The second of the solutions of \nref{Kernel} with $h =3/2$ gives rise to an operator of that dimension, which sits in a multiplet with bosonic partners at dimensions $h=2,1$ \cite{Fu:2016vas}.

An interesting application of the action \nref{SchAct} is the computation of the minimal energy for non-BPS states at a given value of the $R$ charge  \cite{Heydeman:2022lse,Stanford:2017thb} 
\begin{align} \la{Egap}
    E_\mathrm{gap}=\frac{ \mathsf{J}}{8N\alpha_S}\Big(\big|R\big|-\frac{1}{2}\Big)^2\, .
\end{align}

A new feature of our model is the global $SU(2)$ symmetry, and we also expect some low action degrees of freedom from this $SU(2)$ symmetry. This will require understanding operators that are in non-trivial representations of $SU(2)$.  For that purpose, we consider ladder diagrams computing the four point function and combine the two external lines to a low angular momentum representation with spin $\ell $.  In terms of the simple picture presented in section \ref{ArrowsSec}, we expect that these would continue to be melonic, since a small angular momentum should change the directions of the arrows by very little. Some of the details will change. These details only depend on the $SU(2)$ quantum numbers. From the point of view of SU(2) quantum numbers, we can think of the two external lines as joining in a three point vertex that involves a 3j symbol. In other words, we are considering operators of the form  
\be 
 O_{m_3}^\ell = \sum_{m_1,m_2}  (-1)^{m_2} \begin{pmatrix}
        j  & j & \ell \\ m_1 & -m_2 & m_3 
    \end{pmatrix} \psi^\dagger_{m_2} \overset{\leftrightarrow}{\partial} ^{k} b_{m_1} \la{OperL}
    \ee

The diagram that contains just one rung of the ladder, which gives the action of the kernel,  has the form indicated in figure \ref{4ptFunction}(e). This diagram represents just contributions from the SU(2) algebra. In order to compute it, we use the relation in figure \ref{Crossing}, as indicated in figure \ref{4ptFunction}(e), as well as the bubble identity \nref{BubId}. The final result is proportional to the 6j symbol 
\be 
\left\{ \begin{array}{ccc}
        j  & j & j  \\j & j & \ell \end{array} \right\} 
    \ee 
    
 Notice that the singlet case we discussed previously corresponds to the case $\ell =0$. 
 This means that we can define a quantity $\kappa(\ell) $ given by 
 \be \la{TilK}
  \tilde \kappa(\ell) \equiv {\left\{ \begin{array}{ccc}
        j  & j & j  \\j & j & \ell \end{array} \right\}  \over \left\{ \begin{array}{ccc}
        j  & j & j  \\j & j & 0 \end{array} \right\}  } 
\ee  
As an aside, note that $\tilde \kappa(\ell) $ involves essentially the same group theory applied previously to determine $\kappa(h)$ in  \nref{Kernel}, discussed more explicitly in  \cite{Maldacena:2016hyu}. The large $j$ expression for \nref{TilK} is 
\be \la{KapTi}
\tilde \kappa(\ell) = P_{\ell}(-1/2) ~,~~~~~~~~{\rm for } ~~~ j=\infty 
\ee 
where $P_{\ell}$ is the Legendre polynomial \cite{Edmonds:1955fi}.
Some special values are  $\tilde \kappa(0)=1$ and $\tilde \kappa(1) =-\half $. In addition,  $\tilde \kappa(\ell) $ decreases for large $\ell$.  

Using this, we find that operators with angular momentum have dimensions, $h$, which are given by the equations 
\be \la{Kernel2}
1 = \tilde \kappa(\ell) \kappa^s(h) ~,~~~~~~~~~~~ 1 = \tilde \kappa(\ell) \kappa^a(h) 
\ee 
where $\kappa^s(h)$ and $\kappa^a(h)$ are the same as in \nref{Kernel}. 
In general we find that as $\ell $ becomes large $\tilde \kappa(\ell) $ goes to zero and then the dimensions go to their naive generalized free field value given by the sum of the dimensions of the elementary operators. See figure \ref{Dimensions}(b). For example, for the operator in \nref{OperL} we get 
\be \la{AsyDim}
 h^s_{\ell,k} \sim \Delta_\psi + \Delta_b +  2 k + \epsilon_{\ell, k} ~,~~~~~~~\Delta_\psi = { 1 \over 6} ~,~~~~~~~\Delta_b = { 2 \over 3 } 
 \ee 
 with small $\epsilon_{\ell, k}$ when $\ell $ or $k$ are large. 
   This means that operators with large $\ell$ behave as two independent $\psi$ and $b$ operators whose conformal dimension is approximately independent of $\ell$. 
As an aside, note that this is very different from what we get for Kaluza Klein modes for a compactification on $S^2$. However, this is similar to what we expect for fermions moving in a magnetic four dimensional black hole \cite{Maldacena:2020skw}, which also involves fermions in large $SU(2)$ representations.

The case $\ell=1$ is special because we expect to find the conserved $SU(2) $ current in this multiplet. As we mentioned after \nref{KapTi},    $\tilde \kappa (1) = - \half$. Inserting this into \nref{Kernel2}, we get that 
$h =\half$. This mode has superpartner operators at $h = 0,1$ which appear in other ladder kernel, from the operators with a $\psi^\dagger  \psi $ and $b^\dagger b$ in the analog of \nref{OperL}.    We get this doubled because both the symmetric and antisymmetric channels give the same solutions. However, only one of these channels leads to a low action mode, as discussed in \cite{Fu:2016vas}.   In the case of the ordinary SYK-like models, we expect an action which contains terms like $ { N \over {\mathsf J } } \int dt Tr[ (g^{-1} \dot g )^2 ] $, where $g$ is an $SU(2) $ group element. This cannot arise in our case because it would raise the energies of modes with non-zero angular momentum. In contrast,   we saw previously that we have a large set of BPS states with various values of the angular momentum. 

One possibility is that the action instead contains a first order term of the form 
\be 
I = \int dt Tr[ g^{-1} \dot g  \rho ] \la{LLLSpin}
\ee 
where $g$ parametrizes the $SU(2) $ manifold (an $S^3$) and $\rho$ is a new field which is a triplet of $SU(2)$. On shell, $\rho$ becomes constant and equal to the conserved charge $J_a = \rho_a $, $a=1,2,3$. 
This remains a conjecture, and it would be nice to see whether it is true. We leave this question to the future.

\section{Large $U(1)_R$ charge } \label{sec:largeU1}

In this section, we discuss some features of the large $R$ charge states. These are relatively simple since there is a small number of states for extreme values of the $R$ charge.
For this purpose, it is useful to recall the normal ordered expression of the Hamiltonian \nref{HamNor}
\begin{align}\label{eq:Hnormord}
    H=\frac{\mathsf{J}}{3}\Big((2j+1)-3(N_\psi+j+\tfrac{1}{2})+3\sum_m O^\dagger_{j,m}O_{j,m}\Big)
\end{align}
The operators $O_{j,m}$ are simply the fermion bilinears $\psi_{m_1}\psi_{m_2}$ reorganized into a spin-$j$ multiplet. More generally, a 2-fermion spin-$\ell$ multiplet can be defined by 
\begin{align}
O_{\ell,m}=\sqrt{2(2\ell+1)}\sum_{-j\leq m_1<m_2\leq j}\begin{pmatrix}
        j & j & \ell\\
        m_1 & m_2 & -m
\end{pmatrix}\psi_{m_1}\psi_{m_2}\;.
\end{align}
The Hamiltonian in the form \eqref{eq:Hnormord} can be used to read off the energies of the $R=\pm(2j+1)/6$, $R=\pm(2j-1)/6$, and $R=\pm(2j-3)/6$ sectors. For example, the Fock vacuum is straightforwardly seen to have 
\begin{align}
    E_\text{vac}=\frac{ \mathsf{J}}{3}(2j+1), ~~~~~~~R_{\rm vac} = - { ( 2j + 1 ) \over 6 } \;.
\end{align}
We conjecture that this is the maximum energy in the whole spectrum.   We have checked this numerically for $j\leq 11$, see section \ref{sec:exact}.

The $R=-(2j-1)/6= R_{\rm vac} + 1/3$ sector consists of the 1-fermion states $\psi^\dagger_m|0\rangle $, forming a single spin $j$ multiplet. Hence, they are all annihilated by the last term in \eqref{eq:Hnormord}, and the energy is 
\begin{align}
    E_{R=(2j-1)/6}=\frac{2 \mathsf{J}}{3}(j-1)\;.
\end{align}
  
The $R=(2j-3)/6= R_{\rm vac} + 2/3$ sector consists the 2-fermion states which we can organize into definite angular momentum multiplets
\begin{align}
    \Big|\tfrac{2j-3}{6};{\ell,m}\Big\rangle=O^\dagger_{\ell,m}|0\rangle\;,\qquad \ell=1,3,\cdots,2j-1\;.
\end{align}
$O_{j,m}$ annihilates this state unless $\ell=j$. Therefore there are two distinct energies in this sector:
\begin{align}\label{eq:2bitEnergies}
    E_{R=(2j-3)/6}^{(\ell)}=\begin{cases}\displaystyle\frac{ \mathsf{J}}{3}(2j-5)&, \qquad \ell\neq j\vspace{5pt}
    \\
    \displaystyle\frac{ \mathsf{J}}{3}(2j-2)&,\qquad\ell= j
    \end{cases}
\end{align}
In particular, at this R-charge, the $E=\mathsf{J}(2j-5)/3$ energy level is highly degenerate. Namely,  many states with different angular momenta have the same energy.   One might expect there to be similar degeneracies at other states containing more fermions, when we combine the fermions in such a way that for any pair of fermions, we do not have the spin $j$ component.   This is indeed what we find numerically when we do exact diagonalization for some small values of $j$ in \autoref{sec:exact}. On the other hand, we do not expect such degeneracies when we consider relatively small values of $|R|$, where states contain order $j$ fermions. Numerically, we find that the degeneracy stops once there are around $j/2$ fermions, or  $|R|\approx (j+1)/6$. 

One simple way to get some degenerate states is to write states of the form 
\be \la{degstates}
\prod_{i=1}^n \psi^\dagger_{m_i}|0\rangle ~,~~~~~~~~{\rm with}~~~~ m_i \geq (j+1)/2  
\ee 
which are annihilated by $O_{j,m}$ in \nref{eq:Hnormord}. For $j$ large and $n$ sufficiently less than $(j+1)/2$, this yields many degenerate states in different angular momentum representations. 
These are not all the degenerate states; these are  just simple examples. The numerical results contain more states at these energies, which we expect to get by considering fermion wavefunctions that are annihilated by the $O_{j,m}$ in \nref{eq:Hnormord}.

\section{States with the largest spin for a given $U(1)_R$ charge }\la{lgspinperR}

Beyond the sectors discussed in the previous section (and their charge conjugates), we can construct a simple subset of states of the form 
\begin{align} \la{MaxAng}
    \psi^\dagger_{j-n}\cdots \psi^\dagger_{j-1} \psi^\dagger_j|0\rangle\;,\qquad n < j
\end{align}
These are the states with the largest value of $J_3$ (and the largest $SU(2)$ spin) for a given $R$ charge; for this reason, they cannot mix with anything else.   
Their $SU(2)$ spin  and $R$ charges are 
\begin{align} \label{eq:lmax(R)}
 &R = (2n-2j+1)/6\;,\notag\\
 &\ell_{max}^{(R)}=j(n+1)-n(n+1)/2=(N-6R)(N+6R)/8\;.
 \end{align}

  They have a simple description in terms of the fermions moving on the sphere with a magnetic field. This is a set of fermions that are all as close to the north pole as it is allowed by their Fermi statistics. They form a spherical ``cap'' around the north pole, see figure \ref{Hemisphere}(a). The angle from the north pole to the edge of the cap is 
 \be \la{AngleCap}
 \cos \gamma = { 3 R \over j }  ~,~~~~~~~~{\rm for }~~~ R, ~ j \gg 1
 \ee

It is interesting to compute the energy of these states. As shown in appendix \ref{DerEquA}, we get   
\bea \la{FirLi}
  \!\!\!\frac{E(|R|)}{ \mathsf{J}}\!\!\!&=&\!\!\!\frac{2j+1}{3}-(n+1)  = \frac{18|R|-2j-1}{6} ~,~~~~~~~~{\rm for } ~~~n \leq (j-1)/2 ~~~{\rm or } ~~  { j \over 6 } \leq |R| 
  \\
  \label{eq:nonlinear}
    \!\!\!\frac{E\big(|R|\big)}{ \mathsf{J}}\!\!\!&=& (2j+1)\sum_{ -3\lp |R|-\frac{1}{6}\rp \leq m_1 < 0}  \left\{ \sum_{-3\lp |R|-\frac{1}{6}\rp \leq  m_2 \leq -m_1 } \begin{pmatrix}
        j & j & j \\ m_1 & m_2 & -(m_1 +m_2)
    \end{pmatrix}^{2} \right\} ~,~~{\rm for }  ~|R| <  { j \over 6 }~~~~~\;.
  \eea
The simplicity of  \nref{FirLi} arises because for these values of $n$ the operators $O_{j,m}$ in the Hamiltonian \nref{eq:Hnormord} annihilate the state due to angular momentum conservation. This also implies that \nref{FirLi} are the minimal energies for this range of R charges. On the other hand,  the states in \eqref{eq:nonlinear} are \textit{not} the minimum energy states for these values of the $R$ charge, as seen numerically in figure \ref{fig:largespinCFTcomparison}. 
  
  By the way, notice that the change in behavior from \nref{FirLi} to \nref{eq:nonlinear} happens when the angle in \nref{AngleCap} is $2 \gamma =120^\circ$. Therefore, this change in behavior appears when we can have two annihilation operators in $O_{j,m}$ acting on the set of occupied fermions, if we view this operator as involving three vectors as in figure \ref{Arrows}(a).

\section{Large $SU(2)$  charge and the emergent CFT$_2$}
\la{SU2Sec}

In this section, we analyze the model for states with nearly maximal $SU(2) $ charge. In this regime, the model simplifies substantially, and it becomes closely related to a two-dimensional CFT. 
We will first explain intuitively why this should be the case, and then we will perform a more precise analysis. This regime is outside the range of validity of the melonic regime discussed in section \ref{Melonic}.

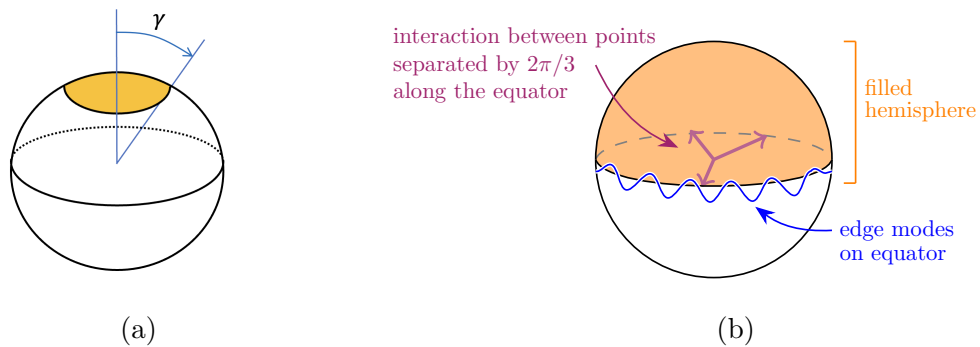
\begin{figure}[h]
~~~~~~~~~~~~~~~~~~~ \includegraphics[scale=.25]{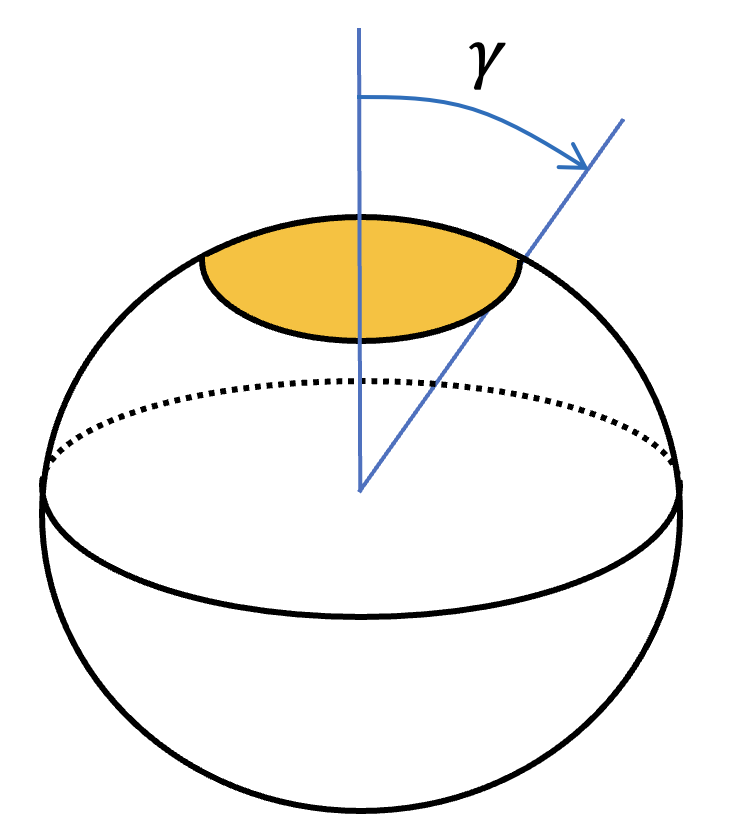}  ~~~~~~~~~~~ 
    \resizebox{!}{3.5cm}{
    \begin{tikzpicture}
    \draw[white,name path=C] (-1,0) .. controls +(0.1,.3) and +(-.1,.3) .. (1,0);
    \draw[white,name path= A] (-1,0) .. controls +(0.1,-.3) and +(-.1,-.3) .. (1,0);
    \draw[white,name path= B] (1,0) arc (0:180:1);
    \draw (-1,0) arc (-180:0:1);
    \node[white] at (-0.95, 0.95) {\small $S^2$};
\fill [orange!50,
          intersection segments={
            of=A and B,
            sequence={L2--R2}
          }];
    \draw[gray,dashed] (-1,0) .. controls +(0.1,.3) and +(-.1,.3) .. (1,0);
    \draw (1,0) arc (0:180:1);
    \draw[-to, line width=0.8pt,RedViolet!70] (0,0) -- (0.45,0.2);
    \draw[-to, line width=0.8pt,RedViolet!70] (0,0) -- (-0.2,0.25);
    \draw[-to, line width=0.8pt,RedViolet!70] (0,0) -- (-.1,-0.23);
    \draw[black] (-1,0) .. controls +(0.1,-.3) and +(-.1,-.3) .. (1,0);
    \draw[snake it,white, line width=1pt] (-1,-0.1) to[bend right = 15] (1,-0.1);
    \draw[snake it,blue] (-1,-0.1) to[bend right = 15] (1,-0.1);
    \node[blue,scale=0.5] at (1.55,-.7) {\shortstack[l]{edge modes\\on equator}};
    \draw[-Stealth,blue] (1, -0.6) to[bend left=20] (0.35,-0.4);
    \draw[orange] (1.1,-0.2) -- (1.2,-0.2);
    \draw[orange] (1.2, -0.2) -- (1.2, 1);
    \draw[orange] (1.2, 1) -- (1.1, 1);
    \node[orange,scale=0.5] at (1.75,0.5) {\shortstack[l]{filled\\hemisphere}};
    \draw[-Stealth,RedViolet] (-1,0.8) to[bend right=20] (-0.3,0.1);
    \node[RedViolet,scale=0.5] at (-1.6,0.8) {\shortstack[l]{interaction between points\\ separated by $2\pi/3$\\
    along the equator}};
\end{tikzpicture}}
\begin{center}~ (a) ~~~~~~~~~~~~~~~~~~~~~~~~~~~~~~~~~~~~~~~~~~~~~~~~~~~~~~~~~(b) 
\end{center}
\caption{  (a) State with maximal angular momentum with a given number of fermions. (b)  Fermions on a sphere in a state with maximal angular momentum fill the northern hemisphere. There are edge modes along the equator that generate the Hilbert space of a chiral 1+1 dimensional CFT. The interaction couples states whose positions differ by $2\pi/3$ along the circle on the equator.     }
    \label{Hemisphere}
\end{figure}

The simplifications in the large $SU(2) $ charge limit are easier to understand when we view the fermions as arising in a quantum Hall system on a sphere. A fermion at some point on the sphere has angular momentum roughly in the direction of that point on the sphere, as we discussed in section \ref{ArrowsSec}. 
For a state to have maximal angular momentum, we want to maximize the angular momentum in the $\hat z $ direction. We can achieve this by adding fermions with angular momenta all pointing up as much as possible. This means that we should fill all states in the northern hemisphere. Then we have a quantum Hall droplet that is completely filled in the northern hemisphere and completely empty in the southern hemisphere, as shown in figure \ref{Hemisphere}. 

In such a system, we have edge modes along the equator. For this reason, for small excitations,  the Hilbert space looks like we have a 1+1 dimensional chiral CFT. This is even before we put any interactions. 

Interestingly, some of this simplicity remains after we include the interactions. The reason is that the interactions involve fermions that are separated by $2\pi/3$ along the circle. Therefore, we can describe the system in terms of three fermions living on a smaller circle with a local interaction. These fermions should have appropriate boundary conditions on the smaller circle. 

In fact, it is a bit more convenient to bosonize the fermions and to think in terms of the boson. The supercharge involves a product of three fermion fields evaluated at positions that differ by $2\pi/3$ along the circle, which only involves boson momentum modes that are zero modulo three. Then the Hamiltonian turns out to be the usual Hamiltonian for a boson on a smaller circle. This Hamiltonian lifts parts of the states and leaves others with zero energy. The ones with zero energy are the BPS states. In this regime, we get a very simple and explicit description of the BPS states. 

Let us now provide a more thorough analysis of this. 
Let us start with some observations about the structure of the Hilbert space around states with maximal $J_3$. There are two states with maximal $J_3$
\be \la{BPSMaxl}
|\tilde 0 \rangle_- =  \psi^\dagger_{1} \psi^\dagger_{2} \cdots \psi^\dagger_j |0 \rangle  ~,~~~~~
|\tilde 0 \rangle_+ =  \psi^\dagger_0 \psi^\dagger_{1} \psi^\dagger_{2} \cdots \psi^\dagger_j |0\rangle 
\ee 
which have $R$ charges $R = \pm {1 \over 6 } $ and $J_3 = j (j+1)/2$. 
We can get other states by acting with $\psi_m $ and $\psi^\dagger_m $ on these states. As long as $|m| \ll j$ we can think of these states as acting on a completely filled relativistic Fermi sea. Then, the operators act as usual fermionic operators on the Ramond ground state. More precisely, defining 
\be 
 \bar \chi_{m} = \psi^\dagger_{m} ~,~~~~~ \chi_{m} = \psi_{-m} 
 \ee 
 Then we find that $\chi_m$ and $\bar \chi_{m}$ look like the usual operators of a free fermion, which annihilate the vacuum, when $m>0$ ($\chi_m |\tilde 0 \rangle_{\pm} = \bar \chi_{m} |\tilde 0 \rangle_{\pm} =0$, for $m>0$). 
This is the Hilbert space of a single complex chiral fermion in the periodic (or Ramond) sector. 
We should note that we can also write 
\be \la{J3CFT}
  J_3^{max} -J_3=   L_0  = \sum_{m } \, \, m: \bar \chi_{m}
\chi_{-m} : ~,~~~~~~~~~~~~~J_3^{max} = { j (j+1 ) \over 2}
\ee 
where the normal ordering constant is precisely the shift of  $J_3^{max}$ in \nref{J3CFT} relative to the form of $J_3$ in \nref{SUTwoCharges}. 

This can also be described in terms of a boson $\phi$. This boson is also defined on the circle, but only a particular subset of momenta or windings is allowed. This can be expressed most simply by writing the equivalence of partition functions 
\be \la{FerPar}
Z_{\rm fermion} =  Tr[ q^{L_0}] =  2 \prod_{n=1}^\infty (1 + q^n)^2 = q^{ - { 1 \over 8 }  }    {1   \over \prod_{n=1}^\infty (1 - q^n ) } \sum_{ p \in Z + \half } q^{ p^2/2}
\ee 
which shows that we only sum over half integer values of the chiral boson ``momentum'', $p$.  

Now, we can use the large $j$ expression of the  $3j$ symbol that is valid for relatively small values of $m_i$ , $|m_i | \ll j $ \cite{PonzanoRegge},  
\begin{align} \la{3jLargej}
      \begin{pmatrix}
            j  & j  & j  \\ m_1 & m_2 & m_3
        \end{pmatrix}  \sim - { 1 \over j } { \sqrt{ 2 \over \pi \sqrt{3} }} \,  \cos\left( \frac{3 \pi j}{2} +  { 2 \pi ( m_1 - m_2 ) \over 3 } \right)  
\end{align}
        
        This means that the expression for the supercharge can be written, up to an overall constant, as 
        \be 
    Q \propto     \int d\varphi \psi(\varphi) \psi\left(\varphi + { 2\pi \over 3 }\right) \psi\left(\varphi - { 2 \pi \over 3}\right) \la{QinCFTf}
        \ee 
Using the bosonized form of the fermions $\psi = e^{ i \phi } $, we can rewrite this as 
\be \la{QinCFT}
Q \propto \int d \varphi \exp\left\{ i \left[\phi(\varphi) + \phi\left(\varphi + {2 \pi \over 3 } \right) +\phi \left(\varphi - { 2 \pi \over 3 } \right)  \right]\right\}
\ee 
We see that only the modes of the boson that are zero modulo three appear in this expression. 
In other words, we can decompose the original bosonic field $\phi$ into the sum of three bosons $\phi = \phi_z + \phi_+ + \phi_-$, with 
\be \la{phidef}
\phi_z = \sum_k  \alpha_{ 3k } e^{ i 3 k \varphi } ~,~~~~~~ \phi_\pm = \sum_k \alpha_{ 3 k \pm 1 } e^{ i (3 k \pm 1  ) \varphi }
\ee 
where now we can think of the bosons as defined on a smaller circle $\tilde \varphi = 3 \varphi$.  

The boson $\phi_z$ is periodic on the smaller circle, $\phi_z(\tilde \varphi + 2 \pi) = \phi_z(\tilde \varphi) $. The other two bosons have twisted boundary conditions on the smaller circle 
\be \la{twistedBC}
\phi_\pm (\tilde \varphi + 2 \pi ) = e^{ \pm i { 2 \pi \over 3 } } \phi_{\pm }(\tilde \varphi ) 
\ee 

The supercharge $Q$ in \nref{QinCFT} involves only $\phi_z$; it is simply an exponential of $\phi_z$. This is the expression of the surpercharge at the ${\cal N}=2$ supersconformal point of the single boson CFT \cite{Ginsparg:1988ui}. This means that the Hamiltonian of our model is given by the usual $ \tilde L_{0}^z \propto (\partial \phi_z)^2$ of the free field $\phi_z$. Importantly, the other two bosons do not appear in the expression for the supercharge or the Hamiltonian.

In other words, the Hilbert space of the theory is the product of two factors 
\be \la{HilSpaz}
{\cal H } ={\cal H}^z \otimes {\cal H }^\pm 
\ee 
where ${\cal H}_z$ describes the $\phi_z$ boson and ${\cal H}_\pm $ the $\phi_\pm $ bosons. The total $L_0 $ can be written as 
\be \la{L0tot}
L_0 = 3 ( \tilde L_0^z + \tilde L_0^\pm ) 
\ee 
This determines the $J_3$ quantum numbers via \nref{J3CFT}. But the actual Hamiltonian of the theory is given only in terms of $\tilde L_0^z$
\be \la{HamLti0}
H =  \mathsf{J} { 6\sqrt{ 3 } \over \pi j } \tilde L_0^{z} ~~~~~~ \tilde L_0^z = {1 \over 3 } { ( p^2 - { 1 \over 4 } ) \over 2 } + \tilde N ~,~~~~~\tilde N \in \mathbb{Z}
\ee 
where $\tilde L_0^z$ is only the energy of the $\phi_z$ boson, rescaled to a circle of radius $2\pi$.  
$\tilde N$ is an integer counting the energy of the oscillators with $n= 3k$ in the partition function \nref{FerPar}.
This means that all states in the Hilbert space of the two $\phi_\pm $ bosons have the same energy. For this reason,  the spectrum is highly degenerate in this approximation. Of course, since the $\phi_\pm$ bosons contribute to the $J_3$ angular momentum, these degenerate states have various values of the angular momentum.

We should also note that the $R$ charge is also given purely in terms of $\phi_z$,  and it is proportional to the usual expression in the CFT, $R = J^R_0$, with $J^R  \propto \partial \phi_z$. The normalization of the $R$ charge is such that 
\be \la{RFerNor}
R = { p \over 3  }~,~~~~~~~~~N_\psi = p 
\ee 
where $p$ is the half integer appearing in \nref{FerPar}, which also determines the fermion number. Note that \nref{J3CFT}, \nref{L0tot}, \nref{HamLti0}, and \nref{RFerNor} together imply that the CFT$_2$ predicts a maximal value of $J_3$ within each R charge sector, $J_3^{(R)} \leq \frac{1}{8}N^{2} - \frac{9}{2}R^{2}$, as expected from \nref{eq:lmax(R)}. 

The two states with maximal angular momentum \nref{BPSMaxl} correspond to the vacuum for the $\phi_\pm $ bosons and the zero energy states 
  of the $\phi_z$ boson, which have no $\phi_z$-oscillators excited and have the lowest values $p= \pm \half $ of ``momentum''. These two states have fermion numbers that differ by one, see \nref{RFerNor}. They also have opposite $R$ charges $R= \pm {1 \over  6} $. 

We get all other BPS states by adding the oscillator modes of the $\phi_\pm$ bosons, while staying always with the two zero energy states of the $\phi_z$ boson.  This gives a very explicit description of the BPS states in this limit. Their partition function is given by 
\be \la{ZBPSbos}
Z_{BPS} = 2 Tr_{{\cal H}^\pm } [ q^{ J^{max}_3 - J_3 } ] =  2 \prod_{k=0}^\infty { 1 \over (1 - q^{ 1 + 3 k} )(1 - q^{ 2 + 3 k } ) }
\ee 
where the trace is only of the Hilbert space ${\cal H}^\pm $ of the $\phi^\pm $ fields, see \nref{HilSpaz}. 
The overall factor of two corresponds to the two zero energy states of the $\phi_z$ boson. The rest spans the Hilbert space ${\cal H}^\pm $ of the $\phi_\pm$ bosons.  Thus, we obtain a family with $R=1/6$ and another set of states with the same $SU(2) $ quantum numbers for $R=-1/6$. 
Note that for large values of $ J^{max}_3 - J_3$ we can use a Cardy-like formula to estimate the number of states 
\be 
 Z_{BPS } = \sum_{\hat N} d_{\hat  N} q^{\hat N} ~,~~~~~~~~~~~ \log d_s \sim  2 \pi \sqrt{ \hat N \over 9  } -{ 3 \over 4 } \log \hat N ~,~~~~~~~~\hat N =  J^{max}_3 - J_3 
 \ee 
  where we also included the logarithmic correction. This agrees with the second line of \nref{BPSspa}. 
This is valid for $\hat  N \ll J_3^{max}$.  If we were to {\it naively extrapolate} this formula all the way to $\hat N = J_3^{max} $ we would get something which is of the same order of magnitude (in the exponent), as a function of $j$, as the number of states at zero $J_3$ that we computed in the first line of \nref{BPSspa}, see also figure \ref{SPplot} in appendix \ref{SPapp}. 

It is also useful to look at the basic fermion field $\psi$. Its conformal dimension receives a contribution only from the boson $\phi_z$ and so it is $1/3$ of the usual one. This means that the dimension is $\Delta =1/6$, which happens to be the same as what we found in the conformal low energy regime at small charges\footnote{It is curious that this fermion and the supercharge $Q$ have the conformal dimensions, respectively,  of the quasiparticle and the full fermion of the edge modes of the 1/3 Laughlin state. This appears to be just a coincidence. }. This apparent coincidence is explained by noticing that in both regimes, we have an emergent superconformal symmetry, and the fermion is a BPS operator with an $R$ charge of 1/3, which determines its dimension.

It is also instructive to understand the expressions for the other SU(2) generators, $J_\pm$. Looking at the expressions in (\ref{Jp},\ref{Jm}), we find that it is convenient to consider the rescaled generators 
\bea \la{GenAp}
\hat J_+ &=&  { J_+\over j}    \sim \sum_m \psi^\dagger_{m+1} \psi_m \propto \alpha_{1} 
\cr 
\hat J_- &=&  { J_- \over j}  \sim \sum_m \psi^\dagger_{m-1} \psi_m \propto \alpha_{-1}
\cr  
[\hat J_+, \hat J_{-} ]  &\sim  & { 2 J_3^{max} \over j^2 } \sim 1 = {\rm constant}
\eea
 where we took the large $j$ limit keeping $m$ fixed. In the last line we have also approximated $J_3$ by its maximal value. 
We also noted that $\hat J_\pm$ are proportional to the first boson creation and annihilation operators in the fields $\phi_\pm$ \nref{phidef}. As is familiar, due to the very large maximal angular momentum we are considering, the SU(2) raising and lowering operators become just ordinary harmonic oscillator creation and annihilation operators. 
 This also means that the $SU(2)$ multiplets are being generated by the action of this first oscillator mode \nref{GenAp}. This mode sits in the $\phi_\pm$ sector.

It is interesting to look at the lowest energy states with a given $R$ charge. Their energies are simply given by \nref{HamLti0} with $\tilde N=0$ 
\be \la{HamLti}
H =  \mathsf{J} { 9 \sqrt{3} \over \pi j }   \left( R^2 - { 1 \over 6^2 } \right)  
\ee 
These are states of the $\phi_z$ boson that have no oscillators excitated, only the $p$ momentum and winding. As for any other state we can tensor any state from the Hilbert space, ${\cal H}^\pm$ of the $\phi_\pm$ bosons. If we choose the vacuum for the  $\phi_\pm$ oscillators we will also have the states with maximal angular momentum with a given $R$ charge. As an interesting cross-check, we can then compare \nref{HamLti} to the formula \nref{eq:nonlinear} as follows.  Starting from \nref{eq:nonlinear}, we insert the large $j$ expression \nref{3jLargej} and perform the sum to obtain \nref{HamLti}.  

The expression \nref{HamLti} can also be compared and contrasted with the minimum energy for a given $R$ charge that was obtained in the Schwarzian regime  \nref{Egap}. We observe that they differ in detail, but both scale as $1/j$ and are quadratic in $R$ for large $R$. 
Note that these two expressions do not need to agree since they are computed in different regimes, for very different values of the angular momentum. 

An important difference with the Schwarzian regime is the large degeneracy that we mentioned earlier.  The spectrum in this CFT approximation is given by \nref{HamLti0}, and each of those states can be tensored with any state of the ${\cal H}^\pm$ Hilbert space, all of which will have the same energy. On the other hand, in the Schwarzian regime of small angular momentum and low energies, we have some exactly degenerate BPS states, but all states above the gap \nref{Egap} form an approximate continuum.  We expect that as we decrease the angular momentum, or consider $1/j$ corrections, the degeneracies found in the CFT approximation will be broken for all non-zero energy states. We will indeed see this for $j=11$ in figure \ref{fig:largespinCFTcomparison}.   So we expect that for all the non-BPS states the degeneracies are progressively broken as we decrease the angular momentum from its maximal value, so as to eventually form the continuum that we see for low values of the angular momentum in the low energy melonic regime. In principle, one could consider the next term in the expansion of \nref{3jLargej}, which are corrections scaling like $m^3/j^2$, and see how the degeneracies are broken.

Let us summarize this discussion. Starting from the corner with maximal value of $J_3$ we get a very simple description of the Hilbert space. We can view it as two decoupled chiral CFTs. The $\phi_\pm $ bosons change the value of $J_3$ but not the energy. The $\phi_z$ boson contributes both to the energy and the value of $J_3$. Starting with a small number of BPS states at the extreme value of $J_3$, this number grows rapidly as we lower the value of $J_3$. In fact, we have a Cardy-like growth as in a CFT  to a number which, if extrapolated naively to $J_3 \sim 0 $, is exponential in $j$,  with the right $j$ dependence, in the exponent, as the number of BPS states discussed more accurately in \nref{BPSspa}. The advantage of the discussion in this corner is that we can give an explicit form of the BPS states. In some sense, we could call all these BPS states monotonous, 
as introduced and defined
in \cite{Chang:2024zqi}.   On the other hand, we do not presently have an explicit description for most of the BPS states, which occur when $J_3 \sim 0$, outside the region where the simple CFT description is valid. This CFT description also gives the near extremal states, which are separated from the BPS states by a gap scaling like $1/j$, which has the same $j$ dependence as in the central, $J_3 \sim 0 $, region where it is well approximated by the melonic approximation and the Schwarzian.   The near extremal entropy in the CFT region is also scaling like $ S \propto \sqrt{ j E } $,  which is similar in its $E$ and $j$ dependence to the  near extermal entropy in the region $J_3\sim 0$, in the melonic or Schwarzian approximations. Here,  it is obtained from the Cardy formula for the CFT that describes the $\phi_z$ boson. An important difference is that the non-zero energy states are very degenerate in the CFT approximation, while they are forming a continuum in the $J_3 \sim 0 $ region described by the Schwarzian. 

Of course, in general systems, it is very common for the description to become simpler at extreme values of some of the charges. For example, in a Heisenberg spin chain, the states near maximal angular momentum are very simply described in terms of magnons. 

These features of BPS states are somewhat reminiscent of some examples of black holes in string theory, where there are sometimes BPS states that can be described quite explicitly in gravity, such as those involving explicit fuzzball-like gravity solutions \cite{Bena:2022ldq},  while others are well approximated by black hole solutions. 

Perhaps a slightly more similar example could be a string with momentum $n$ and winding $w$ along an internal circle. In this situation, we can find BPS states that, at weak coupling, are described by an oscillating string with only left-moving excitations, but not right-moving excitations. This is a BPS configuration. If we added right-moving excitations, we would have non-zero energy above the BPS bound.  This is somewhat analogous to the non-BPS states we encountered earlier. When we include string interactions, in the regime of fixed string coupling and very large $n$ and $w$,  these non-BPS states are expected to form a continuum, somewhat similar to what we had above in the Schwarzian dominated regime. These solutions do not appear to have a trustworthy gravity description near the horizon, see, e.g. \cite{Sen:2005kj} for a related discussion.  These states can have a range of angular momenta in the non-compact directions. The typical state has a small angular momentum. There is a single special state with maximal angular momentum $J_3 =nw$.  It is described by a large circular string in the non-compact directions, which also has momentum and winding in the internal directions; see Appendix \ref{CircularString}.    It is large enough to be weakly coupled, especially for large values of $n$ and $w$. So we can consider excitations consisting of extra left and right moving modes. The left moving excitations are BPS, and the right moving ones take us away from the BPS limit. This is analogous to the CFT that we found near the maximal angular momentum in our case.

\section{Exact Diagonalization}\label{sec:exact}

Recall that Hilbert space dimension for each $j$ is $2^{2j+1}$. In addition, $j$ is odd. So, with each step in $j$,  the Hilbert space grows by a factor of 16. Due to the rapid growth, it is crucial to utilize the SO(3) and R-charge symmetries to separate the Hilbert space into fixed-charge sectors before performing the computations. Therefore, we do not use the Jordan-Wigner method of solving fermionic systems, which is commonly used for SYK models. Instead, we use the normal-ordered form of the Hamiltonian \eqref{eq:Hnormord} and the SO(3) Casimir,
\begin{align}\label{eq:Hnormord2}
    H&=\frac{\mathsf{J}}{3}\Big((2j+1)-3(N_\psi+j+\tfrac{1}{2})+3\sum_m O^\dagger_{j,m}O_{j,m}\Big)\;,\\
\label{eq:casimirnormord}
    J^2&=j(j+1)(N_\psi+j+\tfrac{1}{2})+2\sum_{m<n}mn\,\psi^\dagger_m\psi^\dagger_n\psi_n\psi_m
    \notag
    \\&\qquad
    +\sum_{m,n}\sqrt{(j-m)(j+m+1)(j+n)(j-n+1)}\psi^\dagger_{m+1}\psi^\dagger_{n-1}\psi_n\psi_m\;,
\end{align}
and we diagonalize them in the subspace with fixed $N_\psi$ and $J_3$, which involves a smaller number of states. 

We performed an exact diagonalization up to  $j=11$. At $j=11$, the full Hilbert space dimension is 8,388,608, while the largest fixed-charge block ($R=\pm1/6,\;J_3=0$) has dimension 32,540, so making use of the symmetries gives a significant speedup.\footnote{While one bottleneck of this method of course is the number of states, another bottleneck is the number of terms in $H$, which grows like $N^2$ in our model. For $\mathcal{N}=2$ SYK, it is not advantageous to use this method because the interaction has $N^3$ terms; instead, one typically uses the Jordan-Wigner construction of the fermion operators. The tradeoff is that the Jordan-Wigner method involves multiplying large matrices together to construct the Hamiltonian and other operators, and one cannot straightforwardly fix the fermion number (or any other charges), which is why we prefer the ``normal-ordered method'' for our model.}  

To be more explicit, for fixed $N_\psi=n-j-1/2$ we automated the following procedure:\footnote{We thank Ross Dempsey for suggesting some tricks for the implementation.} 
\begin{enumerate}
    \item Enumerate all Fock states with $N_\psi=n-j-1/2$ and $J_3=0$, a.k.a.\ all possible sequences of $n$ integers $m_i$ such that $-j\leq m_1<m_2<\ldots<  m_{n}\leq j$ and $\sum_{i=1}^nm_i=0$. This produces a basis for the $N_\psi=n-j-1/2$, $J_3=0$ sector of the Hilbert space.
    \item Compute matrix elements of  \eqref{eq:Hnormord2} and \eqref{eq:casimirnormord} between pairs of the Fock states generated in the previous step. Since the operators are normal-ordered, one simply has to read off the coefficient of each term while keeping track of minus signs. (Note that the terms with fewer than 4 fermions are already diagonal in the Fock basis.)  By the end of this step we have generated the sub-matrices of $H$ and $J^2$ in the $N_\psi=n-j-1/2$ and $J_3=0$ sector. 
    \item Calculate the energy eigenvalues and corresponding spins. Because there is some extra degeneracy between states with different spin (see section \ref{sec:largeU1}), one must break this degeneracy by first diagonalizing some generic linear combination of $H$ and $J^2$. This gives the simultaneous eigenvectors, and then one can compute the eigenvalues of those eigenvectors with respect to $H$ and $J^2$ separately.
\end{enumerate}
This gives us the energy spectrum without the charge-conjugation and SO(3) degeneracies. (Since the model is charge conjugation symmetric, we did the computation only for $N_\psi<0$.) The reason it is enough to diagonalize the $J_3=0$ sector is that our Hilbert space only contains integer spin multiplets, so each multiplet contains a $J_3=0$ state. To recover full state counts, we simply put those degeneracies back in by hand. 

In the $j\leq 11$ models, we found a small amount of additional degeneracy in the energy spectrum at fixed $R$, occurring at the energies and R-charges in \eqref{FirLi}. They occur when $O_{j,m}$ completely annihilates a multiplet. One example is the degeneracy for all $\ell\neq j$ in the $|R|=\frac{1}{3}(\frac{N}{2}-2)$ sector (or two fermion sector) which is explained by \eqref{eq:2bitEnergies}. It would be nice to explicitly derive all of these degeneracies by generalizing the discussion of \autoref{sec:largeU1} to smaller $|R|$. We expect that this boils down to writing many-particle wavefunctions which are such that the projectors onto spin $j$ for all two-particle combinations vanish.

\begin{figure}
\centering
    \includegraphics[width=.9\linewidth]{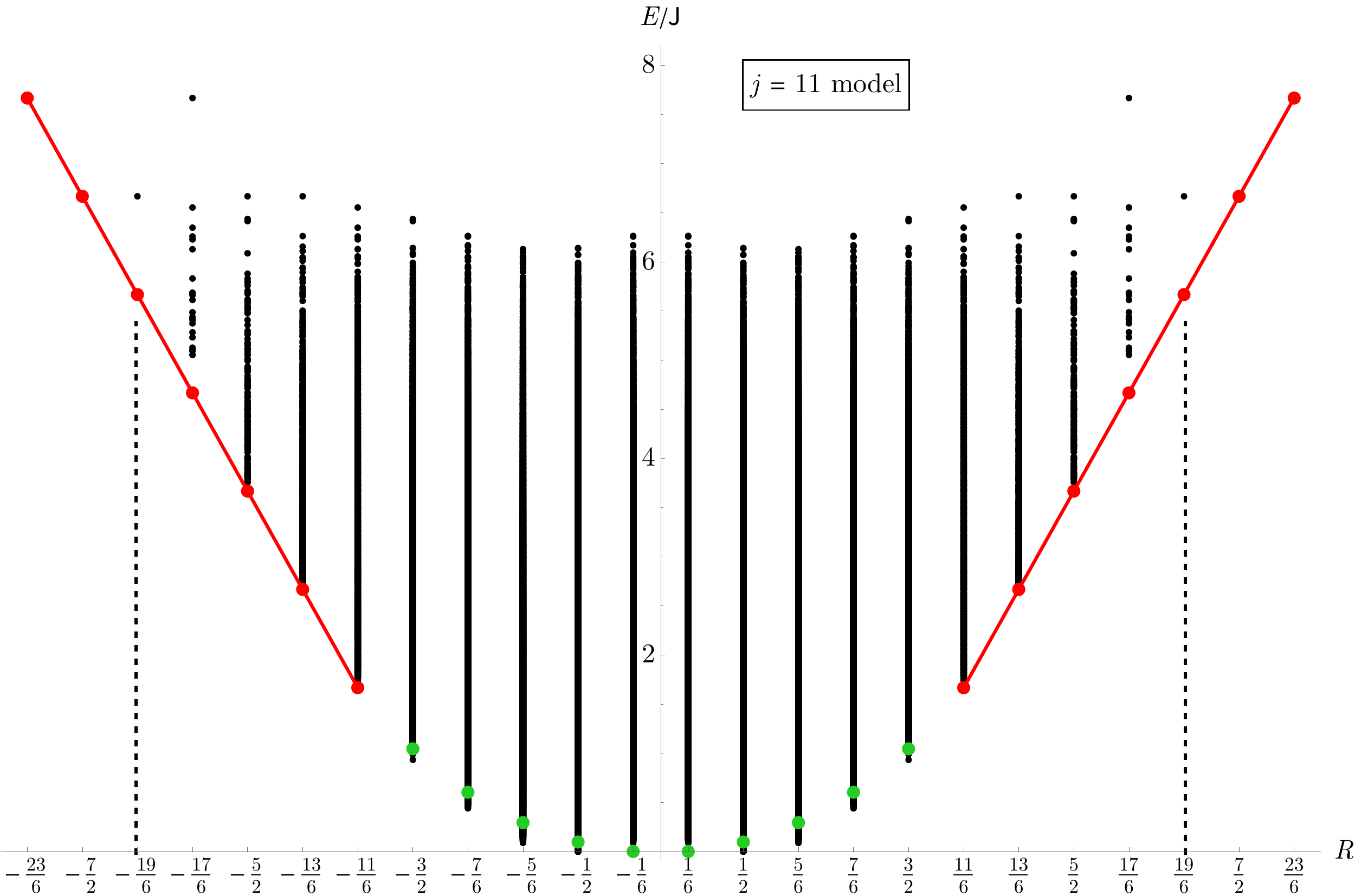}
    \caption{Spectrum vs. R-charge in the $j=11$ model. The red points are the states with $\ell=\ell_{max}^{(R)}$ and  $|R|\geq j/6$   whose energies lie on a linear trajectory \eqref{FirLi}. In particular, at $|R|=19/6$ we have the two fermion energies \eqref{eq:2bitEnergies}, so the red point is the highly degenerate set of $\ell\neq 11$ states while the black point in that sector is the one $\ell=j$ multiplet. 
    It turns out that the red points at $|R|=17/6,5/2,$ and $13/6$ are also highly degenerate across various $\ell$, due to the operator $O_{j,m}$ annihilating many different multiplets. Interestingly, the red point at $|R|=11/6$ is not degenerate. The green points are the maximum spin states with $\ell=\ell_{max}^{(R)}$ and  $|R|< j/6$ with energies on the nonlinear trajectory \eqref{eq:nonlinear}. Note that the green points are \textit{not} the minimum energy states at fixed $R$, while the red points are. 
    The dashed lines at $|R|\approx3.18$  denote the transition \eqref{RcMu} in the spectrum predicted by \cite{Heydeman:2022lse} that we discuss in section \ref{sec:chempot}. }
    \label{fig:j11spectrumvsRcharge}
    \end{figure}

\begin{figure}
    \hspace{2mm}\includegraphics[width=0.65\linewidth]{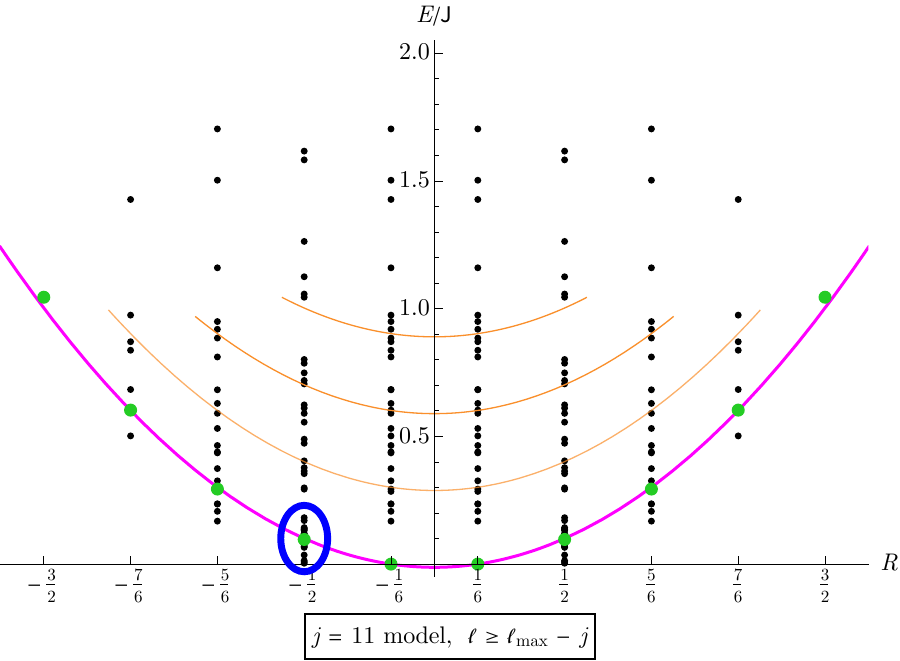}
    \hfill\includegraphics[width=0.24\linewidth]{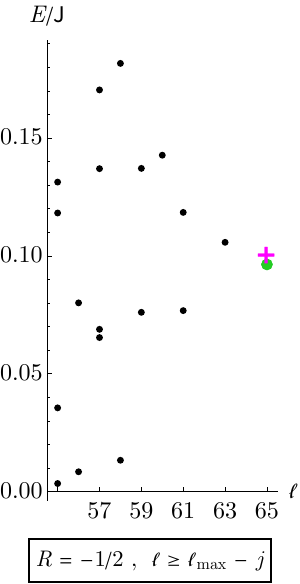}\hspace{2mm}
    \caption{(Left) Spectrum restricted to relatively large angular momentum $\ell\geq\ell_{max}-j$, so as to compare with the CFT results. Note that this is a stronger restriction at large $|R|$ because $\ell_{max}^{(R)}$ decreases as $|R|$ increases, see \eqref{eq:lmax(R)}, hence there are fewer black points at larger $|R|$. 
    We also restrict $|R|<j/6$. The green points are the states with $\ell=\ell_{max}^{(R)}$, which follow a nonlinear trajectory \eqref{eq:nonlinear}. The thick pink line shows the CFT prediction \eqref{HamLti} for the energies of states near the overall maximum spin  $\ell=\ell_{max}$. The CFT predictions \eqref{HamLti0} with $\tilde{N}=1,2,3$ for the higher energy states with $J_3\geq \ell_{max}-j$ are shown by the thin orange lines (recall that $\tilde{N}\mapsto \tilde{N}+1$ corresponds to $J_3\mapsto J_3-3$). One can see that in many places the black points appear to clump around the CFT lines. The remaining sparse black points far above the orange lines are likely an artifact of small $j$ and are not predicted by the CFT.\\
    (Right) We zoom in on the clump of points circled in blue, which are clustered around the maximum spin $R=-1/2$ state. We show the distribution of those states with respect to spin. The pink \textbf{\textcolor{magenta}{+}} is the CFT$_2$ prediction for the maximum spin state (green point), which in the CFT is degenerate with many lower-spin states. In other words, in the CFT limit, all the states shown in this plot should be at the pink energy. The states spread further out from the pink CFT prediction as we decrease the spin, signaling that the large degeneracy of CFT states is increasingly broken. 
    }
    \label{fig:largespinCFTcomparison}
\end{figure}

\begin{figure}
    \centering
    \includegraphics[width=.75\linewidth]{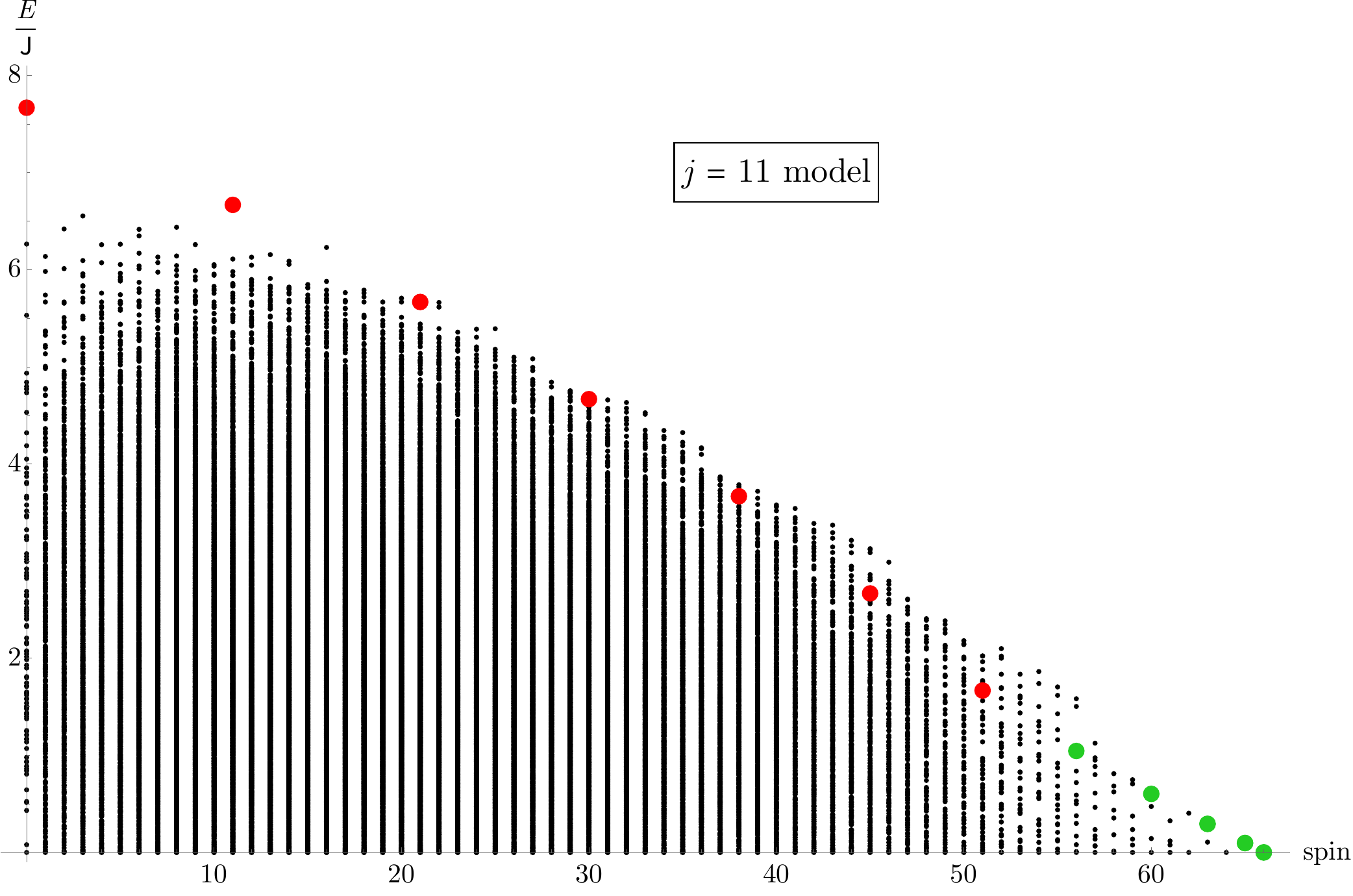}
    \caption{Spectrum vs. spin of the $j=11$ model. The red and green points correspond to the maximum spin states at fixed $|R|$ also shown in Figures \ref{fig:j11spectrumvsRcharge} and \ref{fig:largespinCFTcomparison}. 
    }
    \label{fig:j11spectrumvsspin}
\end{figure}

Figures \ref{fig:j11spectrumvsRcharge}, \ref{fig:largespinCFTcomparison}, and \ref{fig:j11spectrumvsspin} show the spectrum of the $j=11$ model. In each R-charge sector there is a maximum spin \eqref{eq:lmax(R)} $\ell_{max}^{(R)}=(N-6R)(N+6R)/8$. Across all sectors, the \textit{overall} maximum spin is $\ell_{max}=j(j+1)/2$. For large U(1)$_R$ charge $|R|\geq j/6$, the $\ell=\ell_{max}^{(R)}$ states follow the linear trajectory \eqref{FirLi}, and these are also the lowest energy states for a given $R$ charge as we can see from  looking at the red dots in figure \ref{fig:j11spectrumvsRcharge}.  In the small U(1)$_R$ charge and large SU(2) charge regime $\ell \sim \ell_{max} = j(j+1)/2$, the effective CFT$_2$ predicts that the smallest energy for a given $R$ charge is \eqref{HamLti}. These are shown in pink in figure \ref{fig:largespinCFTcomparison}. In the right panel of that figure, we see that we approach the CFT result as we increase the spin, but that the degeneracy predicted by the CFT is split, and that the splitting becomes larger as we decrease the spin.  From these splittings, we also see that the green points, which are the exact energies \nref{eq:nonlinear} of the maximal angular momentum states for a given $R$ charge, are not the minimum energy states for those $R$ charges. In other words, we have black dots below the green dots in figures \ref{fig:j11spectrumvsRcharge} and \ref{fig:largespinCFTcomparison}. 

\begin{figure}
    \centering
    \includegraphics[width=0.8\linewidth]{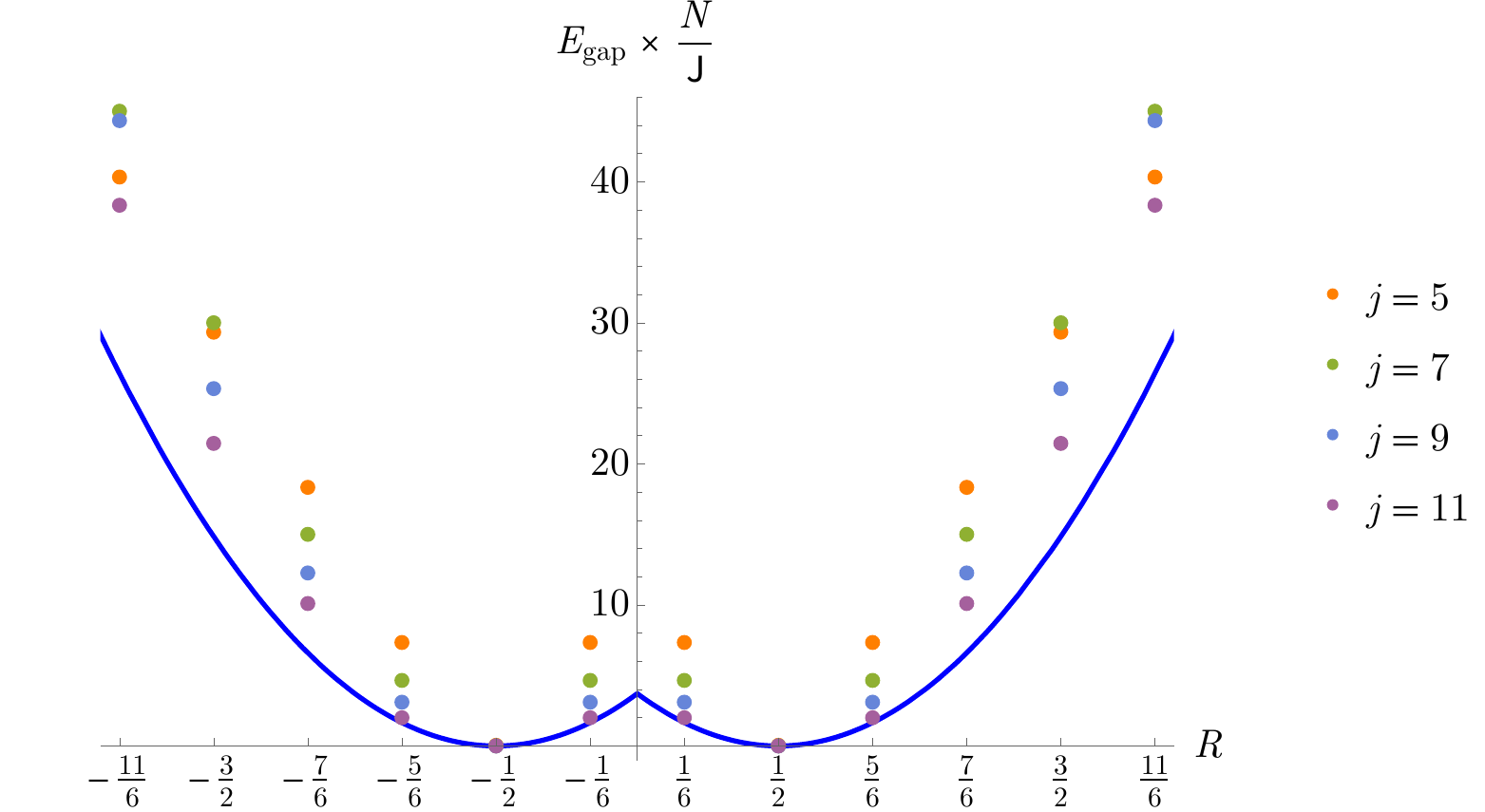}
    \caption{Convergence of the energy gaps as $j$ increases. The blue curve is the Schwarzian prediction \eqref{Egap}. We see that the numerical points are approaching the Schwarzian prediction, and the agreement is better for small $|R|$, as expected. 
    }
    \label{fig:gapconvergence}
\end{figure}
\begin{figure}
    \centering
    \includegraphics[width=0.48\linewidth]{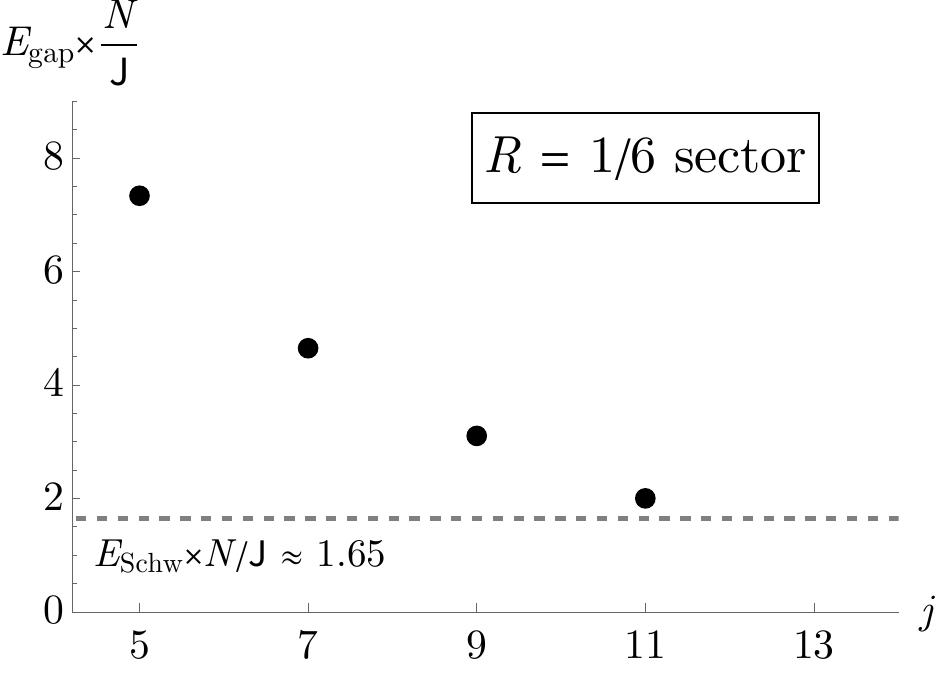}\hfill
    \includegraphics[width=0.48\linewidth]{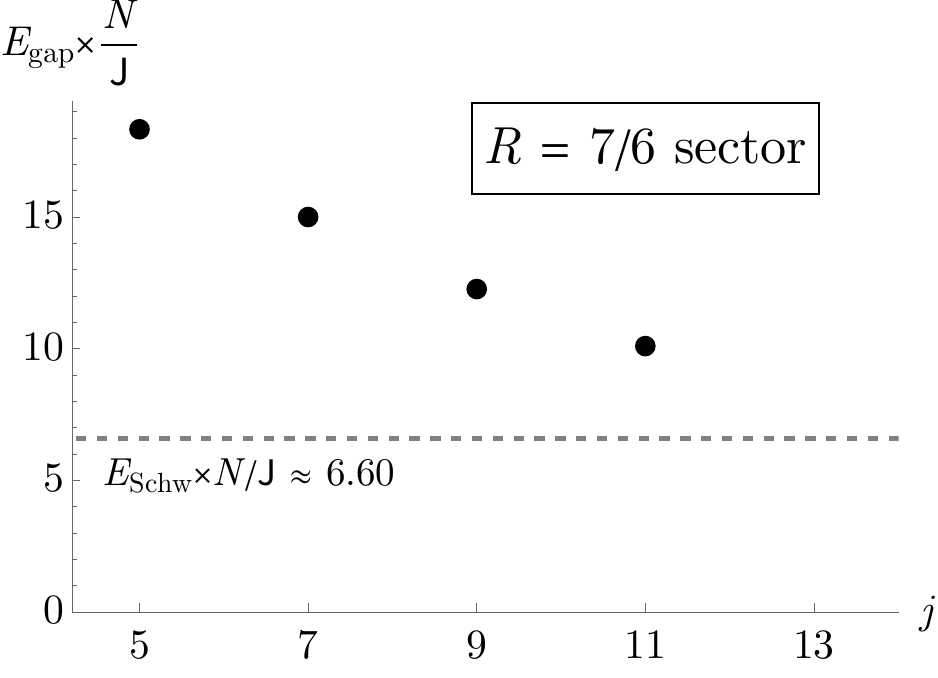}
    \caption{Convergence in $j$ of the energy gaps towards the Schwarzian prediction for the $R=1/6$ (left) and $R=7/6$ (right) sectors. Although we can qualitatively see that the gaps are moving towards the Schwarzian predictions, we could not get a sensible $j\to\infty$ extrapolation. In particular, fitting to a constant $+\,\log(j)/j$ and/or $+\,1/j$ results in a \textit{negative} prediction for $E_\text{gap}$ at $j\to\infty$. In the $R=1/6$ sector, it may be possible that we are actually following a state that becomes BPS at some finite, larger value of $j$, rather than one that approaches a finite $E_\text{gap}$ as $j\to\infty$.  
    This would explain why it seems that the numerical points are on a trajectory that goes below $E_\text{Schw}$ in the plot on the left.  It is also possible that the convergence may oscillate in $j$, as the non-melonic corrections have indefinite sign (see \eqref{12jasmtot} for example). 
    }
    \label{fig:placeholder}
\end{figure}
\begin{figure}
    \centering
    \includegraphics[width=0.8\linewidth]{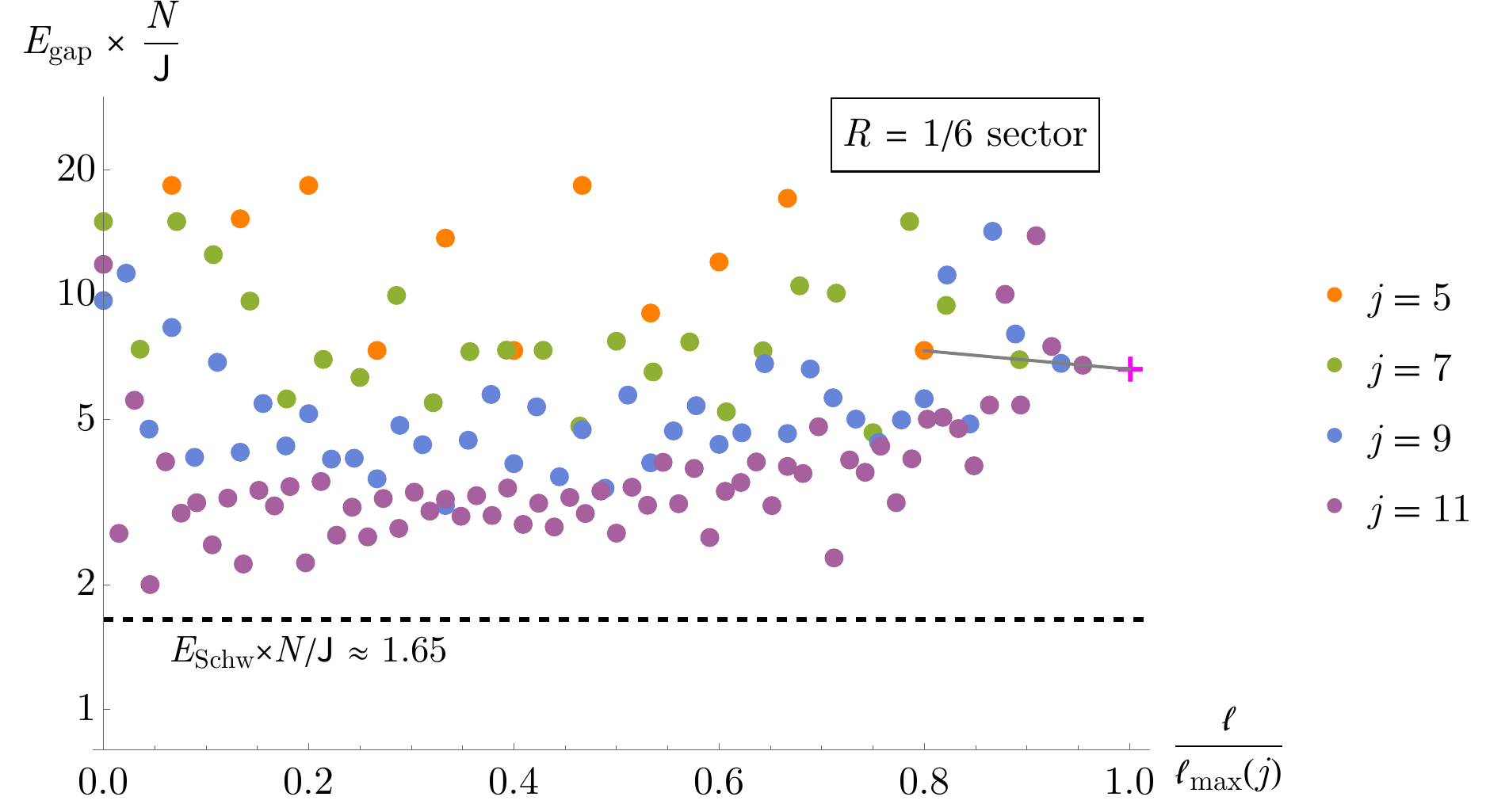}
    \caption{
    $E_{gap}$ as a function of spin in the $R=1/6$ sector. There are three regimes we expect to behave differently: small $\ell\sim \mathcal{O}(1)$, ``medium'' $\ell$, and large $\ell\sim \mathcal{O}(\ell_{max})$. The Schwarzian prediction is valid for $\ell\ll\ell_{max}$. Indeed, the large $\ell$ points are not approaching the dotted line. The convergence seems best for medium $\ell$. It seems not as good for small $\ell$, which we think is because there are fewer multiplets for each $\ell\sim\mathcal{O}(1)$ (see \autoref{SU2plts}). We indicate the large $ j$ CFT$_2$ prediction for the gap by the pink \textbf{\textcolor{magenta}{+}}. The points connected by the solid gray line are converging 
    to the CFT value. 
   }
    \label{fig:Egapsvsspin}
\end{figure}

The Schwarzian prediction of the lowest nonzero energies was given in \nref{Egap}. 
We compare this prediction with the exact energy gaps in the $5\leq j\leq 11$ models in Figure \ref{fig:gapconvergence}. 
 One may notice that the agreement is not as good in our model as it is in $\mathcal{N}=2$ SYK (see figure 9 in \cite{shortpaper}). Recall from \eqref{12jasmtot} that the first non-melonic correction has a leading order term $(\log j)/j$, while in SYK all of the non-melonic diagrams are suppressed by $1/N$, so we should expect slower convergence. 
 It is also possible that in 
 figure \ref{fig:placeholder} we are following a state that will become BPS at some larger value of $j$ (in the $R=1/6$ sector in particular), rather than a state that approaches the $E_\text{gap}$ predicted by \eqref{Egap}. 

Recall that the Schwarzian prediction for the gap \eqref{Egap} formally closes for $|R|= \half$, but we cannot tell from the Schwarzian approximation whether it is really zero or not. The numerical result shows that there are $\mathcal{O}(1)$ BPS multiplets at this R-charge for $j=7,9$ but not $j=3,5,11$. This probably has a symmetry based explanation, which would be interesting to find. It is a different situation from the $\mathcal{N}=2$ SYK model, where the $R=1/2$ sector has $\mathcal{O}(1)$ BPS multiplets for $N=1$ mod 4 \cite{Fu:2016vas}.   
More generally, at some values of $j$, we find a small number of ``sporadic'' BPS multiplets with $|R|\neq1/6$. They are sporadic in the sense that they do not appear where one would naively expect them from the Schwarzian analysis. We tabulate these multiplets, along with their charges and degeneracies  
\begin{align}\label{sporadic}
    \begin{tabular}{c|ccc}
      $ j$  &  $R$ &  $\ell$ & \# \\
    \hline\hline
       5  &  none \\
    \hline
       7  & $+1/2$ & 7 & 1 \\
            & $-1/2$ & 7 & 1 \\
    \hline
       9  &  $+1/2$ & 9 & 2 \\
            &  $-1/2$ & 9 & 2 \\
            & $+5/6$ & 0 & 2 \\
            & $-5/6$ & 0 & 2 \\
    \hline
       11 &  none \\
    \end{tabular}
\end{align}
We have checked that these sporadic states do not violate the index calculated in section \ref{sec:witten}. Of course, that index cannot predict the charges and numbers of these sporadic multiplets because the $\mathbb{Z}_3$ grading allows for cancellations between sectors differing by 1 unit of R charge. For example, $R=\pm1/2$ BPS states cancel with each other since they are charge conjugates so appear with the same degeneracies.  Therefore,  the $R=\pm1/2$ contribution to the index is always zero. Meanwhile, $R=\pm5/6$ BPS states cancel with extra $R=\mp1/6$ BPS states. In other words,   (at $j=9$) the difference in the BPS state counts in these sectors is exactly $2\cdot 3^j$ as predicted by the index.  It would be interesting to see whether these sporadic states only appear for small $j$, especially those with $|R|=5/6$.   Note that in $\mathcal{N}=2$ SYK there are never any $|R|=5/6$ BPS states, so this is a surprising result.

We also looked at the energy gaps as a function of the spin, which we plot in figure \ref{fig:Egapsvsspin} for the $R=1/6$ sector. As discussed around \nref{LLLSpin}, one might have expected that the energy as a function of the spin is proportional to the second Casimir, as in tensor models and in SYK models with global symmetry \cite{Gu:2019jub,Yoon:2017nig,Narayan:2017hvh,Yoon:2017gut,Klebanov:2018nfp,Gaitan:2020zbm}. 
However, we do not find this behavior numerically, suggesting that the effective action is not simply the SU(2) casimir. This is also supported by the fact that there are BPS states with a wide range of spins.  In the figure \ref{fig:Egapsvsspin}, we do see that the gaps for $\ell\ll\ell_{max}$ seem to be converging towards the Schwarzian prediction \eqref{Egap} for that R-charge sector, but the convergence is very non-uniform, especially for $\ell\sim\mathcal{O}(1)$. The reason may be that the $\ell\sim\mathcal{O}(1)$ sectors do not have many multiplets, while the medium-sized $\ell$ are far more populous, as discussed in \autoref{SU2plts} (see also the plots of the degeneracies in \autoref{Repdegs}). 

\begin{figure}[h]
    \centering
    \includegraphics[width=0.7\linewidth]{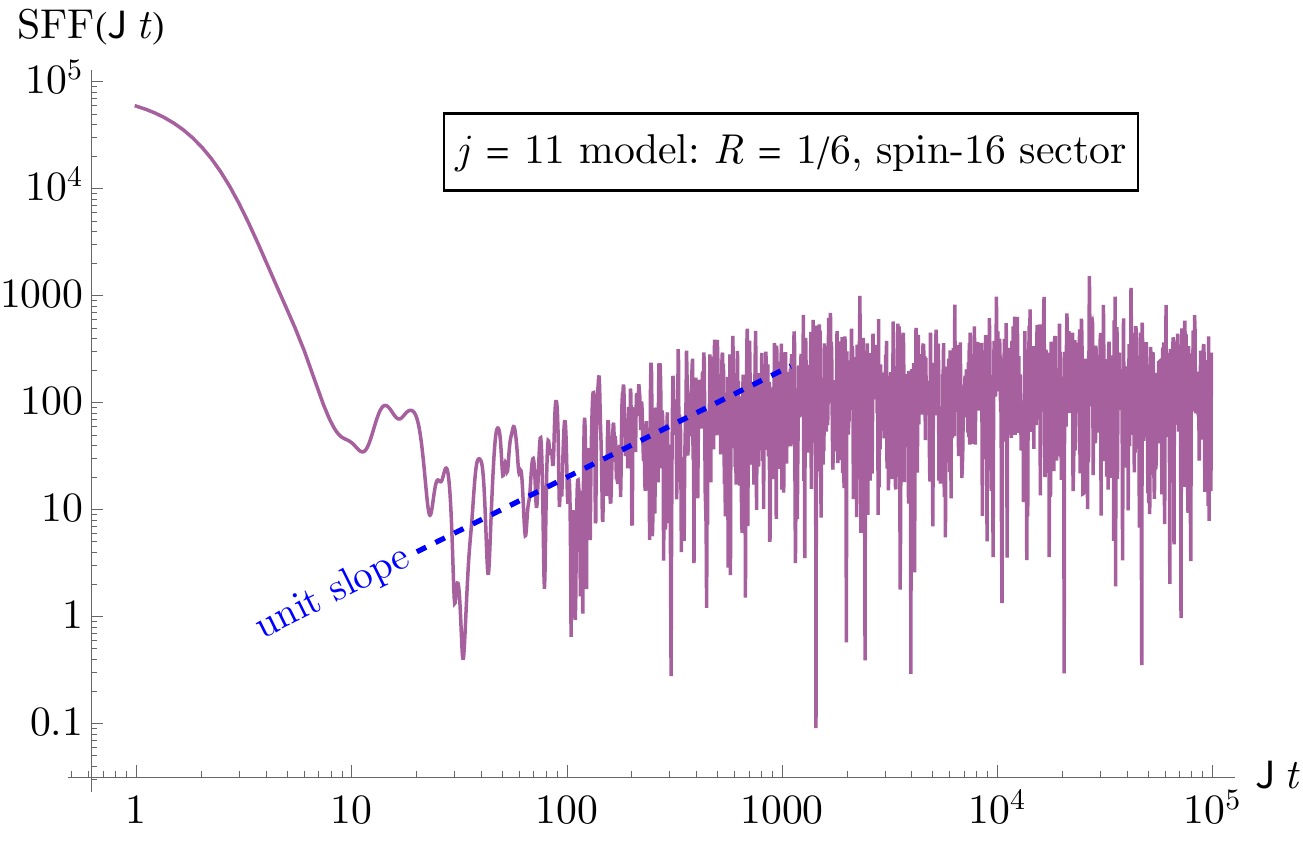}
    \caption{Spectral form factor (SFF) in the $R=1/6,\;\ell=16$ sector of the $j=11$ model (with a Gaussian filter of width $\mathsf{J}$), plotted on log-log axes. There are 1078 non-BPS multiplets in this sector.  We overlay a line of unit slope in dashed blue for comparison.  We find a ramp and plateau behavior in the SFF, suggesting that the model in this sector is chaotic. If one assumes GOE (Gaussian Orthogonal Ensemble) statistics, one should find  $\log[\text{SFF}(\mathsf{J}\,t)]=\log(w\,\mathsf{J}\,t)-\log(2\sqrt{2\pi})$. The blue line compares the slope, but not the intercept. 
    Because $N=2j+1=23$ is not particularly large, and because this is the smallest $j$ for which there are enough states to get a sensible SFF, we are not able to decide whether the ramp starts at $\mathsf{J}\,t\sim\mathcal{O}(1)$ or $\sim \mathcal{O}(N)$. 
    }
    \label{fig:sff}
\end{figure}

Since in the melonic limit the model is the same as $\mathcal{N}=2$ SYK, we expect it to have a maximally chaotic sector at large $j$. We computed the Gaussian-filtered spectral form factor
\begin{align}
    \mathrm{SFF}(\mathsf{J}\,t)=\bigg|\sum_{\lambda_i}e^{-\lambda_i^2/w^2-i\mathsf{J}\,t}\bigg|^2\;,
\end{align}
where $\lambda_i$ are the non-BPS eigenvalues of $H$ in a particular R-charge and SU(2)-charge sector, and $w\sim \mathsf{J}$ is some choice of width for the Gaussian filter. This is displayed in Figure \ref{fig:sff} for $R=1/6,\;\ell=16$, and $w=\mathsf{J}$. We find the ramp-and-plateau behavior that is characteristic of a chaotic system, but we are not able to classify it as strongly or weakly chaotic \cite{Kos:2017zjh,Cotler:2016fpe,Saad:2018bqo,Chen:2024oqv,Gharibyan:2018jrp,Nosaka:2018iat}. 

One could also look for signatures of chaos in the BPS sector via the LMRS criterion \cite{shortpaper}. We did not attempt this because at $j=11$, the most populated fixed-charge sector only has $\mathcal{O}(100)$ multiplets, which makes it difficult to see chaos.

\section{Conclusions and Discussion } \label{sec:conclusions}

\subsection{Conclusions}

 We considered a supersymmetric model with a supercharge given by a 3j symbol of SU(2) for large $j$, inspired by the models discussed in \cite{Amit:1979ev,Benedetti:2020iku}. The model has a global $SU(2) $ symmetry. 
 
 An interesting aspect of these models is that their free energy is dominated by melon diagrams (for small enough chemical potentials). In section \ref{ArrowsSec}, we gave an intuitive explanation in terms of a certain kinematic enhancement for melon diagrams, once we take into account the  $SU(2)$ angular momentum constraints. We have also given an estimate of the leading corrections to the melonic expansion which are suppressed by an extra factor of $(\log j)/j$. It would be nice to give a more intuitive explanation for the appearance of this $\log j$ prefactor. 

 One question we can study in these models is the set of BPS states. Given the relatively simple structure of this theory, one might expect that it could be possible to give an explicit description of the BPS states. 
 We analyzed the theory in a specially solvable corner, which displays a dramatic simplification. This is the region where the angular momentum is close to its maximum.  In this corner, the full theory simplifies and reduces to a simple 1+1 dimensional CFT, which is the product of two sectors, with one of the sectors having zero Hamiltonian. This last sector spans the BPS states and also produces large degeneracies for all excited states. 
 We expect, and have checked numerically for $j=11$, that these degeneracies (for non-BPS states) are split by further corrections. 

We have also examined various other extreme values of the quantum numbers, where the model becomes simple, such as the largest $R$ charge or the largest angular momentum for a given $R$ charge.

We have also performed numerical simulations of the model for low values of $j$, up to $j=11$, and have verified some of the analytic expectations. In addition, these numerical results display some features that we have not explained. For example, there are a small number of sporadic BPS states at $R$ charge $|R|=1/2$, and surprisingly also $|R|=5/6$, where we expected no BPS states from the Schwarzian analysis.   One would therefore like to understand whether this feature persists for larger values of $j$.

\subsection{Possible Generalizations for future work}

One simple generalization would be a model with different types of fermions, each with a different angular momentum. Then the supercharge is given by 
\be 
 Q \sim \sum_{m_i} \begin{pmatrix}
            j_1 & j_2 & j_3 \\ m_1 & m_2 & m_3 
        \end{pmatrix} \psi^1_{m_1} \psi^2_{m_2} \psi^3_{m3}
        \ee 
where $j_1, j_2, j_3$ are all large. The arguments discussed in section \ref{Melonic} imply that this model should also have a melonic expansion. The only difference is that now the arrows in figure \ref{Arrows} would have different lengths and the angles in the analog of figure \ref{Arrows}(a) would not all be equal. Of course, one still needs $j_1,j_2,j_3$ to be far from saturating the triangle inequality, otherwise there will not be an enhancement for melon diagrams coming from a large degeneracy of arrow configurations.

Another simple generalization would be to change the group from $SU(2)$ to $SU(N)$, and consider representations with large quantum numbers or large Dynkin labels.\footnote{The model with $Q\sim Tr[\psi^3]$, with $\psi$ in the adjoint, discussed in \cite{Chen:2025sum},  can also be written in terms of 3j symbols for the adjoint. However, for the adjoint case, we do not expect melon diagrams, since the quantum numbers are not large. } 

In the fuzzy sphere picture, our model bears some resemblance to the problem of the ABJM theory \cite{Aharony:2008ug} in the presence of a magnetic field, as described in the context of magnetic black holes in the bulk \cite{Benini:2015eyy}. One important difference is that in the ABJM case, the interaction is local on the sphere.  

Though we discussed the supersymmetric case here, it should also be possible to consider non-supersymmetric models where the Hamiltonian is given by 
\be 
H = \mathsf{J} \sum_{m} O_{j_3 m}^\dagger  O_{j_3 m} ~,~~~~~~~O_{j_3 m} =\begin{pmatrix}
            j_1 & j_2 & j_3 \\ m_1 & m_2 & m 
        \end{pmatrix} \psi_{j_1 m_1} \psi_{j_2 m_2}
\ee 
where $j_1$ and $j_2$ could be equal or different. We expect these cases to have somewhat similar physics, though not precisely the same. They are also expected to give rise to melonic diagrams.  Note that the particular case of $j_3=0$, and  $j_1=j_2$, reduces to a $O(N)$ model, which looks reminiscent of the one encountered in the BCS theory, as discussed by \cite{Shankar:1993pf}. 

This type of model could also exist in other dimensions. In fact, they were originally introduced as bosonic models in six dimensions \cite{Amit:1979ev}. A particularly interesting case might be the model in $1+1 $ dimensions. Similar models were discussed in \cite{Murugan:2017eto}. There, it was noted that global symmetries, such as our SU(2) symmetry, might imply that the model does not really have conformally invariant IR fixed point. Such 1+1 dimensional models are somewhat similar to what we would expect for a Fermi surface with rotational SU(2) symmetry.  If we interpret the sphere as arising from the momentum directions, then perhaps the strange non-local interactions we discussed could arise as local interactions induced by other fields in position space.

\section*{ Acknowledgments }

We would like to thank K. Budzik, J. Cotler, R. Dempsey, D. Gaiotto, A. Herderschee, M. Heydeman, I. Klebanov, L. Pando Zayas, X.L. Qi, E. Witten, and G.Y. Zhou for discussions. In particular we thank L. Pando Zayas for pointing out the original model of \cite{Amit:1979ev}, leading to this work.

The work of JM was supported in part by U.S. Department of Energy grant
DE-SC0009988. The work of LLL was supported by a Graduate Research Fellowship from the National Science Foundation under grant DGE-2241144 and a Rackham Merit Fellowship from the University of Michigan. LLL thanks the Institute of Advanced Study for hospitality during part of this collaboration. 

This research was supported in part through computational resources and services provided by Advanced Research Computing at
the University of Michigan, Ann Arbor. 

    \appendix
    
\section{Saddle point approximation for $SU(2)$ state counts}\la{SPapp}

In this appendix, we explain the saddle point approximations for the large $j$ count of states in each $J_{3}$ sector. See equations  \nref{gauss}, \nref{cardyd}, and \nref{BPSspa}. First, we will consider the entire Hilbert space, and then restrict to the BPS subspace.

Changing variables in  \nref{dnexact} to $x = - i \theta j$,
\begin{align}
    d_{m} &=\frac{1}{2 \pi i} \frac{1}{j} \int dx \exp \lp \frac{x m}{j} + \sum_{n=-j}^{j} \ln (1+e^{-\frac{x n}{j}})\rp  \\
    &\approx \frac{1}{2 \pi i}\frac{1}{j} \int_{\gamma-i\infty}^{\gamma+i\infty}  dx \exp\lp j S(x)\rp ~~~\text{ where } ~~~S(x) = x \tilde m + \int_{-1}^{1} dy \ln(1+e^{-x y}), ~~~~\tilde m = \frac{m}{j^{2}}\la{ac}
\end{align}
In the second equality we have used the Euler-Maclaurin formula to approximate the sum by an integral. The integration contour runs along the vertical line $\Re(x) = \gamma$ where $\gamma$ is a nonzero real number. 
The saddle point equation is
\begin{align}\la{SPE}
    S'(x) = 0 \quad \RA \quad \tilde m = -\frac{1}{x}\lp \ln(1+e^{-x}) + \ln(1+e^{x}) \rp  + \frac{1}{x^{2}} \lp \text{Li}_{2}(-e^{-x}) -\text{Li}_{2}(-e^{x}) \rp 
\end{align}
For $0< |\tilde m| < \frac{1}{2}$, \nref{SPE} has a real solution. 
\nref{SPE} can be numerically inverted and plugged back into \nref{ac} to find the classical action $S_{0}(\tilde m)$. This exact solution to the saddle point equation is plotted as the gray curve in figure \ref{SPplot}. 

We can also derive approximate closed form expressions for $S_{0}(\tilde m)$ in various limits. \nref{SPE} implies that  as  $|x_0| \to \infty$, $|\tilde m| \to 1/2$, and as $|x_0| \to 0$, $|\tilde m| \to 0$. We first consider the regime of small $x_0$ and $\tilde m$. In this case, the saddle point equation becomes  $\tilde m \approx -\frac{x}{6}$, and we find the classical action\footnote{In this limit, it is convenient to include also the first $1/j$ correction to the classical action, because it contributes an overall factor of 2 to $d_{m}$ that is necessary to find agreement with the integral formula \nref{J3count}. More precisely, the first $1/j$ correction to the action is $\tilde S(x) = \frac{1}{2}\ln(2+e^{x}+e^{-x})$, which in the $x \ll 1$ limit becomes $\tilde S_{0} \approx \log(2)$.}
\begin{align}\la{approx1}
     S_{0} \approx  \log(4) -3 \tilde m^{2}~~~~~~~~~~|\tilde m| \ll 1/2
\end{align}
 On the other hand, for $|x_0| \gg 1$, the saddle point equation reduces to 
 $ |\tilde m| = \frac{1}{2}-\frac{\pi^{2}}{6 x^{2}}$, which when plugged back into the classical action gives
\begin{align}\la{approx2}
     S_{0} \approx  2 \pi \sqrt{\frac{1}{6}\lp \frac{1}{2}-|\tilde m| \rp }~~~~~~~~ \frac{1}{2}-|\tilde m| \ll \frac{1}{2}
\end{align}
Including the gaussian integrals around these saddle points gives equations \nref{gauss} and \nref{cardyd}. 

For large $j$, per \nref{Drel} we can approximate $D_{\ell}$ by the first derivative of $d_n$, $ D_{\ell} \approx -\frac{d}{d\ell}d_{\ell} $. So we have, in addition, the following saddle point approximations for the number of spin $\ell$ representations\footnote{Near $\ell = 0$, the first derivative of $d_{\ell}$ is zero, so we must expand \nref{Drel} to next order, giving $D_{0} \approx \frac{1}{2}2^{2j+1}\frac{1}{\sqrt{2 \pi}\sigma^{3}}$.} 
\begin{align}\la{Dlspa}
   \hspace{-9mm}
 D_{\ell}  \approx \begin{cases}2^{2j+1}\frac{\ell}{\sigma^{3}\sqrt{2 \pi}}\exp(-\frac{1}{2}\frac{\ell^{2}}{\sigma^{2}}) ~, ~~~~~~~\sigma = \sqrt{\frac{j^{3}}{6}}~~~~~~~~~~~~~~~~~~~~~~~~~~~~~~~~~~~~~~~~~~~~~~ \ell \ll \ell_{max}\\ 
    \frac{1}{2}\frac{1}{6^{1/4}}(\ell_{max}-\ell)^{-5/4}\lp \frac{\pi}{\sqrt{6}} - \frac{3}{4}(\ell_{max}-\ell)^{-1/2}\rp  \exp(2 \pi \sqrt{\frac{\ell_{max} - \ell}{6}}) ~~~~1 \ll \ell_{max}- \ell \ll \ell_{max} 
    \end{cases}
\end{align}
The calculation is similar when we restrict to BPS states. To leading order in $j$, we have
\begin{align}
    d_{m}^{BPS} \approx  \frac{1}{2 \pi i}\frac{1}{3}\frac{1}{j}\int dx ~e^{j \mathcal{S}(x)} ~~~~~~~~~~~\mathcal{S}(x) = x(\tilde m+1/2) + \int_{0}^{1}dy \log(1+e^{-x y} + e^{-2 xy})
\end{align}
In the regimes of large and small $|\tilde m|$, we find the classical actions
\begin{align}
    \mathcal{S}_0 &\approx -\frac{9 \tilde m^{2}}{4} + \log(3)~~~~~~~~~~~~~~~~~~~~|\tilde m| \ll 1/2 \la{BPSa1}\\
    \mathcal{S}_{0}& \approx    2 \pi \sqrt{\frac{1}{9}\lp\frac{1}{2}-|\tilde m |\rp}~~~~~~~~~~~~~~~~\frac{1}{2}-|\tilde m|\ll \frac{1}{2}\la{BPSa2}
\end{align}
and the integral around the saddle points in the Gaussian approximation produces \nref{BPSspa}. As before, the saddle point approximation for the number of spin $\ell$ multiplets in the BPS subspace is given by $D_{\ell}^{BPS} \approx -\frac{d}{d\ell}d_{\ell}^{BPS}$.

\begin{center}
\includegraphics[width=\linewidth]{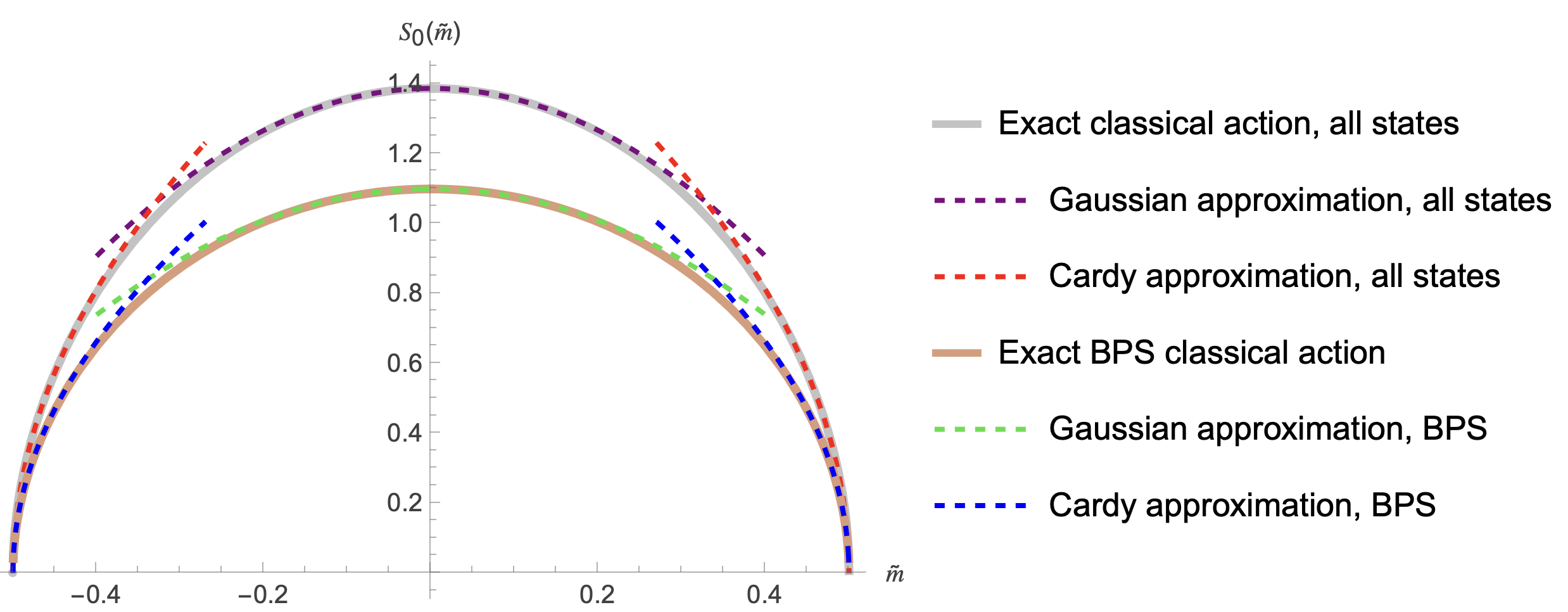}
\vspace{1mm}
\includegraphics[width=0.49\linewidth]{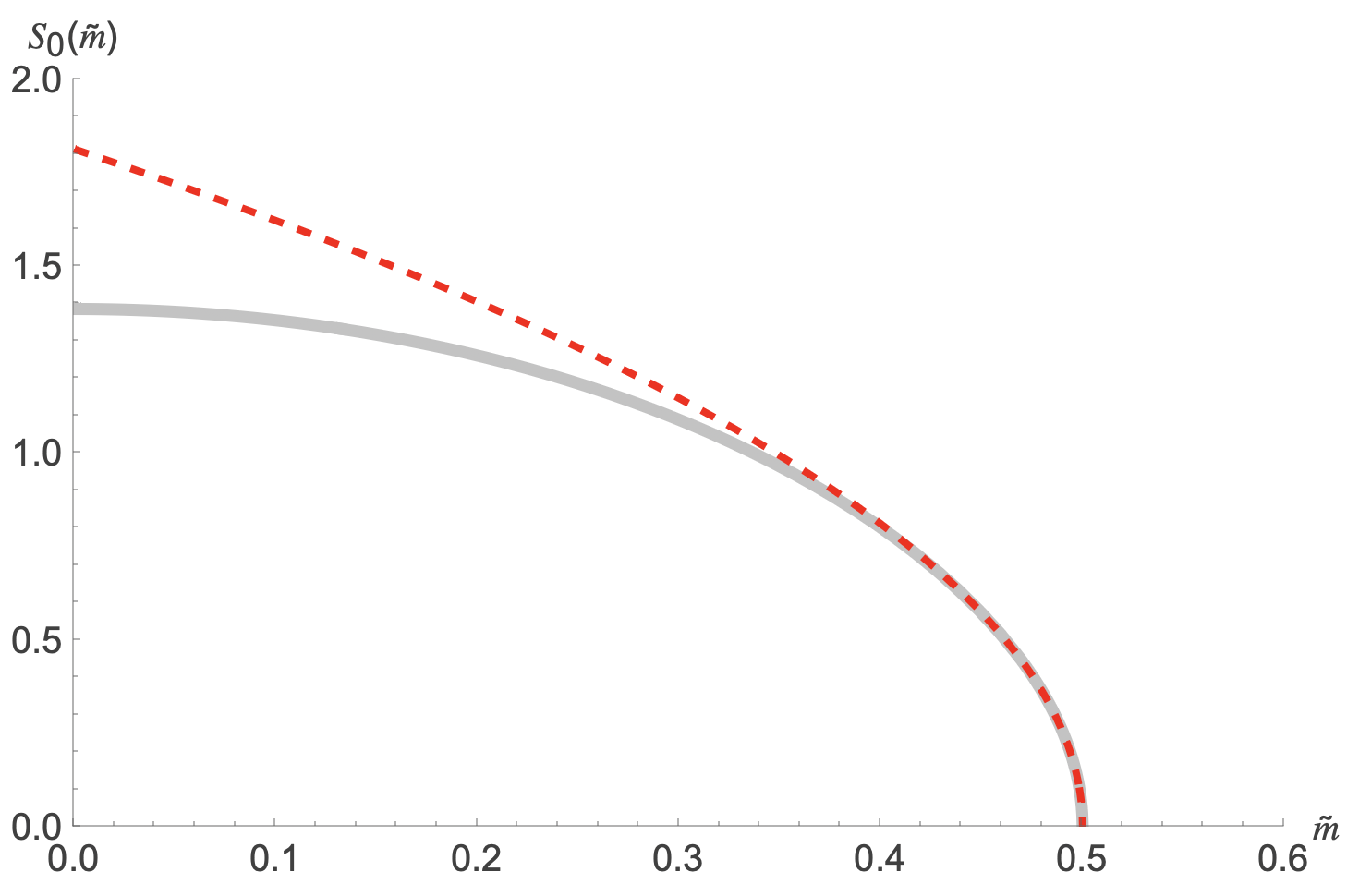}\hspace{1mm}
\includegraphics[width=0.49\linewidth]{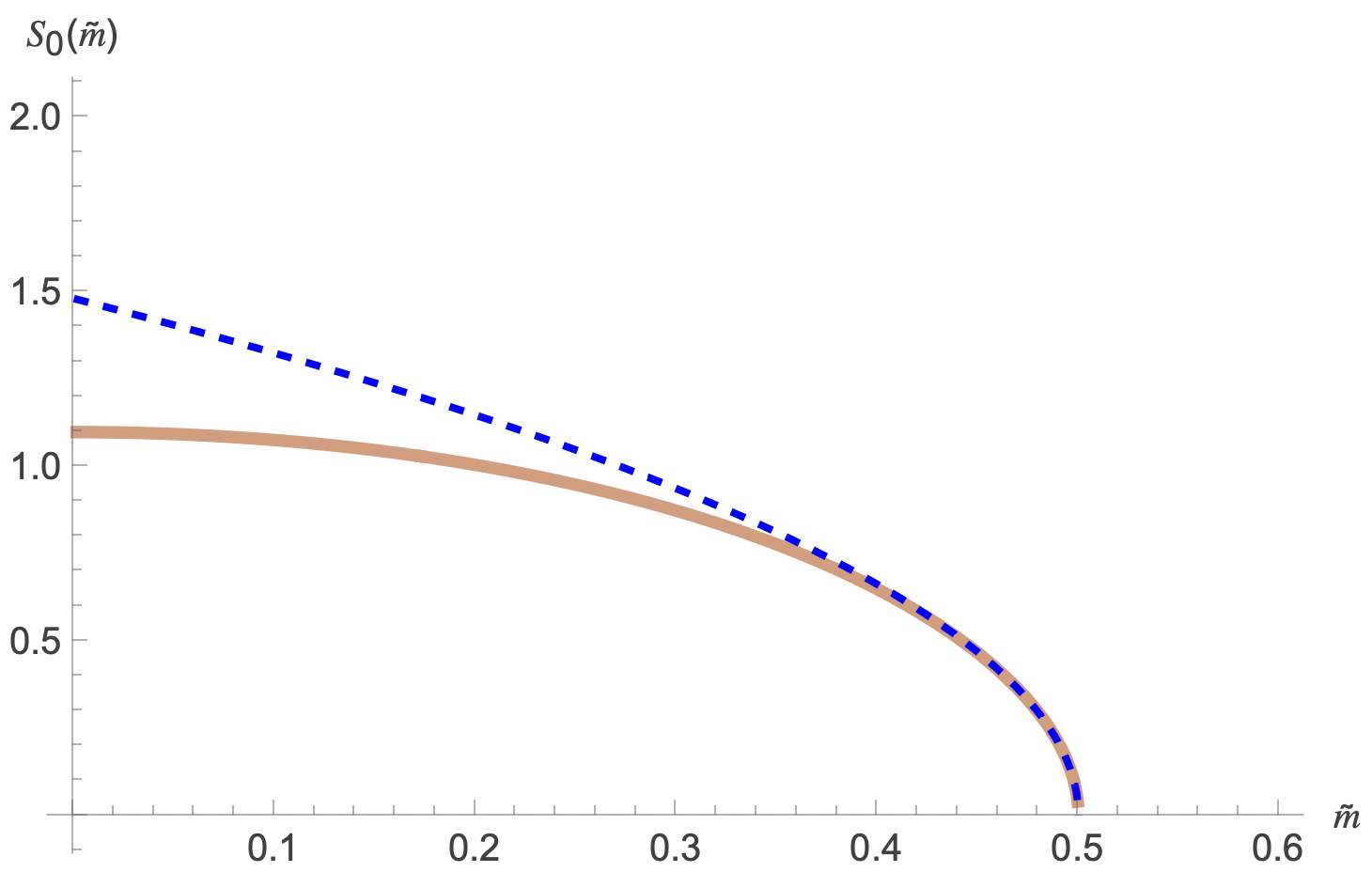}
\captionof{figure}{The classical action on the entire Hilbert space is shown in gray, and the action of the BPS states in brown. The approximate expressions in the large and small $|\tilde m|$  regimes, given in \nref{approx1}, \nref{approx2}, \nref{BPSa1}, and \nref{BPSa2} are plotted in purple, red, green, and blue, respectively. The bottom diagram is the same plot, but we separately consider all states (left) and BPS states only (right) in order to show the extrapolation of the Cardy approximations \nref{approx2} and \nref{BPSa2} to the $\tilde m \sim 0$ region.}
\la{SPplot}
\end{center}

\section{Plots of $SU(2)$ state counts at finite $j$}\la{SU2plts}

In this appendix, we plot the probability distributions corresponding to the counts $d_{n}$ and $D_{\ell}$ for various values of $j$, both on the entire Hilbert space and restricted to the BPS subspace. These distributions correspond to the integral formulas \nref{dnexact} and \nref{DBPSL}. $D_{\ell}$ is then obtained by \nref{Drel}, or equivalently, by writing the partition function as \begin{align}
   Z= \sum_{\ell}D_{\ell}\chi_{\ell}~~~~~~ \text{ where } ~~~~~\chi_{\ell} =  \sum_{m = -\ell}^{\ell}e^{i m \theta} = \frac{\sin((2 \ell + 1)\frac{\theta}{2})}{\sin(\frac{\theta}{2})}
\end{align}
and using the orthogonality of $SU(2)$ characters. 
\begin{center}
\centerline{\includegraphics[width=1.1\linewidth]{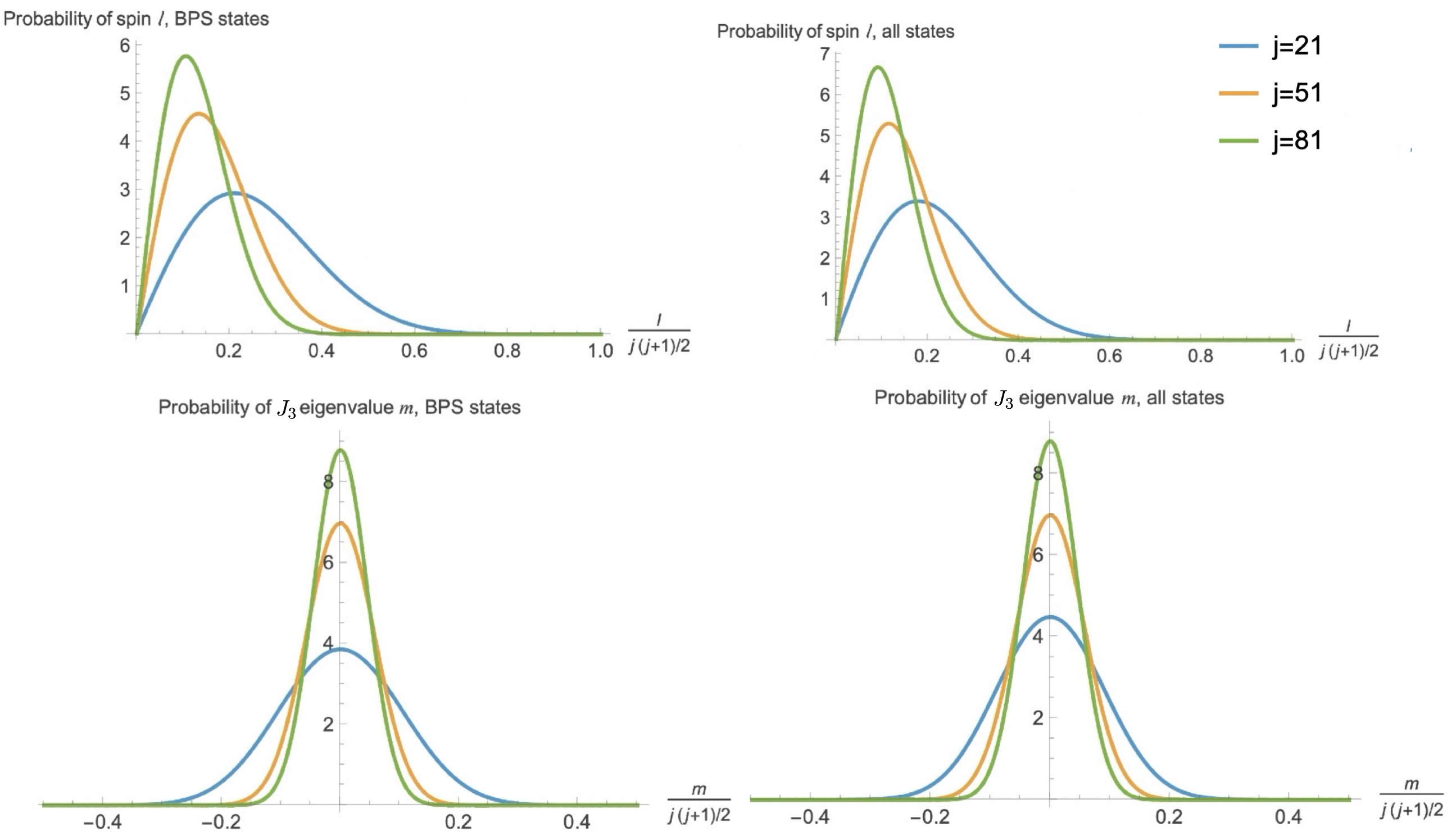}}
    \captionof{figure}{The probability distributions corresponding to the $J^{2}$ and $J_{3}$ quantum numbers of BPS states (left) and of all states in the Hilbert space (right) at various values of $j$. See equations  \nref{J3count} and \nref{DBPSL}. As $j$ increases, the states become increasingly concentrated near $J_{3}  = 0$. The most probable spin representation among the BPS states occurs at $\ell = \sqrt{ 2 j^{3}/9}$ and among all states at $\ell = \sqrt{j^{3}/6}$.}
    \label{Repdegs}
\end{center}
Note that $d_{n}$ is symmetric around $n = 0$ and has a maximum there. Since $D_{\ell}$ is roughly the derivative of $d_{n}$, $D_{\ell}$ is maximized at at a finite value of $\ell$ which scales as $\sim j^{3/2}$.

\section{Comments on BPS states beyond the CFT$_2$ regime}

In section \ref{SU2Sec}, we described the BPS states in the limit of large $SU(2)$ charge. In this limit, the BPS states are generated by acting on the CFT$_2$ vacuum $|\tilde 0\rangle_{\pm}$ with the oscillator modes of the bosons $\phi_{\pm}$. In terms of the original fermions, these modes correspond to the operators
\begin{align}
    \alpha_{-n} = \sum_{m}\psi^{\dagger}_{-n+m}\psi_{m}~~~~~~~ n = \pm 1 \text{ mod } 3
\end{align}
Indeed, one can check that $\alpha_{-n}$ commute with the supercharges in the $j \to \infty$ limit where $Q$ takes the form \nref{QinCFTf}. In this regime, it is clear that the number of states in each $J_3$ sector does not depend on $j$. 

We could wonder how much we need to ``un-fill'' the Fermi sea, or equivalently, how much we need to lower $J_3$ relative to $|\tilde 0\rangle_{\pm}$, before the state counts become $j$-dependent. Recall that the finite $j$ partition function for BPS states is given by (see eq. \nref{DBPSL})
\begin{align}
    Z_j^{BPS} = 2\prod_{m=1}^{j}(1+q^m + q^{2m})= \sum_{s \geq 0}d_s(j) q^{s}~~~~~~~~ s = J_3^{max}-J_3
\end{align}
where $d_s(j)$ is the number of states at ``energy $s$'' above the CFT$_2$ vacuum in the model with $N = 2j+1$ fermions. This partition function can be thought of as imposing an energy cutoff on the CFT$_2$ describing the bosons $\phi_{\pm}$. Now, from $Z_j^{BPS}$ we can derive a simple recursion relation in $j$ for $d_s(j)$,
\begin{align}\la{recur}
    d_s(j) - d_s(j-1) =  d_{s-j}(j-1) +d_{s-2j}(j-1)
\end{align}
The left hand side of this formula is the number of additional BPS states we will get at $J_3 = J_3^{max}-s$ when we increase $j$ by one.\footnote{Note that our model is only defined for $j$ odd, a fact which the partition function $Z_{BPS}^{j}$ is not sensitive to. So, the proper comparison would be between $d_s(j)$ and $d_s(j-2)$. However, this detail is irrelevant for the qualitative points we will make here.} Note that $s = 0$ corresponds to the CFT$_2$ vacuum, so $d_s(j) = 0$ when $s < 0$. So, \nref{recur} is saying that when $0 \leq s \leq j-1$, the number of BPS states is independent of $j$. This is the regime where we expect the CFT$_2$ description to be valid.  The fact that the first correction appears at $s=j$ is related to the following fact. We can consider single particle excitations of the state in figure \ref{Hemisphere}b. This corresponds to adding a fermion or a hole. When this fermion or hole reaches the south or north pole, it realizes that the Fermi sea is not infinitely deep.  

On the other hand, when $s \geq j$, the number of states is $j$-dependent. At $s \gtrsim j$, we get an order one number of new BPS states each time we increase $j$. For example,
\begin{align}
    d_j(j) - d_j(j-1) = d_0(j-1) = 2
\end{align}
When $s\sim j^2/2$, the righthand side of \nref{recur} involves state counts at $J_3 \sim j$, which is where the gaussian approximation to the BPS density of states in \nref{BPSspa} is valid. So, in the sectors near $J_3 \sim 0$, we should see the number of BPS states increase by an amount exponentially large in $j$ each time we increase $j$. Of course, the total increase in the number of BPS states is accounted for by the changing size of the Hilbert space, since the total number of BPS states is $\sim 2 \times 3^{j}$. Here we are simply saying that this increase occurs in the middle of the $J_3$ spectrum, while the number of states in $J_3$ sectors which are an order $j$ amount away from $J_3^{max}$ are fixed.

Another way to describe the deviation of the Cardy density from the state counts in the finite $j$ theory or ``cutoff CFT$_2$'' is to write down a finite $j$ version of the formula  \nref{ZBPSbos} for the partition function  in terms of bosonic oscillator modes: 
\begin{align}
 Z_{BPS}^{j} =2 \prod_{m=1}^{j}(1+q^{m} + q^{2m})=\frac{2\prod_{m=j/3+1}^{j}(1-q^{3m})}{\prod_{m=0}^{j/3-1}(1-q^{3m+1})(1-q^{3m+2})}  ~~~~~~j = 0 \text{ mod } 3
\end{align}
In order to give the simplest example, here we have written the formula assuming $j$ is a multiple of 3. From this expression we can again see that infinite $j$ and finite $j$ state counts will agree up to $s \sim j$, after which point ``wrong sign'' modes in the numerator begin to contribute negatively.

Here we have described only how to enumerate the BPS states beyond the CFT$_2$ regime. We could wonder whether there is a modification of the CFT$_2$ picture which would allow us to describe them explicitly. This is of interest because the BPS states at lower values of $J_3$, where the model has a Schwarzian description in the infrared, we expect to exhibit random matrix statistics, or ``BPS chaos'' in the sense of \cite{shortpaper,Chen:2024oqv}, and have a holographic interpretation as black hole microstates. We leave this as a direction for the future.

\section{Showing that the 9j symbol diagram is zero} \la{9jZero}

The 9j symbol can be expressed as the contraction of six 3j symbols, corresponding to the diagram \nref{9jdiag}.
\begin{align}\la{9jdef}
    \left\{ \begin{array}{ccc} j & j & j\cr  j & j & j \cr j & j & j\end{array} \right\} &= \sum_{\text{ all }m_{rs}}\begin{pmatrix}
        j & j & j \\ m_{11} & m_{12} & m_{13}
    \end{pmatrix}\begin{pmatrix}
        j & j & j \\ m_{21} & m_{22} & m_{23}
    \end{pmatrix}\begin{pmatrix}
        j & j & j \\ m_{31} & m_{32} & m_{33}
    \end{pmatrix}\\
    &\times \begin{pmatrix}
        j & j & j \\ m_{11} & m_{21} & m_{31}
    \end{pmatrix}\begin{pmatrix}
        j & j & j \\ m_{12} & m_{22} & m_{32}
    \end{pmatrix}\begin{pmatrix}
        j & j & j \\ m_{13} & m_{23} & m_{33}\nonumber
    \end{pmatrix}
\end{align}
We can see that \nref{9jdef} is zero for odd integer $j$ as follows. We first exchange the first two indices of the first three 3j symbols. This introduces an overall sign $(-1)^{3} = -1$. Next, we relabel three pairs of indices: $m_{11} \leftrightarrow m_{12}$, $m_{21} \leftrightarrow m_{22},$ $m_{31} \leftrightarrow m_{32}$. After these manipulations, we have an expression which is identical to the first up to an overall minus sign, so we conclude that \nref{9jdef} is zero. These steps are written explicitly below.

\footnotesize
\begin{widerequation}
    \sum_{\text{ all }m_{rs}}\begin{pmatrix}
        j & j & j \\ m_{11} & m_{12} & m_{13}
    \end{pmatrix}\begin{pmatrix}
        j & j & j \\ m_{21} & m_{22} & m_{23}
    \end{pmatrix}\begin{pmatrix}
        j & j & j \\ m_{31} & m_{32} & m_{33}
    \end{pmatrix} \begin{pmatrix}
        j & j & j \\ m_{11} & m_{21} & m_{31}
    \end{pmatrix}\begin{pmatrix}
        j & j & j \\ m_{12} & m_{22} & m_{32}
    \end{pmatrix}\begin{pmatrix}
        j & j & j \\ m_{13} & m_{23} & m_{33}\end{pmatrix}\\
        =-\sum_{\text{ all }m_{rs}}\begin{pmatrix}
        j & j & j \\ m_{12} & m_{11} & m_{13}
    \end{pmatrix}\begin{pmatrix}
        j & j & j \\ m_{22} & m_{21} & m_{23}
    \end{pmatrix}\begin{pmatrix}
        j & j & j \\ m_{32} & m_{31} & m_{33}
    \end{pmatrix} \begin{pmatrix}
        j & j & j \\ m_{11} & m_{21} & m_{31}
    \end{pmatrix}\begin{pmatrix}
        j & j & j \\ m_{12} & m_{22} & m_{32}
    \end{pmatrix}\begin{pmatrix}
        j & j & j \\ m_{13} & m_{23} & m_{33}\end{pmatrix}\\
        =-\sum_{\text{ all }m_{rs}}\begin{pmatrix}
        j & j & j \\ m_{11} & m_{12} & m_{13}
    \end{pmatrix}\begin{pmatrix}
        j & j & j \\ m_{21} & m_{22} & m_{23}
    \end{pmatrix}\begin{pmatrix}
        j & j & j \\ m_{31} & m_{32} & m_{33}
    \end{pmatrix} \begin{pmatrix}
        j & j & j \\ m_{12} & m_{22} & m_{32}
    \end{pmatrix}\begin{pmatrix}
        j & j & j \\ m_{11} & m_{21} & m_{31}
    \end{pmatrix}\begin{pmatrix}
        j & j & j \\ m_{13} & m_{23} & m_{33}\end{pmatrix}\\
        =0~~~~~~~~~~~~~~~~~~~~~~~~~~~~~~~~~~~~~~~~~~~~~~~~~~~~~~~~~~~~~~~~~~~~~~~~~~~~~~~~~~~~~~~~~~~~~~~~~~~~~~~~~~~~~~~~~~~~~~~~~~~~~~~~~~~~~~~~~~~~~~~~~~~~~~~~~~~~~~~~~~~~~~~~~~~~
\end{widerequation}
\normalsize

\section{12j symbol normalization}\la{12jnorm}

The 12j symbol of the second kind with all equal $j$, which for brevity we will denote by $12j_{(II)}$, is defined as \cite{QuantumTheoryOfAngularMomentum}
\begin{align}\la{12j2}
    12j_{(II)} = \sum_{x=0}^{2j}(2x+1)  \left\{ \begin{array}{ccc} j & j &  j \cr  j & j & x \end{array} \right\}^{4}
\end{align}
The authors of \cite{Garoufalidis:2009vi} compute the large $j$ asymptotics of a function $a_n$ which is closely related to \nref{12j2}, but which differs by an $j$-dependent normalization which we will now explain. The quantities denoted $a_{n,k}$ and $a_n$ in equation (23) of \cite{Garoufalidis:2009vi} can be expressed as
\begin{align}
    a_{n,k} &= \sum_{j \in \Z}(-1)^{j}\begin{pmatrix}
         k \\ j-3n
    \end{pmatrix}^{2} \begin{pmatrix}
        2n -k \\ 4n-j
    \end{pmatrix}\begin{pmatrix}
        j+1 \\ 2n+k+1
    \end{pmatrix} = \frac{(3n+1)!}{(n!)^{3}} \left\{ \begin{array}{ccc} n & n &  n \cr  n & n & k \end{array} \right\} \\
    a_n &=\sum_{k=0}^{2n}(2k+1)a_{n,k}^{4} = \lb \frac{(3n+1)!}{(n!)^{3}} \rb^{4}\sum_{k=0}^{2n}(2k+1) \left\{ \begin{array}{ccc} n & n &  n \cr  n & n & k \end{array} \right\}^{4}\la{an}
\end{align}
Per \nref{an}, the difference in normalization between $12j_{(II)}$ and $a_n$ is
\begin{align}
    a_j &= \lb \frac{(3j+1)!}{(j!)^{3}} \rb^{4} \times [12j_{(II)}]\\
    &\simeq \frac{3^{6}}{(2 \pi)^{4}}3^{12 j}\times [12j_{(II)}]~~~~~~~~~~ j \gg 1
\end{align}
where in the second step we have used Stirling's approximation.

\section{Energy of the state with maximal angular momentum for a given R charge }
\la{DerEquA}

We consider states of the form
\begin{align}\label{eq:lmax(n)}
    \big|\ell_{max}^{(n)}\big\rangle=\psi^\dagger_{j-n}\psi^\dagger_{j-n+1}\cdots\psi^\dagger_{j-1}\psi^\dagger_j|0\rangle\;,\qquad n<j\;.
\end{align}
It has R-charge $(2n-2j+1)/6$. We now compute the energy of these states. The normal-ordered Hamiltonian was given in \nref{HamNor}, which we copy here for convenience:
\begin{align}\label{eq:Hexplicit}
 &H=\frac{\mathsf{J}}{3}\Big[(2j+1)-3(N_\psi+j+\tfrac{1}{2})+3\sum_m O^\dagger_{j,m}O_{j,m}\Big] ~~,~~ \text{ where }\\
 &\sum_m O^\dagger_{j,m}O_{j,m}=2(2j+1) \sum_{\substack{m_1<m_2\\m_3<m_4}}\begin{pmatrix}
        j&j&j\\
        m_1&m_2&-(m_1+m_2)
    \end{pmatrix}\begin{pmatrix}
        j&j&j\\
        m_3&m_4&-(m_3+m_4)
    \end{pmatrix}(\psi_{m_1}\psi_{m_2})^\dagger(\psi_{m_3}\psi_{m_4})\;.\nonumber
 \end{align}
Let us focus on the $\sum O^{\dagger}O$ contribution. Note that because the state we are considering is the maximal $J_3$ state in its $R$-charge sector, and $H$ conserves $J_3$ and $R$, only the terms where $m_1=m_3,\,m_2=m_4$ can act nontrivially. 
Furthermore, the terms with $m_1$ and/or $m_2<j-n$ will annihilate $\big|\ell_{max}^{(n)}\big\rangle$ because the state only contains fermions with $m\geq j-n$.

From \eqref{eq:lmax(n)}, the minimum possible value of $m_1+m_2$ is $(j-n)+(j-n+1)=2j-2n+1$. If this is greater than $j$, then  $\sum O^{\dagger}O$ will annihilate $\big|\ell_{max}^{(n)}\big\rangle$ due to the selection rules of the 3j symbol. Therefore we can separate the states into two trajectories based on whether it gets an energy contribution from the 4-fermion term. For $n\leq(j-1)/2$, only the first two terms of \eqref{eq:Hexplicit} contribute to the energy, giving rise to a linear dependence of $E$ on $n$. For $n>(j-1)/2$, there is a nonzero contribution from $\sum O^{\dagger}O$,  and so the energy dependence on $n$ is nonlinear. Putting together all of the contributions, and converting $n$ to $R$-charge, we have
\bea \la{apFirLi}
  \!\!\!\frac{E(|R|)}{ \mathsf{J}}\!\!\!&=&\!\!\!\frac{2j+1}{3}-(n+1)  = \frac{18|R|-2j-1}{6} ~,~~~~~~~~{\rm for } ~~~n \leq (j-1)/2 ~~~{\rm or } ~~  { j \over 6 } \leq |R| 
  \\
  \label{eqap:nonlinear}
    \!\!\!\frac{E\big(|R|\big)}{ \mathsf{J}}\!\!\!&=&\!\!\!\frac{-18|R|-2j-1}{6}+2(2j+1)\;\sum_{\vphantom{\big(}\mathclap{3|R|+j-1/2\leq m_1<m_2\leq j}}\;\;\begin{pmatrix}
        j & j& j\\
        m_1 & m_2 & -(m_1+m_2)
    \end{pmatrix}^2 ~,~~{\rm for }  ~|R| <  { j \over 6 }\;
  \eea
  where we have used charge conjugation symmetry to write $R\mapsto |R|$, and we have replaced $n=3R+j-1/2$.

In particular,  $E\big(|R|\big)=0$ exactly for $R=\pm 1/6$, which correspond to the largest spin BPS states discussed in more detail in section \ref{SU2Sec}. This is not obvious from the expression \nref{eqap:nonlinear}. We will first prove that $E\big(1/6\big)=0$ and then explain how to rewrite \nref{eqap:nonlinear} in the form \nref{eq:nonlinear} where this is manifest.

For $|R| = 1/6$, \nref{eqap:nonlinear} can be rewritten as
\begin{align}
    \frac{E(\frac{1}{6})}{\mathsf{J}} = -\frac{j+2}{3} + (2j+1)\sum_{0 \leq m_1,m_2}\begin{pmatrix}
        j & j & j \\ m_1 & m_2 & -(m_1+m_2)
    \end{pmatrix}^{2}
\end{align}
We would like to evaluate the sum over 3j symbols. Note that this is different from the sum that appears in the orthogonality identity \nref{BubId} because $m_1, m_2$ are restricted to be nonnegative. The sum can be broken up into terms where $m_1$ or $m_2$ are equal to zero and those where $m_1$ and $m_2$ are both greater than zero.
\begin{align}\la{sum}
   \sum_{0 \leq m_1,m_2}\begin{pmatrix}
        j & j & j \\ m_1 & m_2 & -(m_1+m_2)
    \end{pmatrix}^{2} = 2\sum_{m>0}\begin{pmatrix}
        j & j & j \\ 0 & m & -m
    \end{pmatrix}^{2} + \sum_{0 < m_1,m_2}\begin{pmatrix}
        j & j & j \\ m_1 & m_2 & -(m_1+m_2)
    \end{pmatrix}^{2}
\end{align}
The first term can be evaluated by the orthogonality identity \nref{BubId},
\begin{align}
    \sum_{m>0}\begin{pmatrix}
        j & j & j \\ 0 & m & -m
    \end{pmatrix}^{2} = \frac{1}{2}\frac{1}{2j+1}
\end{align}
where the factor of $\frac{1}{2}$ is due to the fact that we are summing over only positive values of $m$. Note that $m=0$ vanishes due to the antisymmetry of the 3j symbol for odd $j$. 
Now we address the second term of  \nref{sum}. This involves a sum over 3j symbols in which all $m_i \neq 0$. Recall that the sum over all possible values of the $m_i$ gives
\begin{align}
    \sum_{m_1, m_2, m_3} \begin{pmatrix}
        j & j & j \\ m_1 & m_2 & m_3
    \end{pmatrix}^{2} = 1
\end{align}
If we remove all terms in which one of the $m_i = 0$, then we have
\begin{align}
    \sum_{m_1, m_2, m_3 \neq 0} 
        \begin{pmatrix}
        j & j & j \\ m_1 & m_2 & m_3
    \end{pmatrix}^{2} = 1-\frac{3}{2j+1}
\end{align}
The remaining terms, in which all $m_i$ are nonzero, can be grouped by the signs of the $m_i$. In particular, there are 6 possible sign combinations:
\begin{align}
    (+ + -),~~ (+-+), ~~(-++), ~~(--+),~~ (-+-), ~~(+--)
\end{align}
The term we wish to evaluate in \nref{sum} corresponds to the first combination, $(++-)$. All six sectors are related by permuting the columns or flipping the signs of all three $m_i$ simultaneously. So, they contribute with equal weight to the overall sum.  We conclude that
\begin{align}
    \sum_{0 < m_1,m_2}\begin{pmatrix}
        j & j & j \\ m_1 & m_2 & -(m_1+m_2)
    \end{pmatrix}^{2} = \frac{1}{6} \lp 1-\frac{3}{2j+1}\rp =\frac{1}{3}\frac{j-1}{2j+1}
\end{align}
In total, we have
\begin{align}
    \frac{E(\frac{1}{6})}{\mathsf{J}} =-\frac{j+2}{3} + (2j+1) \lb \frac{1}{2j+1} + \frac{1}{3}\frac{j-1}{2j+1}\rb = -\frac{j+2}{3}+\frac{3+j-1}{3}=0
\end{align}

We conclude that $E(|R| = \frac{1}{6}) = 0$. Given this, we can rewrite $E(|R|)$ by including only terms that contribute nonzero energy as we move away from $|R| = 1/6$. In other words, we write the energy as a function of $|R|-\frac{1}{6}$ and use the fact that $E(|R| = 1/6) = 0$ to simplify the resulting expression. This gives
\begin{align}
   E(|R|)/\mathsf{J} &= -3\lp |R|-\frac{1}{6}\rp  + (2j+1)\sum_{-3\lp |R|-\frac{1}{6}\rp \leq m_1, m_2 \leq 0}\begin{pmatrix}
        j & j & j \\ m_1 & m_2 & -(m_1 +m_2)
    \end{pmatrix}^{2}\\
    &+2(2j+1)\sum_{\substack{-3\lp |R|-\frac{1}{6}\rp \leq m_1 < 0\\
    0 < m_2 }} \begin{pmatrix}
        j & j & j \\ m_1 & m_2 & -(m_1 +m_2)
    \end{pmatrix}^{2}\la{secline}
\end{align}
Finally, we can rewrite the  term in \nref{secline} so as to cancel the first term linear in $|R|$. By re-indexing and using the symmetries of the 3j symbol, we have
\begin{align*}
\sum_{ 0 < m_2 } \begin{pmatrix}
        j & j & j \\ m_1 & m_2 & -(m_1 +m_2)
    \end{pmatrix}^{2}&=  \sum_{-j \leq  m_2 < -m_1} \begin{pmatrix}
        j & j & j \\ m_1 & m_2 & -(m_1 +m_2)
    \end{pmatrix}^{2}\\
    &=\lb \sum_{-j \leq  m_2 \leq j} - \sum_{-m_1 \leq  m_2 \leq j}  \rb \begin{pmatrix}
        j & j & j \\ m_1 & m_2 & -(m_1 +m_2)
    \end{pmatrix}^{2} \\
    &=\frac{1}{2}\frac{1}{2j+1} + \frac{1}{2} \lb \sum_{-j \leq  m_2 < -m_1} - \sum_{-m_1 \leq  m_2 \leq j}\rb \begin{pmatrix}
        j & j & j \\ m_1 & m_2 & -(m_1 +m_2)
    \end{pmatrix}^{2} \\
    &=\frac{1}{2}\frac{1}{2j+1} + \frac{1}{2} \lb \sum_{-j \leq  m_2 < -m_1} - \sum_{-j \leq  m_2 \leq 0}\rb \begin{pmatrix}
        j & j & j \\ m_1 & m_2 & -(m_1 +m_2)
    \end{pmatrix}^{2} \\
     &=\frac{1}{2}\frac{1}{2j+1} + \frac{1}{2} \sum_{0 <m_2 < -m_1} \begin{pmatrix}
        j & j & j \\ m_1 & m_2 & -(m_1 +m_2)
    \end{pmatrix}^{2} 
\end{align*}
Plugging this into \nref{secline}, we are left with
\begin{align}
   E(|R|)/\mathsf{J} &=  (2j+1)\sum_{-3\lp |R|-\frac{1}{6}\rp \leq m_1, m_2 \leq 0}\begin{pmatrix}
        j & j & j \\ m_1 & m_2 & -(m_1 +m_2)
    \end{pmatrix}^{2}   \\
    &+(2j+1)\sum_{ -3\lp |R|-\frac{1}{6}\rp \leq m_1 < 0}  \left\{ \sum_{0 < m_2 < -m_1 } \begin{pmatrix}
        j & j & j \\ m_1 & m_2 & -(m_1 +m_2)
    \end{pmatrix}^{2} \right\} \la{Esfinal}
\end{align}
This can also be written as 
\begin{align}\la{Efinal}
   E(|R|)/\mathsf{J} &=   (2j+1)\sum_{ -3\lp |R|-\frac{1}{6}\rp \leq m_1 < 0}  \left\{ \sum_{-3\lp |R|-\frac{1}{6}\rp \leq  m_2 \leq -m_1 } \begin{pmatrix}
        j & j & j \\ m_1 & m_2 & -(m_1 +m_2)
    \end{pmatrix}^{2} \right\} 
\end{align}

\section{Counting the degrees of freedom of the cube diagram}
\la{CubeDiag}

In this appendix, we discuss the large $j$ asymptotic scaling of the 8-vertex cube diagram in figure \ref{fig:6j}. 

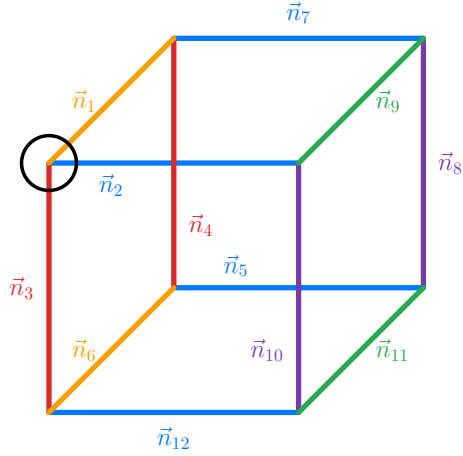
\begin{figure}[h]
\centering
\resizebox{!}{6cm}{
\begin{tikzpicture}[
  line cap=round,
  line join=round,
  edge/.style={line width=3pt},
  lab/.style={font=\Large, inner sep=1pt}
]

\def\L{5.0}                         
\pgfmathsetmacro{\d}{\L/2}     

\coordinate (A) at (0,0);      
\coordinate (B) at (\L,0);     
\coordinate (C) at (\L,\L);    
\coordinate (D) at (0,\L);     

\coordinate (E) at ($(A)+(\d,\d)$);
\coordinate (F) at ($(B)+(\d,\d)$);
\coordinate (G) at ($(C)+(\d,\d)$);
\coordinate (H) at ($(D)+(\d,\d)$);

\definecolor{myblue}{RGB}{0,120,255}
\definecolor{myred}{RGB}{220,40,40}
\definecolor{myorange}{RGB}{245,160,0}
\definecolor{mypurple}{RGB}{120,60,170}
\definecolor{mygreen}{RGB}{40,170,80}

\draw[edge,myblue] (A)--(B) node[midway,below=6pt,lab] {$\vec n_{12}$};
\draw[edge,myblue] (H)--(G) node[midway,above=6pt,lab] {$\vec n_{7}$};
\draw[edge,myblue] (D)--(C) node[pos=0.25,below=4pt,lab] {$\vec n_{2}$};
\draw[edge,myblue] (F)--(E) node[pos=0.75,above=4pt,lab] {$\vec n_{5}$};

\draw[edge,myred] (A)--(D) node[midway,left=6pt,lab] {$\vec n_{3}$};
\draw[edge,myred] (H)--(E) node[pos=0.75,right=6pt,lab] {$\vec n_{4}$};

\draw[edge,myorange] (D)--(H) node[midway,left=6pt,lab] {$\vec n_{1}$};
\draw[edge,myorange] (A)--(E) node[midway,left=6pt,lab] {$\vec n_{6}$};

\draw[edge,mypurple] (F)--(G) node[midway,right=6pt,lab] {$\vec n_{8}$};
\draw[edge,mypurple] (B)--(C) node[pos=0.25,left=6pt,lab] {$\vec n_{10}$};

\draw[edge,mygreen] (C)--(G) node[midway,right=6pt,lab] {$\vec n_{9}$};
\draw[edge,mygreen] (B)--(F) node[midway,right=6pt,lab] {$\vec n_{11}$};

\draw[black, line width=2pt] (D) circle[radius=0.55];
\end{tikzpicture}}
\caption{The cube diagram. Each line is assigned a vector $\vec n_i$. We consider a configuration where lines with the same color correspond to the same vectors. }
\label{fig:colcube}
\end{figure}

There are 8 angular momentum constraint equations, one for each vertex. Labeling the unit vectors $\vec{n}_i$ as in figure \ref{fig:colcube}, they are
\begin{align*}
    \vec{n}_1 + \vec{n}_2 + \vec{n_3} = 0~~~~~~\vec{n}_2 + \vec{n}_9 + \vec{n}_{10} = 0~~~~~~\vec{n}_{10} + \vec{n}_{11} + \vec{n}_{12} = 0 ~~~~~~\vec{n}_3 + \vec{n}_{12} + \vec{n}_{6} = 0\\
    \vec{n}_1 + \vec{n}_7 + \vec{n}_4 = 0 ~~~~~~~ \vec{n}_7 + \vec{n}_8 + \vec{n}_9 = 0 ~~~~~~~~ \vec{n}_5 + \vec{n}_8 + \vec{n}_{11} = 0 ~~~~~~~~ \vec{n}_4 + \vec{n}_5 + \vec{n}_6 = 0
\end{align*}
A naive counting of the variables versus constraints would suggest that these equations admit a discrete set of solutions. We will explain that there is, in fact, a continuous 1-parameter family of solutions. 

The presence of this extra degree of freedom can be understood by considering the solution of the constraint equations depicted in figure \ref{fig:colcube}. First, suppose we use our freedom to rotate all of the vectors uniformly inside the sphere to fix  $\vec{n}_1, \vec{n}_2, \vec{n}_3$. Angular momentum conservation forces these vectors into an equilateral triangle, but the plane of the triangle may be rotated. The first  constraint equation, corresponding to the circled vertex, is then trivially satisfied. 

We may now satisfy the constraint equations corresponding to the remaining 3 vertices of the left face by setting 
\begin{align}
    \vec{n}_3 = \vec{n}_4, ~~~~\vec{n}_6 = \vec{n}_1,~~~~ \vec{n}_5 = \vec{n}_7 = \vec{n}_{12} = \vec{n}_2
\end{align}
The edges of the cube have been colored to illustrate which vectors have been set equal.

Now consider the vertex between $\vec{n}_7, \vec{n}_8$, and $\vec{n}_{9}$. $\vec{n}_7 = \vec{n}_2$ is fixed, so this vertex forces $\vec{n}_8$ and $\vec{n}_9$ to lie at an angle of $2 \pi/3$ relative to $\vec{n}_7$ and relative to each other. In other words, there is an $S^1$ worth of solutions, corresponding to picking any vector $\vec{n}_8$ which satisfies $$\vec{n}_2 \cdot \vec{n}_8 = -\frac{1}{2}$$ and then $\vec{n}_9$ is fixed to be $$\vec{n}_9 = -\vec{n}_2 -\vec{n}_8$$ Note that this also implies that $\vec n_2 \cdot \vec n_8=\vec n_8 \cdot \vec n_9 = -\half$.  Now, we still need to satisfy the constraint equations of every other vertex on the right hand face, which we can do by setting
\begin{align}
    \vec{n}_{10} = \vec{n}_8, ~~ \vec{n}_{11} = \vec{n}_9
\end{align}

In conclusion, this diagram has one more redundant vertex constraint in comparison to the generic non-melonic diagram. This redundancy leads to a rotational degree of freedom in the solution space of possible arrow configurations, accounting for the $\sqrt{j}$ enhancement discussed in section \ref{ArrowsSec}.

\section{Recalling the circular string solution }
\la{CircularString}

We consider the metric $ds^2 = -dt^2 + dx_9^2 + dx_1^2 + dx_2^2 + \cdots $, where  $x_1$, $x_2$ are two non-compact directions and $x_9 \sim x_9 + 2 \pi R$ is a compact direction. 
Then we consider the solution (in conformal gauge)
\be 
t = \left( { \alpha' n \over R } + w R \right)\tau ~,~~~~~~~x_9 = { \alpha' n \over R} \tau - w R \sigma  ~,~~~~~ x_1 + i x_2 = a e^{ i (\tau + \sigma ) } ~,~~~~~a^2 = \alpha' n w
\ee 
where $n$ is the momentum and $w$ is the winding along the circle. The angular momentum is $|J_{12}| = nw$ and $1/(2 \pi \alpha')$ is the string tension. We see that for very large $n$ and $w$ we get a very large and weakly coupled string. This is related to a circular superconducting string configuration stabilized by a constant current (of KK momentum and winding in this case).

\eject 

\bibliographystyle{apsrev4-1long}
\bibliography{GeneralBibliography.bib}

@article{Yoon:2017gut,
    author = "Yoon, Junggi",
    title = "{Supersymmetric SYK Model: Bi-local Collective Superfield/Supermatrix Formulation}",
    eprint = "1706.05914",
    archivePrefix = "arXiv",
    primaryClass = "hep-th",
    doi = "10.1007/JHEP10(2017)172",
    journal = "JHEP",
    volume = "10",
    pages = "172",
    year = "2017"
}

@article{Yoon:2017nig,
    author = "Yoon, Junggi",
    title = "{SYK Models and SYK-like Tensor Models with Global Symmetry}",
    eprint = "1707.01740",
    archivePrefix = "arXiv",
    primaryClass = "hep-th",
    doi = "10.1007/JHEP10(2017)183",
    journal = "JHEP",
    volume = "10",
    pages = "183",
    year = "2017"
}

@article{Chang:2024zqi,
    author = "Chang, Chi-Ming and Lin, Ying-Hsuan",
    title = "{Holographic covering and the fortuity of black holes}",
    eprint = "2402.10129",
    archivePrefix = "arXiv",
    primaryClass = "hep-th",
    month = "2",
    year = "2024"
}

@article{Choi:2023znd,
    author = "Choi, Sunjin and Kim, Seok and Lee, Eunwoo and Lee, Siyul and Park, Jaemo",
    title = "{Towards quantum black hole microstates}",
    eprint = "2304.10155",
    archivePrefix = "arXiv",
    primaryClass = "hep-th",
    doi = "10.1007/JHEP11(2023)175",
    journal = "JHEP",
    volume = "11",
    pages = "175",
    year = "2023",
    note = "[Erratum: JHEP 03, 091 (2025)]"
}

@article{Choi:2023vdm,
    author = "Choi, Jaehyeok and Choi, Sunjin and Kim, Seok and Lee, Jehyun and Lee, Siyul",
    title = "{Finite N black hole cohomologies}",
    eprint = "2312.16443",
    archivePrefix = "arXiv",
    primaryClass = "hep-th",
    reportNumber = "SNUTP23-002, KIAS-P23070, LCTP-23-20, SNUTP23-002; KIAS-P23070; LCTP-23-20;",
    doi = "10.1007/JHEP12(2024)029",
    journal = "JHEP",
    volume = "12",
    pages = "029",
    year = "2024"
}

@article{Chang:2024lxt,
    author = "Chang, Chi-Ming and Chen, Yiming and Sia, Bik Soon and Yang, Zhenbin",
    title = "{Fortuity in SYK models}",
    eprint = "2412.06902",
    archivePrefix = "arXiv",
    primaryClass = "hep-th",
    doi = "10.1007/JHEP08(2025)003",
    journal = "JHEP",
    volume = "08",
    pages = "003",
    year = "2025"
}

@article{Chen:2024oqv,
    author = "Chen, Yiming and Lin, Henry W. and Shenker, Stephen H.",
    title = "{BPS chaos}",
    eprint = "2407.19387",
    archivePrefix = "arXiv",
    primaryClass = "hep-th",
    doi = "10.21468/SciPostPhys.18.2.072",
    journal = "SciPost Phys.",
    volume = "18",
    number = "2",
    pages = "072",
    year = "2025"
}

@article{Chen:2025sum,
    author = "Chen, Yiming",
    title = "{Fortuity with a single matrix}",
    eprint = "2511.00790",
    archivePrefix = "arXiv",
    primaryClass = "hep-th",
    month = "11",
    year = "2025"
}

@article{Klebanov:2016xxf,
    author = "Klebanov, Igor R. and Tarnopolsky, Grigory",
    title = "{Uncolored random tensors, melon diagrams, and the Sachdev-Ye-Kitaev models}",
    eprint = "1611.08915",
    archivePrefix = "arXiv",
    primaryClass = "hep-th",
    reportNumber = "PUPT-2514",
    doi = "10.1103/PhysRevD.95.046004",
    journal = "Phys. Rev. D",
    volume = "95",
    number = "4",
    pages = "046004",
    year = "2017"
}

@article{Witten:2016iux,
    author = "Witten, Edward",
    title = "{An SYK-Like Model Without Disorder}",
    eprint = "1610.09758",
    archivePrefix = "arXiv",
    primaryClass = "hep-th",
    doi = "10.1088/1751-8121/ab3752",
    journal = "J. Phys. A",
    volume = "52",
    number = "47",
    pages = "474002",
    year = "2019"
}

@article{Bulycheva:2018qcp,
    author = "Bulycheva, Ksenia",
    title = "{$ \mathcal{N}=2 $ SYK model in the superspace formalism}",
    eprint = "1801.09006",
    archivePrefix = "arXiv",
    primaryClass = "hep-th",
    doi = "10.1007/JHEP04(2018)036",
    journal = "JHEP",
    volume = "04",
    pages = "036",
    year = "2018"
}

@article{Peng:2017spg,
    author = "Peng, Cheng and Spradlin, Marcus and Volovich, Anastasia",
    title = "{Correlators in the $\mathcal{N}=2$ Supersymmetric SYK Model}",
    eprint = "1706.06078",
    archivePrefix = "arXiv",
    primaryClass = "hep-th",
    reportNumber = "BROWN-HET-1716",
    doi = "10.1007/JHEP10(2017)202",
    journal = "JHEP",
    volume = "10",
    pages = "202",
    year = "2017"
}

@article{Heydeman:2022lse,
    author = "Heydeman, Matthew and Turiaci, Gustavo J. and Zhao, Wenli",
    title = "{Phases of $ \mathcal{N} $ = 2 Sachdev-Ye-Kitaev models}",
    eprint = "2206.14900",
    archivePrefix = "arXiv",
    primaryClass = "hep-th",
    doi = "10.1007/JHEP01(2023)098",
    journal = "JHEP",
    volume = "01",
    pages = "098",
    year = "2023"
}

@article{Maldacena:2020skw,
    author = "Maldacena, Juan",
    title = "{Comments on magnetic black holes}",
    eprint = "2004.06084",
    archivePrefix = "arXiv",
    primaryClass = "hep-th",
    doi = "10.1007/JHEP04(2021)079",
    journal = "JHEP",
    volume = "04",
    pages = "079",
    year = "2021"
}

@article{Shankar:1993pf,
    author = "Shankar, R.",
    title = "{Renormalization group approach to interacting fermions}",
    eprint = "cond-mat/9307009",
    archivePrefix = "arXiv",
    reportNumber = "PRINT-93-0200-REV (YALE), PRINT-93-0200 (YALE)",
    doi = "10.1103/RevModPhys.66.129",
    journal = "Rev. Mod. Phys.",
    volume = "66",
    pages = "129--192",
    year = "1994"
}

@article{Maldacena:2016hyu,
    author = "Maldacena, Juan and Stanford, Douglas",
    title = "{Remarks on the Sachdev-Ye-Kitaev model}",
    eprint = "1604.07818",
    archivePrefix = "arXiv",
    primaryClass = "hep-th",
    doi = "10.1103/PhysRevD.94.106002",
    journal = "Phys. Rev. D",
    volume = "94",
    number = "10",
    pages = "106002",
    year = "2016"
}

@article{Haldane:1983xm,
    author = "Haldane, F. D. M.",
    title = "{Fractional quantization of the Hall effect: A Hierarchy of incompressible quantum fluid states}",
    doi = "10.1103/PhysRevLett.51.605",
    journal = "Phys. Rev. Lett.",
    volume = "51",
    pages = "605--608",
    year = "1983"
}

@article{Narayan:2017hvh,
    author = "Narayan, Prithvi and Yoon, Junggi",
    title = "{Supersymmetric SYK Model with Global Symmetry}",
    eprint = "1712.02647",
    archivePrefix = "arXiv",
    primaryClass = "hep-th",
    doi = "10.1007/JHEP08(2018)159",
    journal = "JHEP",
    volume = "08",
    pages = "159",
    year = "2018"
}

@article{Fu:2016vas,
    author = "Fu, Wenbo and Gaiotto, Davide and Maldacena, Juan and Sachdev, Subir",
    title = "{Supersymmetric Sachdev-Ye-Kitaev models}",
    eprint = "1610.08917",
    archivePrefix = "arXiv",
    primaryClass = "hep-th",
    doi = "10.1103/PhysRevD.95.026009",
    journal = "Phys. Rev. D",
    volume = "95",
    number = "2",
    pages = "026009",
    year = "2017",
    note = "[Addendum: Phys.Rev.D 95, 069904 (2017)]"
}

@article{Murugan:2017eto,
    author = "Murugan, Jeff and Stanford, Douglas and Witten, Edward",
    title = "{More on Supersymmetric and 2d Analogs of the SYK Model}",
    eprint = "1706.05362",
    archivePrefix = "arXiv",
    primaryClass = "hep-th",
    doi = "10.1007/JHEP08(2017)146",
    journal = "JHEP",
    volume = "08",
    pages = "146",
    year = "2017"
}

@article{shortpaper,
    author = "Lin, Henry W. and Maldacena, Juan and Rozenberg, Liza and Shan, Jieru",
    title = "{Holography for people with no time}",
    eprint = "2207.00407",
    archivePrefix = "arXiv",
    primaryClass = "hep-th",
    month = "7",
    year = "2022"
}

@article{Stanford:2017thb,
    author = "Stanford, Douglas and Witten, Edward",
    title = "{Fermionic Localization of the Schwarzian Theory}",
    eprint = "1703.04612",
    archivePrefix = "arXiv",
    primaryClass = "hep-th",
    doi = "10.1007/JHEP10(2017)008",
    journal = "JHEP",
    volume = "10",
    pages = "008",
    year = "2017"
}

@article{Saad:2018bqo,
    author = "Saad, Phil and Shenker, Stephen H. and Stanford, Douglas",
    title = "{A semiclassical ramp in SYK and in gravity}",
    eprint = "1806.06840",
    archivePrefix = "arXiv",
    primaryClass = "hep-th",
    month = "6",
    year = "2018"
}

@article{Bena:2022ldq,
    author = "Bena, Iosif and Martinec, Emil J. and Mathur, Samir D. and Warner, Nicholas P.",
    title = "{Snowmass White Paper: Micro- and Macro-Structure of Black Holes}",
    eprint = "2203.04981",
    archivePrefix = "arXiv",
    primaryClass = "hep-th",
    month = "3",
    year = "2022"
}

@article{Benedetti:2020iku,
    author = "Benedetti, Dario and Delporte, Nicolas",
    title = "{Remarks on a melonic field theory with cubic interaction}",
    eprint = "2012.12238",
    archivePrefix = "arXiv",
    primaryClass = "hep-th",
    doi = "10.1007/JHEP04(2021)197",
    journal = "JHEP",
    volume = "04",
    pages = "197",
    year = "2021"
}

@article{Gu:2019jub,
    author = "Gu, Yingfei and Kitaev, Alexei and Sachdev, Subir and Tarnopolsky, Grigory",
    title = "{Notes on the complex Sachdev-Ye-Kitaev model}",
    eprint = "1910.14099",
    archivePrefix = "arXiv",
    primaryClass = "hep-th",
    doi = "10.1007/JHEP02(2020)157",
    journal = "JHEP",
    volume = "02",
    pages = "157",
    year = "2020"
}

@article{Sen:2005kj,
    author = "Sen, Ashoke",
    title = "{Stretching the horizon of a higher dimensional small black hole}",
    eprint = "hep-th/0505122",
    archivePrefix = "arXiv",
    doi = "10.1088/1126-6708/2005/07/073",
    journal = "JHEP",
    volume = "07",
    pages = "073",
    year = "2005"
}

@article{Benini:2015eyy,
    author = "Benini, Francesco and Hristov, Kiril and Zaffaroni, Alberto",
    title = "{Black hole microstates in AdS$_{4}$ from supersymmetric localization}",
    eprint = "1511.04085",
    archivePrefix = "arXiv",
    primaryClass = "hep-th",
    reportNumber = "IMPERIAL-TP-2015-FB-03",
    doi = "10.1007/JHEP05(2016)054",
    journal = "JHEP",
    volume = "05",
    pages = "054",
    year = "2016"
}

@article{Aharony:2008ug,
    author = "Aharony, Ofer and Bergman, Oren and Jafferis, Daniel Louis and Maldacena, Juan",
    title = "{N=6 superconformal Chern-Simons-matter theories, M2-branes and their gravity duals}",
    eprint = "0806.1218",
    archivePrefix = "arXiv",
    primaryClass = "hep-th",
    reportNumber = "WIS-12-08-JUN-DPP",
    doi = "10.1088/1126-6708/2008/10/091",
    journal = "JHEP",
    volume = "10",
    pages = "091",
    year = "2008"
}

@article{Amit:1979ev,
    author = "Amit, D. J. and Roginsky, D. V. I.",
    title = "{Exactly soluble limit of phi**3 field theory with internal Potts symmetry }",
    doi = "10.1088/0305-4470/12/5/017",
    journal = "J. Phys. A",
    volume = "12",
    pages = "689--713",
    year = "1979"
}

@article{Garoufalidis:2009vi,
    author = "Garoufalidis, Stavros and van der Veen, Roland and Zagier, with an appendix by Don",
    title = "{Asymptotics of classical spin networks}",
    eprint = "0902.3113",
    archivePrefix = "arXiv",
    primaryClass = "math.GT",
    doi = "10.2140/gt.2013.17.1",
    journal = "Geom. Topol.",
    volume = "17",
    pages = "1--37",
    year = "2013"
}

@article{Klebanov:2018nfp,
    author = "Klebanov, Igor R. and Milekhin, Alexey and Popov, Fedor and Tarnopolsky, Grigory",
    title = "{Spectra of eigenstates in fermionic tensor quantum mechanics}",
    eprint = "1802.10263",
    archivePrefix = "arXiv",
    primaryClass = "hep-th",
    reportNumber = "PUPT-2552",
    doi = "10.1103/PhysRevD.97.106023",
    journal = "Phys. Rev. D",
    volume = "97",
    number = "10",
    pages = "106023",
    year = "2018"
}

@article{Gaitan:2020zbm,
    author = "Gaitan, Gabriel and Klebanov, Igor R. and Pakrouski, Kiryl and Pallegar, Preethi N. and Popov, Fedor K.",
    title = "{Hagedorn Temperature in Large $N$ Majorana Quantum Mechanics}",
    eprint = "2002.02066",
    archivePrefix = "arXiv",
    primaryClass = "hep-th",
    reportNumber = "PUPT-2610",
    doi = "10.1103/PhysRevD.101.126002",
    journal = "Phys. Rev. D",
    volume = "101",
    number = "12",
    pages = "126002",
    year = "2020"
}

@inproceedings{Ginsparg:1988ui,
    author = "Ginsparg, Paul H.",
    title = "{Applied Conformal Field Theory}",
    booktitle = "{Les Houches Summer School in Theoretical Physics: Fields, Strings, Critical Phenomena}",
    eprint = "hep-th/9108028",
    archivePrefix = "arXiv",
    reportNumber = "HUTP-88-A054",
    month = "9",
    year = "1988"
}

@book{Edmonds:1955fi,
    author = "Edmonds, A. R.",
    title = "{Angular momentum in quantum mechanics}",
    isbn = "9780691025896",
    publisher = "Princeton University Press",
    year = "1957"
}

@article{PhysRev.93.318,
  title = {Symmetry Properties of the Wigner $9j$ Symbol},
  author = {Jahn, H. A. and Hope, J.},
  journal = {Phys. Rev.},
  volume = {93},
  issue = {2},
  pages = {318--321},
  numpages = {0},
  year = {1954},
  month = {Jan},
  publisher = {American Physical Society},
  doi = {10.1103/PhysRev.93.318},
  url = {https://link.aps.org/doi/10.1103/PhysRev.93.318}
}

@article{Roberts:1998zka,
    author = "Roberts, Justin",
    title = "{Classical 6j-symbols and the tetrahedron}",
    eprint = "math-ph/9812013",
    archivePrefix = "arXiv",
    reportNumber = "G{\&}T-MIGRATION-1999-2",
    doi = "10.2140/gt.1999.3.21",
    journal = "Geom. Topol.",
    volume = "3",
    number = "1",
    pages = "21--66",
    year = "1999"
}

@article{PonzanoRegge,
author = {Ponzano, G. and Regge, T},
title = {Semiclassical Limit of Racah Coefficients},
journal = {Spectroscopy and Group Theoretical Methods in Physics},
pages = {1-58},
year = {1968}
}

@article{Williams:1991cd,
    author = "Williams, Ruth M. and Tuckey, Philip A.",
    title = "{Regge calculus: A Bibliography and brief review}",
    reportNumber = "CERN-TH-6211-91",
    doi = "10.1088/0264-9381/9/5/021",
    journal = "Class. Quant. Grav.",
    volume = "9",
    pages = "1409--1422",
    year = "1992"
}

@book{QuantumTheoryOfAngularMomentum,
author = {Varshalovich, D.A. and Moskalev, A.N. and Khersonskii, V.K.},
booktitle = {Quantum Theory of Angular Momentum},
chapter = {10},
pages = {333-411},
doi = {10.1142/9789814415491_0011}
}

@article{Gharibyan:2018jrp,
    author = "Gharibyan, Hrant and Hanada, Masanori and Shenker, Stephen H. and Tezuka, Masaki",
    title = "{Onset of Random Matrix Behavior in Scrambling Systems}",
    eprint = "1803.08050",
    archivePrefix = "arXiv",
    primaryClass = "hep-th",
    doi = "10.1007/JHEP07(2018)124",
    journal = "JHEP",
    volume = "07",
    pages = "124",
    year = "2018",
    note = "[Erratum: JHEP 02, 197 (2019)]"
}

@article{Cotler:2016fpe,
    author = "Cotler, Jordan S. and Gur-Ari, Guy and Hanada, Masanori and Polchinski, Joseph and Saad, Phil and Shenker, Stephen H. and Stanford, Douglas and Streicher, Alexandre and Tezuka, Masaki",
    title = "{Black Holes and Random Matrices}",
    eprint = "1611.04650",
    archivePrefix = "arXiv",
    primaryClass = "hep-th",
    reportNumber = "SU-ITP-16-19, SU-ITP-16/19, YITP-16-124",
    doi = "10.1007/JHEP05(2017)118",
    journal = "JHEP",
    volume = "05",
    pages = "118",
    year = "2017",
    note = "[Erratum: JHEP 09, 002 (2018)]"
}

@article{Kos:2017zjh,
    author = "Kos, Pavel and Ljubotina, Marko and Prosen, Tomaz",
    title = "{Many-body quantum chaos: Analytic connection to random matrix theory}",
    eprint = "1712.02665",
    archivePrefix = "arXiv",
    primaryClass = "nlin.CD",
    doi = "10.1103/PhysRevX.8.021062",
    journal = "Phys. Rev. X",
    volume = "8",
    number = "2",
    pages = "021062",
    year = "2018"
}

@article{Nosaka:2018iat,
    author = "Nosaka, Tomoki and Rosa, Dario and Yoon, Junggi",
    title = "{The Thouless time for mass-deformed SYK}",
    eprint = "1804.09934",
    archivePrefix = "arXiv",
    primaryClass = "hep-th",
    reportNumber = "KIAS-P18037",
    doi = "10.1007/JHEP09(2018)041",
    journal = "JHEP",
    volume = "09",
    pages = "041",
    year = "2018"
}

@article{ReggeCalc,
author = "Regge, T.",
title = "General relativity without coordinates",
journal = "Nuovo Cim ",
volume = "19",
year = "1961",
pages = "558-571",
doi = "10.1007/BF02733251"}

\end{document}